\documentclass[aps,pra,twocolumn,showpacs,amsmath,amssymb,preprintnumbers,superscriptaddress,10pt,longbibliography]{revtex4-1}

\usepackage{graphicx}
	\graphicspath{{figures/}}
\usepackage{SIunits}
\usepackage{hyphenat}
\usepackage[bookmarks=false]{hyperref}
\usepackage{color}
\usepackage[capitalise]{cleveref}
	\crefname{section}{Sec.}{Secs.}
	\Crefname{section}{Section}{Sections}
\usepackage{xspace}
\usepackage{makecell}
\usepackage{booktabs}
\usepackage{longtable,needspace,hhline}
\usepackage{xfrac}
\usepackage{braket}

\usepackage{array}
\newcolumntype{L}[1]{>{\raggedright\let\newline\\\arraybackslash\hspace{0pt}}p{#1}}
\newcolumntype{C}[1]{>{\centering\let\newline\\\arraybackslash\hspace{0pt}}p{#1}}
\newcolumntype{R}[1]{>{\raggedleft\let\newline\\\arraybackslash\hspace{0pt}}p{#1}}

\newcommand{\loss}[1]{\ensuremath{\alpha_\text{#1}}}

\newcommand{\Hr}{{\textbf{\textcolor{red}{H}}}\xspace}
\newcommand{\Mr}{{\textbf{\textcolor{blue}{M}}}\xspace}
\newcommand{\Lr}{{\textbf{\textcolor{blue}{L}}}\xspace}

\newcommand{\riskev}[2]{\textit{Risk evaluation:}~#1~(#2).}
\newcommand{\suggestion}[1]{\textit{Further suggestions:}~#1}

\definecolor{pink}{RGB}{255,0,255}

\definecolor{ss_color}{rgb}{0.5,0,0.5}

\definecolor{pc_color}{RGB}{0,180,220}

\definecolor{darkorange}{RGB}{255,120,0} 

\definecolor{green}{RGB}{52,201,36}

\definecolor{ginger}{RGB}{255,150,0}

\definecolor{msg}{RGB}{60,179,113}

\definecolor{blue}{RGB}{0,0,205}

\definecolor{darkgreen}{RGB}{0,170,0}

\definecolor{darkred}{RGB}{140,5,0}

\definecolor{eggplant}{RGB}{140,0,80}

\definecolor{RED}{RGB}{255,0,0}

\begin{document}

\title{Preparing a commercial quantum key distribution system for certification against implementation loopholes}

\author{Vadim~Makarov}
\email{makarov@vad1.com}
\affiliation{Russian Quantum Center, Skolkovo, Moscow 121205, Russia}
\affiliation{Vigo Quantum Communication Center, University of Vigo, Vigo E-36310, Spain}
\affiliation{NTI Center for Quantum Communications, National University of Science and Technology MISiS, Moscow 119049, Russia}

\author{Alexey~Abrikosov}
\affiliation{Russian Quantum Center, Skolkovo, Moscow 121205, Russia}
\affiliation{NTI Center for Quantum Communications, National University of Science and Technology MISiS, Moscow 119049, Russia}

\author{Poompong~Chaiwongkhot}
\affiliation{Department of Physics, Faculty of Science, Mahidol University, Bangkok, 10400 Thailand}
\affiliation{Institute for Quantum Computing, University of Waterloo, Waterloo, ON, N2L~3G1 Canada}
\affiliation{Department of Physics and Astronomy, University of Waterloo, Waterloo, ON, N2L~3G1 Canada}
\affiliation{Quantum technology foundation (Thailand), Bangkok, 10110 Thailand}

\author{Aleksey~K.~Fedorov}
\affiliation{Russian Quantum Center, Skolkovo, Moscow 121205, Russia}
\affiliation{QRate, Skolkovo, Moscow 143026, Russia}

\author{Anqi~Huang}
\affiliation{\mbox{Institute for Quantum Information \& State Key Laboratory of High Performance Computing, College of Computer Science} \mbox{and Technology, National University of Defense Technology, Changsha 410073, People's Republic of China}}

\author{Evgeny~Kiktenko}
\affiliation{Russian Quantum Center, Skolkovo, Moscow 121205, Russia}
\affiliation{NTI Center for Quantum Communications, National University of Science and Technology MISiS, Moscow 119049, Russia}
\affiliation{Steklov Mathematical Institute, Russian Academy of Sciences, Moscow 119991, Russia}

\author{Mikhail~Petrov}
\affiliation{Vigo Quantum Communication Center, University of Vigo, Vigo E-36310, Spain}
\affiliation{Russian Quantum Center, Skolkovo, Moscow 121205, Russia}
\affiliation{atlanTTic Research Center, University of Vigo, Vigo E-36310, Spain}
\affiliation{NTI Center for Quantum Communications, National University of Science and Technology MISiS, Moscow 119049, Russia}

\author{Anastasiya~Ponosova}
\affiliation{Russian Quantum Center, Skolkovo, Moscow 121205, Russia}
\affiliation{NTI Center for Quantum Communications, National University of Science and Technology MISiS, Moscow 119049, Russia}

\author{Daria~Ruzhitskaya}
\affiliation{Russian Quantum Center, Skolkovo, Moscow 121205, Russia}
\affiliation{NTI Center for Quantum Communications, National University of Science and Technology MISiS, Moscow 119049, Russia}

\author{Andrey~Tayduganov}
\affiliation{NTI Center for Quantum Communications, National University of Science and Technology MISiS, Moscow 119049, Russia}
\affiliation{QRate, Skolkovo, Moscow 143026, Russia}

\author{Daniil~Trefilov}
\affiliation{Vigo Quantum Communication Center, University of Vigo, Vigo E-36310, Spain}
\affiliation{Russian Quantum Center, Skolkovo, Moscow 121205, Russia}
\affiliation{atlanTTic Research Center, University of Vigo, Vigo E-36310, Spain}
\affiliation{NTI Center for Quantum Communications, National University of Science and Technology MISiS, Moscow 119049, Russia}
\affiliation{\mbox{School of Telecommunication Engineering, Department of Signal Theory and Communications,} \mbox{University of Vigo, Vigo E-36310, Spain}}
\affiliation{National Research University Higher School of Economics, Moscow 101000, Russia}

\author{Konstantin~Zaitsev}
\affiliation{Vigo Quantum Communication Center, University of Vigo, Vigo E-36310, Spain}
\affiliation{Russian Quantum Center, Skolkovo, Moscow 121205, Russia}
\affiliation{atlanTTic Research Center, University of Vigo, Vigo E-36310, Spain}
\affiliation{NTI Center for Quantum Communications, National University of Science and Technology MISiS, Moscow 119049, Russia}
\affiliation{\mbox{School of Telecommunication Engineering, Department of Signal Theory and Communications,} \mbox{University of Vigo, Vigo E-36310, Spain}}

\date{12 October 2024}
	
\begin{abstract}
A commercial quantum key distribution (QKD) system needs to be formally certified to enable its wide deployment. The certification should include the system's robustness against known implementation loopholes and attacks that exploit them. Here we ready a fiber-optic QKD system for this procedure. The system has a prepare-and-measure scheme with decoy-state BB84 protocol, polarisation encoding, qubit source rate of 312.5~MHz, and is manufactured by QRate. We detail its hardware and post-processing. We analyse the hardware for known implementation loopholes, search for possible new loopholes, and discuss countermeasures. We then amend the system design to address the highest-risk loopholes identified. We also work out technical requirements on the certification lab and outline its possible structure.
%
\end{abstract}

\maketitle

\section{Introduction}
\label{sec:intro}

Over the past three decades, quantum key distribution (QKD) has progressed from a proof-of-principle tabletop demonstration \cite{bennett1992b} to commercial deployment in fiber networks in many countries \cite{chen2021,vedovato2021,mehic2020}. Cryptographic systems must ensure reliable and secure operation, and therefore undergo a formal certification procedure \cite{langer2009,alleaume2014}. This involves analysing the system's robustness against known vulnerabilities that exploit the imperfections in its hardware \cite{lo2014,dixon2017,etsi2018,tomita2019,xu2020,sun2022,marquardt2023}. While both national and international certification standards for QKD have been developed \cite{iso23837-2023,a-etsi2021}, the full certification ecosystem for it is not yet established. It's thus a high priority to implement the certification procedures, both by QKD vendors and by future certification labs.


Preparing a QKD system for certification involves (i)~documenting the system in sufficient detail for it to be analysed, (ii)~analysing it, (iii)~patching the security loopholes found \cite{sajeed2021}, and (iv)~proposing the requirements for future certification tests. These four steps should be completed by the developer of the QKD system and possibly involve an external security analysis team. Here we perform them for a commercial system from QRate, utilising the latest developments in vulnerabilities, countermeasures, and security proofs. This is to be followed by (v)~the actual implementation of certification, however in Russia this last step is classified, thus our paper probably constitutes all we can publicly disclose about this system's preparation to it. While Russia follows its own certification standards, we expect the process described here to be broadly similar to preparation of any QKD system for the international certification.

The paper is organised as follows. In \Cref{sec:risk-scale}, we define a risk factor that tells the manufacturer whether a given vulnerability is easily exploitable and thus must be closed by a countermeasure before the system is passed to the formal certification. In \Cref{sec:unified-proof}, we decide how to combine existing security proofs for systems with imperfections. We describe the QKD system under evaluation in \cref{sec:system}, including a fairly detailed disclosure of its optical scheme and post-processing protocol. We discuss every potential vulnerability in this system and possible countermeasures to them in \cref{sec:vulnerabilities} and summarise this initial analysis in \cref{sec:summary}. \Cref{sec:countermeasures} reports how the manufacturer has subsequently addressed the high-risk vulnerabilities. We outline the test capabilities the certification lab should have in \cref{sec:certification} and conclude in \cref{sec:conclusion}.

\section{Risk evaluation scale}
\label{sec:risk-scale}

The company should prioritise patching security issues that are more easily exploitable in practice \cite{sajeed2021}. We thus need to score each issue identified. The cryptography community commonly ranks attacks by their likelihood of success, time and other resources needed to execute them, evaluating these relatively precisely. The recent ISO standard for QKD attempts to follow this practice \cite{iso23837-2023}. It uses a set of factors to evaluate an attack potential that follows the standard evaluation method for security products. However, hacking and security vulnerabilities in QKD remain an active research area, with many relative unknowns and new results published often. For example, no explicit attack scheme is known for many QKD imperfections, but it may appear suddenly. Guessing the precise amount of time and resources required for an exploit is thus very unreliable. Also, we find that the possible standard values currently suggested of each factor are not suitable for QKD yet. All the vulnerabilities we discuss in our paper score as either ``highly resistant to attack'' or ``beyond-high'' in the ISO scale. I.e.,\ developing a working exploit for a vulnerability regarded as easy today in the QKD community still requires a multiple-experts team, longer than six months of work, and bespoke equipment (of which a good example is \cite{gerhardt2011}). It is then difficult to differentiate between the vulnerabilities, as they tend to be off that scale. At the very least, the values of each factor in the ISO standard need to be adjusted before they become applicable to QKD.

Meanwhile, we temporarily adopt an alternative risk evaluation scale that seems to better accommodate research uncertainties and essentially spans the difficulties of exploit higher than the ISO scale. This allows us to compare the risk of vulnerabilities. Our empirical scale is the following. If the security issue has been eliminated or addressed sufficiently well such that it no longer presents a security risk, its overall risk factor is set to `solved'. For those issues where this is not the case, we first evaluate the severity of the issue by three parameters.

\begin{itemize}

\item {\em Loophole likelihood:} How likely is it that the particular loophole exists in the system, according to our present knowledge? If its existence has been confirmed or suspected to be likely, this parameter has value 1. If the loophole is considered to be possible in principle but not very likely (and we have not tested the system yet to find out for sure if the particular imperfection exists), the value is 0. The idea here is that security problems known to be more likely to exist should be more important for the manufacturer to address.

\item {\em Future or current technology:} If the loophole may be exploited with today's technology (i.e.,\ all the components for building a full security exploit can be purchased or easily developed), the value is 1. If it would need future technology that does not exist yet, the value is 0. For example, if the exploit requires an adversary Eve to use 95\% efficient single-photon detectors in her setup, these are available commercially today. If the exploit however requires Eve to use a lossless optical communication line, it is of course possible in principle (physics doesn't prohibit lossless lines) but not available today. In the latter case, Eve would need to wait until 
optical fiber with much lower loss than exists today is produced \cite{fokoua2023}, or until high-quality quantum repeaters are built so that she can use them to implement lossless quantum teleportation over today's lossy optical fibers. Both are long shots for Eve in practice, thus she would not be able to build the exploit today and the security problem is less urgent for the company to address.

The value is also 0 if it is presently not known how to construct an attack.

\item {\em Amount of key leakage:} If the attack provides Eve full or nearly full information about the secret key, the value is 1. If the attack can only provide Eve a minor partial information about the secret key, the value is 0. For example, most intercept-resend type attacks give Eve 100\% (or close to that) information about the secret key \cite{gerhardt2011}. It would then be relatively easy for her to attack a classical cryptographic algorithm that subsequently uses this compromised key. However if the attack only results in the leakage of partial key information, this presents Eve two additional practical challenges. First, she would need to construct her exploit apparatus very carefully such that it works almost perfectly and does not introduce side effects (such as additional errors in the key) that would make Eve's eavesdropped key information zero. Second, she needs to solve a non-trivial classical cryptanalytic task when attacking the classical cryptographic scheme with only partial key information. Although these problems have not been explored, we feel that vulnerabilities that deliver Eve full or nearly full key information should be more urgent for the company to address.

\end{itemize}

We add up the values of the three parameters and evaluate the overall risk factor. If the sum is 0 or 1, the overall risk is low (\Lr); 2, medium (\Mr); 3, high (\Hr). As the reader will see, these three rough risk grades are evenly distributed across today's QKD vulnerabilities, giving us a usable ranking.

\section{Securing the system in the absence of a unified security proof}
\label{sec:unified-proof}

\begin{figure*}
	\includegraphics{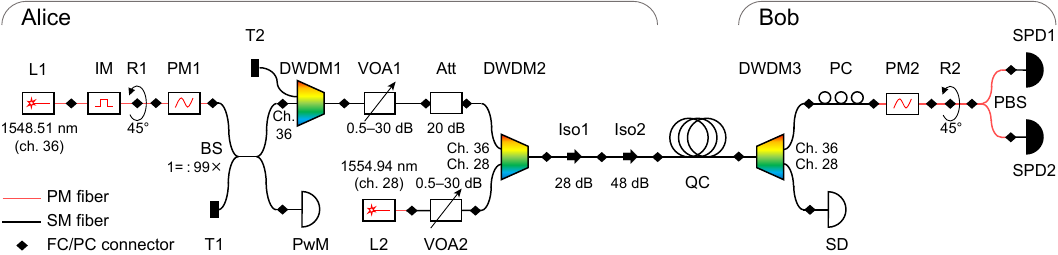}
	\caption{Optical scheme of the QKD system under evaluation. L, lasers [L1:\ Nolatech DFB-1550-5PM; L2:\ Shengshi Optical SWLD-1554.94-FC/PC-05-PM(DFB)]; IM, intensity modulator (iXblue MX-LN-10); R, FC/PC connector with $45\degree$ rotation (custom-made by QRate based on bulkhead adapter Opneti AD-FC/SM-SP04); PM, phase modulator (iXblue MPZ-LN-10); T, optical terminator; BS, beamsplitter (with its splitting ratio noted; Opneti CP-S-P-1x2-1550-1/99-900-1-0.3-FC-3x54); PwM, power meter (Thorlabs PM101 with S154C sensor); DWDM, dense-wavelength-division multiplexer (DWDM1 and DWDM3:\ Opneti DWDM-1-100-36-900-1-0.3-FC; DWDM2:\ Opneti DWDM-1-100-28-900-1-0.3-FC); VOA, variable optical attenuator (Opneti SVOA-B-1550-30-5.2250-1-1-FC); Att, fixed attenuator (Opneti FOA-P-1-20-FC); Iso, polarisation-independent isolator (Iso1:\ Opneti IS-S-P-1550-900-1-0.3-FC-5.5x35; Iso2:\ Opneti IS-D-P-1550-900-1-0.3-FC-5.5x35); QC, quantum channel; SD, synchronisation detector (Fujitsu FRM5W232BS); PC, polarisation controller (General Photonics MPC-4X-7-P-FC/PC); PBS, polarising beamsplitter (Opneti PBS-1x2-P-1550-900-1-0.8-FC); SPD, single-photon detector. Note that the components used in the system at the time of the initial analysis may be replaced with other similar models before the final certification, especially because some of the original components may no longer be available in Russia.
\bigskip}
	\label{fig:setup}
\end{figure*}

At least five attacks in this report require updating the key rate formula according to available security proofs taking each individual attack into account. However, there is no unified security proof that takes into account all these attacks simultaneously and offers a general key rate formula simultaneously accounting for the effects of several attacks. Having said that, we have to mention recent attempts in this direction. In order to take into account various source flaws and side channels, the so-called loss-tolerant QKD protocol was proposed by Tamaki and his coworkers \cite{tamaki2014}. The three-state ($\ket{0_Z},\ket{1_Z},\ket{0_X}$) loss-tolerant protocol with imperfect state preparation is studied together with either intensity fluctuations \cite{mizutani2015,mizutani2019} or Trojan-horse attacks \cite{pereira2019,pereira2020}, and can simultaneously account for correlations among the source pulses \cite{mizutani2019,pereira2020}. Similarly, the four-state ($\ket{0_Z},\ket{1_Z},\ket{0_X},\ket{1_X}$) loss-tolerant protocol with imperfect state preparation has been investigated as well, combined with the Trojan-horse attacks \cite{pereira2023,curras-lorenzo2023} and correlations among the source pulses \cite{pereira2023}. One has to point out that only a single-photon source is considered in \cite{pereira2019,pereira2020,pereira2023,curras-lorenzo2023}. The ``standard'' decoy-state BB84 protocol with four encoding states and three intensities is also studied in \cite{molotkov2020,molotkov2021,sun2021}. There, the Trojan-horse attack is considered along with the vulnerabilities of detector backflash \cite{molotkov2020} or detector efficiency mismatch \cite{molotkov2021}, and multiple source imperfections are treated together with the detector efficiency mismatch \cite{sun2021}. However, several new attacks, such as light injection and induced photorefraction, seem to be not included in the security proofs, which is a subject for future research. Thus, we conclude that no complete security proof currently exists that takes into account all the potential imperfections and side channels we list in \cref{sec:vulnerabilities}. Deriving such security proof is an open academic question, and a very non-trivial one.

Without this theoretical treatment, we are in the realm of guessing. We still, however, need to make a practical decision how to treat these vulnerabilities in QRate's system. Our first idea was to sum algebraically the key rate corrections owing to the different vulnerabilities. We discussed this idea with theoreticians and, while they conceded it might turn out to be approximately correct, no one really liked it.

Our second idea is to use hardware countermeasures (filters, isolators, etc.)\ to minimise the key rate reduction of \emph{each and every vulnerability considered alone} to a negligible level. This means that for every individual vulnerability for which a security proof is available, the hardware is characterised, then reinforced and improved until the proof gives a very small correction to the key rate and maximum transmission distance comparing to the case of a perfect hardware. We arbitrarily suggest that a reduction of the secret the key rate by less than 1\% can be considered negligible. We deem it exceedingly unlikely that Eve could gain any advantage of this, because she would have to implement a complex and perfect optimal attack on the entire key exchange to access this 1\% information in the already-privacy-amplified key, which is unrealistic. An incidental advantage of this is that the key rate formula for the perfect hardware can be used in the system.

\begin{figure*}
	\includegraphics{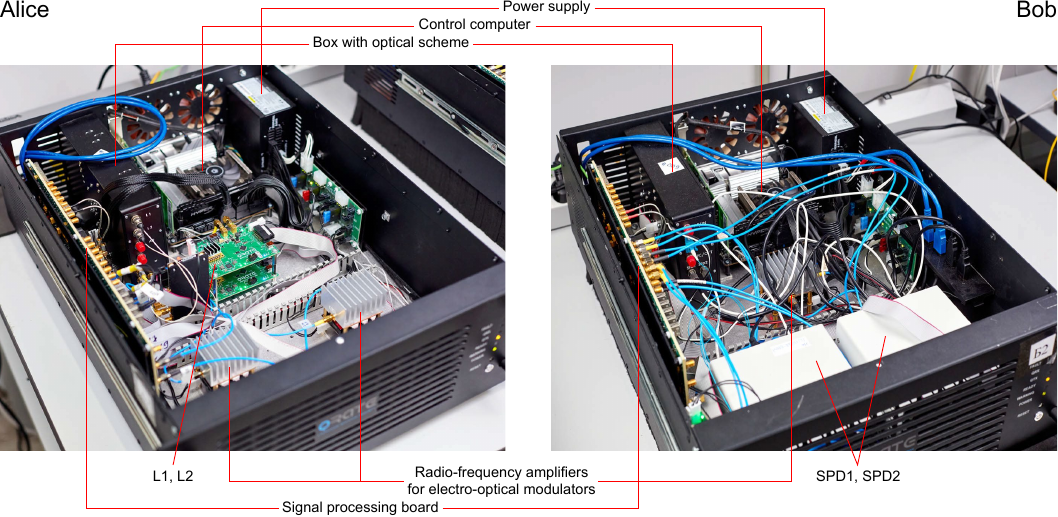}
	\caption{Quantum key distribution system under evaluation (a prototype built in 2021), with covers removed from Alice and~Bob.}
	\label{fig:setup-photo}
\end{figure*}

We hope that the latter approach turns out to be robust, and suggest to use it to claim the QRate's system is secure. Again, there is no strict proof of that, but this is a reasonable best-practice approach we can currently do. The rest of this report adopts this approach.

\section{System under evaluation}
\label{sec:system}

The QKD system we study is an industrial prototype under development at QRate. It has a prepare-and-measure scheme and uses a decoy-state Bennett-Brassard 1984 (BB84) protocol with polarisation-encoded states at approximately $1550~\nano\meter$ wavelength and $312.5~\mega\hertz$ clock rate. The optical scheme is shown in \cref{fig:setup} and photos in \cref{fig:setup-photo}. Further details can be found in a Russian-language Ph.D.\ thesis \cite{duplinskiy2019}; see also \cite{duplinskiy2017,duplinskiy2018}.

The system manufacturer has also provided us a Design specification sheet of the overall scheme (dated 2018-11-07) that contains a high-level description of the hardware and software structure, as well as later documents on extensive changes and updates made by the end of 2021. We have received further oral information and written notes on various aspects of design and manufacturing from the company engineers. At this evaluation stage, we have not yet tested the system hardware for most vulnerabilities (with a few exceptions that will be noted through the text).

The system software uses the post-processing procedure containing the following standard steps.
\begin{enumerate}
	\item {\it Sifting}. Bob announces the positions of registered pulses and their measurement bases. Alice announces whether her basis matches and they discard all events with incompatible bases. For matching bases, Alice also announces the type of each pulse (signal or decoy).
	\item {\it Information reconciliation}. Alice's sifted key is taken as a reference, while Bob attempts to find and eliminate the discrepancies (errors) between the keys. For this purpose the low-density parity-check (LDPC) codes are used \cite{kiktenko2017,borisov2022}.
	\item {\it Verification and parameter estimation}. The identity of error-corrected keys is verified using a modified PolyP32 hash function \cite{krovetz2001,fedorov2018}.
	\item {\it Estimation of the level of eavesdropping}. Bob estimates the total amount of information leaked to Eve during the quantum phase and the previous post-processing steps, and computes the secret key length $\ell_\text{sec}$ \labelcref{eq:l_sec} according to the model from \cite{trushechkin2017} taking into account the finite-key-size effects. If $\ell_\text{sec}\leq0$, Alice and Bob abort the protocol and proceed to the next generated raw key block.
	\item {\it Privacy amplification}. In order to get rid of Eve's residual information about the verified key, it is compressed using a 2-universal hash function from the Toeplitz family \cite{krawczyk1994,krawczyk1995}. As a result, Alice and Bob obtain a common shorter key of length $\ell_\text{sec}$, Eve's information about which is now negligible.
\end{enumerate}
A more detailed description of the post-processing is given in \cref{sec:post-processing}.

\begin{table*}
	\vspace{-0.7em} 
	\caption{{\bf Summary of potential security issues in QRate \boldmath{$312.5~\mega\hertz$} QKD system found at its initial evaluation (completed in January 2022).} $Q$,~system implementation layers involved (see \cite{sajeed2021} or \cref{sec:Q-reprint}). Risk evaluation lists the overall assessment and the values of the three parameters (see \cref{sec:risk-scale}).}
	\label{tab:attacks}
	\footnotetext[1]{All the high-risk issues identified have been addressed by QRate before publication of this report, see \cref{sec:countermeasures}.}
	\footnotetext[2]{Issue added in mid-2022 when we learned about a recent study \cite{ye2023}.}
	\begin{tabular}[t]{L{24mm}L{10mm}L{16mm}L{107mm}L{15mm}}
	\hline\hline
	\makecell{Potential\\ security issue} & 
	\makecell{$Q$} & 
	\makecell{Target\\ component} & 
	\makecell{Action recommended to the company} & 
	\makecell{Risk\\ evaluation} \\
	\hline
Choice of QKD protocol & Q5 & Protocol & None. & Solved \\

Superlinear detector control & Q1--5,7 & SPDs & The development of the photocurrent-measurement countermeasure should continue at the company. It should be tested in our lab. & \Hr (1,1,1)\footnotemark[1]\\

Detector efficiency mismatch & Q1--5 & SPDs, Bob's PM & Update the key rate equation. Spectrally characterise Bob's components. Discuss countermeasures to timing attacks. & \Hr (1,1,1)\footnotemark[1] \\

Detector deadtime & Q1,2,5 & SPDs & Supplement the hardware simultaneous deadtime with implementing it in post-processing. & \Hr (1,1,1)\footnotemark[1] \\

Trojan-horse & Q1,2 & Alice's optics & Characterise Alice's components in a wide spectral range. Install additional isolators and, possibly, spectral filters. & \Lr (0,0,0) \\

Laser seeding & Q1,2 & Laser & None. & Solved \\

Light injection into Alice's power meter & Q1--3 & IM & Characterise Alice's components in a wide spectral range. Install additional isolators and, possibly, spectral filters. & \Lr (1,0,0) \\

Induced photorefraction\footnotemark[2] & Q1--3 & Alice's IM and PM & Characterise Alice's components in a wide spectral range. Optical measurements should be done in our lab. & \Mr (0,1,1) \\

Laser damage & Q1 & Alice's \& Bob's optics & Install an additional sacrificial isolator at Alice's exit. & \makecell[tl]{\Mr (1,0,1)\\ \Mr (0,1,1)} \\

APD backflash & Q1,2 & SPDs & Characterise Bob's components in a wide spectral range. Measure backflash photon emission probability of the SPD. & \Mr (1,1,0) \\

Intersymbol interference & Q1--3 & Alice's active components & Optical measurements should be done in our lab. & \Lr (1,0,0) \\

Imperfect state preparation & Q1--3,5 & Alice's optics & Optical measurements should be done in our lab. & \Lr (1,0,0) \\

Calibration via channel Alice--Bob & Q1--5 & SPDs, IM, PM & The analysis team did not know how to solve this and proposed to discuss with QRate. QRate has subsequently found solutions acceptable for the manufacturing process, see \cref{sec:countermeasures}. & \Hr (1,1,1)\footnotemark[1] \\

Quantum random number generator & Q5 & Protocol & Implement the quantum random number generator and integrate it into the system. & \Lr (1,0,0) \\

Compromised supply chain & All & Any & Learn mitigation strategies from the national cryptography licensing authority. & \Mr (0,1,1) \\
	\hline\hline
	\end{tabular}
\end{table*}

\section{Potential vulnerabilities}
\label{sec:vulnerabilities}

In order to simplify the task of security evaluation and for ease of understanding, we have subdivided the system implementation into several layers according to the hierarchical order of information flow \cite{sajeed2021} (recapped in \cref{sec:Q-reprint}). In this work, we perform a complete security analysis of the bottom four layers (Q1--Q4) that correspond to optics, analog electronics, driver and calibration algorithms, and operation cycle of the system. For these layers, we aim to examine all suspected implementation security issues according to the current knowledge. For higher layers Q5 and up (from QKD protocol post-processing and up), we cannot perform a complete security evaluation as they lay outside the expertise of most of the authors. Nevertheless, we point out a few issues in the layer Q5 and we include its fairly detailed description in \cref{sec:post-processing} to aid any independent analysis.

Based on the information received about the system, we have identified a number of potential security issues that might be exploitable by Eve. A summary of them is given in \cref{tab:attacks}. Note that QRate has subsequently addressed all the high-risk issues, as detailed later in \cref{sec:countermeasures}. We now explain the identified issues.

\subsection{Choice of QKD protocol}
\label{sec:protocol-choice}

The choice of the QKD protocol and scheme is one of the most important decisions a designer makes. It affects the product through its lifetime.

QRate has chosen the best understood and most widely studied scheme and protocol: the prepare-and-measure (one-way) scheme and BB84 protocol with decoy states of three intensities (vacuum, decoy, signal). This choice has the advantage that complete general security proofs are available that have been widely scrutinised for correctness. An additional advantage crucial for our analysis is that modifications of these security proofs that take into account various hardware imperfections are often also available. We cite these through this report.

We remark that not every company has made the same choices. Sometimes the motivation of developing its own intellectual property prevails and a less studied protocol that lacks the general security proof is chosen. This often raises questions. For instance, the excellent current commercial QKD system by ID~Quantique (Switzerland) \cite{idq-Cerberis-XG,walenta2014} implements a coherent-one-way protocol \cite{stucki2005}. This protocol lacked the general proof at the time of its initial commercialisation in 2014. Subsequently, quantum attacks on the original coherent-one-way protocol have been discovered that severely limit the key rate and communication distance \cite{trenyi2021}. Although the latest version of the system \cite{idq-Cerberis-XG} uses a modified protocol, the general security proof for it is also not available. In other examples, the subcarrier-wave QKD system being commercialised by Smarts-Quanttelecom (Russia) \cite{sajeed2021} still has its security proof in development \cite{gaidash2022,kozubov2019}. For the system with a geometrically-uniform-coherent-states QKD protocol \cite{kulik2007} commercialised by Infotecs (Russia) \cite{infotecs-Quandor}, the integrity of its security proof is being debated in the scientific community \cite{kronberg2023}. Most partial security proofs available for these systems do not incorporate device imperfections, which may further hinder their analysis.

\textit{Risk evaluation:} Solved.

\suggestion{None. Decoy-state BB84 protocol is the safest available choice for the prepare-and-measure QKD scheme.}

\subsection{Superlinear detector control}
\label{sec:detector-control}

Superlinear detector control attacks are based on three phenomena. First, most single-photon detectors (SPDs) are threshold detectors, which means that they cannot resolve the number of photons in a pulse. When they produce a detection event, called a click, they do not distinguish whether it has been caused by one or multiple photons. Second, the SPD's detection efficiency of multiphoton pulses may exhibit a so-called superlinearity effect \cite{lydersen2011b}. SPDs are usually characterised by their quantum efficiency $\eta$, which is the probability to detect a single photon ($\eta\sim10\%$ for QRate's SPD). For a multiphoton pulse the detection probability can be estimated as
\begin{equation}
	\label{eq:detector-linear-expectation}
	p_\text{det}(n)=1-(1-\eta)^n,
\end{equation}
where $n$ is the number of photons in the pulse. An SPD whose multiphoton detection probability is higher than \cref{eq:detector-linear-expectation} exhibits superlinear behavior. The third phenomenon is a threshold level shift, which is the ability of the detector to reduce its quantum efficiency partially or completely to zero. Engineers exploit the latter effect in a gated regime to decrease the detector's dark count rate \cite{cova2004}. The reverse-bias voltage at an avalanche photodiode is lowered between the gates, so that the detector is insensitive to single photons ($\eta=0$) in between the gates. It then behaves as a normal photodiode and may only respond to bright light pulses at this time, with a classical threshold on the pulse energy \cite{wiechers2011}.

Several attacks that exploit these and other phenomena in the SPDs have been developed and multiple countermeasures to them have been proposed. This is arguably the most difficult group of vulnerabilities in today's QKD. For readers not familiar with these developments, we survey them in \cref{sec:superlinear-detector-control-survey}.

\textit{Features of the QKD system under analysis:} The detection system is developed by QRate with the use of the avalanche photodiode (APD) PGA-025u-1550TF based on InGaAs/InP structure from Princeton Lightwave. From our discussion with QRate's engineers, we have found that no measures have been taken to prevent the superlinearity detector control attacks. As our preliminary detector tests show, the detector is blinded with continuous-wave (cw) light of $3~\micro\watt$ ($-25~\deci\bel\milli$) power. It allows total control at $250~\micro\watt$ ($-6~\deci\bel\milli$) blinding power and trigger pulse energies $E_\text{never} = 12~\femto\joule$ and $E_\text{always} \gtrsim 22~\femto\joule$ (see \cref{sec:test-of-detector-control}).

In QRate implementation shown in \cref{fig:setup}, synchronisation detector (SD) can be used as a watchdog (see Countermeasure \labelcref{item:Add-a-watchdog} in \cref{sec:superlinear-detector-control-survey}). However we think this would be a bad idea, for the following reasons. First of all, the SD is not sufficiently sensitive, its threshold starting at a few microwatt ($-20$ to $-30~\deci\bel\milli$) level. The presence of demultiplexer (DWDM3) adds about $35~\deci\bel$ to this level (at the particular wavelength of $1548.5~\nano\meter$). Secondly, the SD's sensitivity can be controllably reduced by the laser damage attack \cite{bugge2014,makarov2016}. Thirdly, putting extra functionality on the SD would complicate synchronisation routines that are already far from perfect (see \cref{sec:efficiency-mismatch}).

A more promising approach is to add a photocurrent measurement to the SPDs (see Countermeasure \labelcref{item:Observe-the-observer} in \cref{sec:superlinear-detector-control-survey}). QRate has implemented this measurement at a stand-alone sinusoidally-gated SPD, but haven't integrated it as a countermeasure into the system yet. Our preliminary tests of this implementation in a setup from \cref{sec:test-of-detector-control} show a countermeasure readout (roughly proportional to a logarithm of averaged APD photocurrent) of $400$--$1200$ arbitrary units under single-photon pulses, depending on the count rate. Under the blinding attack, the readout is $2100$--$2400$ arbitrary units. There is a clear separation between the normal operation and blinding, which is encouraging. However this countermeasure needs to be tested with a pulsed blinding \cite{gras2020,wu2020,bulavkin2022} and be fully integrated into the QKD system. We treat this problem further in \cite{acheva2023,kuzmin2023a}.

The after-gate attack is probably possible in the current implementation, especially given that Eve may control the timing synchronisation inside Bob and that Bob registers clicks with a coarse $3.2$-$\nano\second$ resolution corresponding to one bit period (\cref{sec:efficiency-mismatch}). One possible countermeasure would be to make the phase modulator pulse shorter than the detector gate, i.e.,\ shorter than $400$ to $800~\pico\second$. This however can be difficult to implement and may lead to less accurate state preparation (see \cref{sec:state-preparation}). Another possible countermeasure is a precise click time measurement, however the detector jitter and timing drift may make this difficult to implement.

\riskev{\Hr}{1 vulnerability is likely exploitable, 1 with current technology, 1 might give Eve high key information}
 
\suggestion{We suggest the company to finish the implementation of the photocurrent-measurement countermeasure (which has been built and tested preliminarily). Our lab will test it in a stand-alone detector against the blinding, after-gate, and falling-edge attacks. We will then possibly repeat the tests in the complete QKD system. This should, at least, allow the company to claim that the system is protected against the detector blinding attack.

If the detector's vulnerability to the after-gate and falling-edge attacks is experimentally confirmed, countermeasures against them would require a discussion with QRate engineers. We are unsure what solutions are practical given the high time precision and calibration requirements on the PM pulse.

Developing a measurement-device-independent or twin-field commercial system \cite{lo2012,lucamarini2018,wang2018,wang2022,fan-yuan2022} is a radical alternative that may be considered, as this would remove all the detector vulnerabilities. However this is a major business decision influenced by many factors.}

We remark that the work currently progresses according to the above suggestions, as detailed in \cref{sec:countermeasures,sec:certification}. This note applies to every potential vulnerability from here on.

\subsection{Detector efficiency mismatch}
\label{sec:efficiency-mismatch}

In a theoretical security proof it is assumed that Bob's SPDs are identical \cite{rice2009}. For real-world SPDs that are not identical, there are three possible mismatches that have to be included in the security proof.
\begin{enumerate}

\item Static efficiency mismatch. The average photon detection probability in the SPDs is $10\%$ \cite{duplinskiy2019}. If we assume that one SPD has $9\%$ efficiency and another $11\%$, the probability ratio of bits detected would be $45\!:\!55$ instead of $50\!:\!50$, with no Eve's influence. This asymmetry gives Eve some {\em a priori} information about the raw key. While the current QRate's firmware ignores this issue and assumes the equal $50\!:\!50$ probabilities, the company plans to update the key rate equation to one that takes into account unequal static probabilities, according to the security proof \cite{bochkov2019,trushechkin2022}.
	
A simpler solution that does not require the modification of \cref{eq:l_sec} is the ``four-state measurement'' scheme, originally proposed as a countermeasure against the time-shift attack \cite{qi2007}. Bob randomly chooses not only his basis but also bit-0 and bit-1 assignment of his detectors. In such setup, even if Eve has the information about which detector clicks, she still does not know Bob's bit value since she is not aware about which detector corresponds to bit-0. During the sifting communication rounds Bob announces the bit positions when the detectors were ``swapped'', and Alice performs a bit-flip in respective positions on her side. In this way, the distribution of zeros and ones becomes uniform. The potential loophole of the four-state measurement method is that Eve may try to read out Bob's detector assignments by injecting a strong pulse like in the Trojan-horse attack \cite{vakhitov2001,jain2014,sajeed2017}.

\item Time mismatch. The detectors are sinusoidally-gated and are sensitive to single photons for about $800~\pico\second$ out of the $3.2~\nano\second$ gate period. Any gated detectors are likely vulnerable to time-shift attacks (TSA) \cite{makarov2006,zhao2008}.

\item Wavelength mismatch. Characteristics of all optical components depend on the wavelength, which often leads to loopholes. On Bob's side an attack is in principle possible using wavelength dependence of the polarising beamsplitter (PBS) and SPDs \cite{li2011a,huang2015}. A combination of this attack with other attacks should also be considered. Spectral characterisation of Bob's components that is necessary for further study of this attack is discussed in \cref{sec:spectral-testing}.

\end{enumerate}

\riskev{\Hr}{1 vulnerability is likely exploitable, 1 with current technology, 1 might give Eve high key information}

\suggestion{Although the security proofs \cite{bochkov2019,trushechkin2022} derive the key rate equation that accounts for the static efficiency mismatch, they are not applicable to the mismatch in the time and wavelength domain that Eve can dynamically control. This leaves us with the only realistic option to solve this problem by implementing a four-state Bob (i.e.,\ Bob who randomly swaps or not swaps his detectors' assignment to bit values 0 and 1 by applying or not applying an additional $\pi$ phase shift at his PM \cite{qi2007}). This eliminates all the efficiency mismatches and corresponding corrections to the key rate equation. However, Bob then needs to additionally guarantee a certain amount of isolation against the Trojan-horse attack on him, which becomes necessary because the detectors' assignment has to remain secret. A security proof that estimates the required amount of the latter isolation is not available in the literature. It needs to be developed and \cref{eq:E_1_u} amended by including a Trojan-horse leakage term.

We remind that measurement-device-independent and twin-field QKD systems do not suffer from the detector vulnerabilities. They may be considered as an alternative solution.}

\subsection{Detector deadtime attack}
\label{sec:deadtime}

The security proof requires that both Bob's detectors are sensitive to photons when Bob registers a click. If one detector remains sensitive and clicks from it are accepted as valid while the other detector is having a deadtime, an attack becomes possible \cite{weier2011}.

\begin{figure}
	\includegraphics{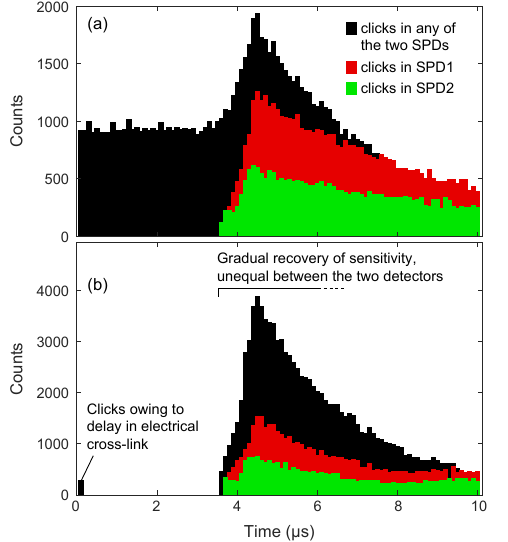}
	\caption{Test of simultaneous deadtime of Bob's two SPDs (performed by QRate). The histograms show click rate versus time after a click (a)~in independent detectors without simultaneous deadtime and (b)~in interlinked detectors with simultaneous deadtime. The detectors are illuminated with Alice's light typical for QKD operation.}
	\label{fig:deadtime-test}
\end{figure}

In QRate's system, whenever one detector clicks, a simultaneous deadtime of about $4.5~\micro\second$ is introduced to both detectors, via electrical cross-links between the detector units \cite{wiechers2011}. \Cref{fig:deadtime-test} shows the effect of the simultaneous deadtime on cross-correlation between the detectors' clicks. While for the detector that has clicked the deadtime begins instantly, the other detector starts it a few $3.2$-$\nano\second$ gating periods later, owing to the delay in the electrical cross-link. We thus see a few cross-clicks early in the deadtime, in which only one detector remains sensitive to single photons. This would present a loophole if these clicks are accepted into the raw key \cite{makarov2006}. The gradual recovery from the deadtime is also uneven between the detector units, with a significant efficiency mismatch visible in the time range starting at $3.4~\micro\second$ and extending roughly to $6$--$9~\micro\second$. Similarly, this may leave Eve possibilities to construct attacks.

\riskev{\Hr}{1 vulnerability is likely exploitable, 1 with current technology, 1 might give Eve high key information}

\suggestion{Implementing the simultaneous deadtime precisely in post-processing should be sufficient to close this vulnerability. In the case of QRate system, this would supplement the hardware deadtime that already prevents the majority of the unusable clicks, thus reducing their impact on the key rate. The software should then discard all clicks that occur fewer than a fixed number of gates after any click in either detector (corresponding to at least $6~\micro\second$, exact time to be determined by a more accurate cross-correlation measurement). Note that if a click is being discarded, it also renews the discard time period. This should close this vulnerability.

Here we assume that the system does not implement the four-state Bob countermeasure (\cref{sec:efficiency-mismatch}). If it does, the detector deadtime attack should be re-evaluated.}

\subsection{Trojan-horse attack}
\label{sec:THA}

In this section, we consider the Trojan-horse attack (THA) on Alice's phase and intensity modulators \cite{vakhitov2001,gisin2006,jain2014}. In this attack, Eve attempts to read Alice's IM and PM settings by injecting light, called Trojan photons, into her apparatus. The outbound photons that have passed Alice's PM and IM will thus contain the secret information about the phase and intensity encoded into them. There are several security proofs for the decoy-state BB84 protocol that take this information leakage into account \cite{lucamarini2015,tamaki2016,wang2018a}. Here we use the latest proof to calculate the required isolation values in the finite-key regime \cite{wang2018a}. For this, we need to upper-bound the intensity (conventionally called `intensity' in QKD but actually meaning energy) of the leaked signals
\begin{equation}
	\label{eq:Imax}
	I_\text{max} = 10^{-\frac{\loss{A}}{10}} I_\text{in},
\end{equation}
where $\loss A$ is the total loss of the Trojan photons in Alice (in decibel) and
\begin{equation}
	\label{eq:IinValue}
	\begin{aligned}
	I_\text{in} &= \frac{W_\text{in}}{f_p}\frac{\lambda}{hc} = \frac{100~\watt}{312.5~\mega\hertz}\times\frac{1550~\nano\meter}{1.99\times10^{-25}~\joule\cdot\meter}\\
	&= 2.5\times10^{12} \text{ photons per pulse},
	\end{aligned}
\end{equation}
where $f_p$ is the qubit repetition rate and $W_\text{in}$ the maximum optical power that can be transmitted through the standard telecommunication optical fiber (assumed here to be $100~\watt$). To estimate $\loss A$ we need to know the losses inside Alice; her component parameters are given in \cref{tab:component-loss}. Taking into account that the Trojan photons pass each component twice, we obtain
\begin{align}
	\label{eq:AliceTrojanLoss}
	\loss A = &~ 2 \, \big(	\loss{IM} + \loss{PM1} + \loss{BS} + \loss{DWDM1} + \loss{VOA1} \nonumber \\
						&	+ \loss{Att} + \loss{DWDM2} \big)	+ \loss{Iso1rev} + \loss{Iso2rev} \nonumber \\
						& + \loss{Iso1forw} +\loss{Iso2forw}.
\end{align}
This formula incorporates the following assumptions.
\begin{enumerate}

\begin{table}
	\vspace{-0.7em} 
	\caption{{\bf Optical insertion loss \boldmath{$\alpha$} of system components,} in the quantum signal path (L1--QC--SPDs), at the system operating wavelength of $1548.51~\nano\meter$. The values are taken from component data sheets. The values at other wavelengths are not specified and may differ considerably. Connector loss (typically $0.3~\deci\bel$) is neglected.}
	\label{tab:component-loss}
	\begin{tabular}[t]{L{20mm}@{\hspace{5.5mm}}C{10mm}C{10.5mm}L{20mm}@{\hspace{5.5mm}}C{10mm}}
	\cmidrule[0.4pt]{1-2}\cmidrule[0.4pt]{4-5}
	\morecmidrules
	\cmidrule[0.4pt]{1-2}\cmidrule[0.4pt]{4-5}
	\makecell{Alice's\\ component} & \makecell{$\alpha$ ($\deci\bel$)} && \makecell{Bob's\\ component} & \makecell{$\alpha$ ($\deci\bel$)} \\
	\cmidrule[0.4pt]{1-2}\cmidrule[0.4pt]{4-5}
	IM & 2.7 && DWDM3 & 1 \\
	PM1 & 2.5 && PC & 0.05 \\
	BS & 20 && PM2 & 2.5 \\
	DWDM1 & 1 && PBS & 0.5 \\
	VOA & 0.5--30 \\
	Att & 20 \\
	DWDM2 & 1 \\
	Iso1 reverse / forward & 28 / 0.35 \\
	Iso2 reverse / forward & 48 / 0.4 \\
	\cmidrule[0.4pt]{1-2}\cmidrule[0.4pt]{4-5}
	\morecmidrules
	\cmidrule[0.4pt]{1-2}\cmidrule[0.4pt]{4-5}
	\end{tabular}
\end{table}

\item Here we assume that Eve's Trojan photons have $1548.51~\nano\meter$ (i.e.,\ channel~36) wavelength. Thus both DWDMs have $1~\deci\bel$ insertion loss and the insertion loss values of the other components can be taken from their data sheets. This allows us to make a quick estimate but is in no way sufficient to treat this vulnerability \cite{jain2015,sajeed2017}. Eve is, of course, not limited to this wavelength. She may use any other wavelength if the combined loss at it is lower. None of the components in the QRate system have been characterised in a sufficiently wide spectral range. This data is never available from the component manufacturers, because it is not needed for normal applications, not measured, and not guaranteed. We must thus perform a wide spectral characterisation of all the components ourselves in $\sim 350$--$2400~\nano\meter$ range (see \cref{sec:spectral-testing}), then find the minimum of $\loss A$ over this entire spectral range.

\item The Trojan pulses experience losses and reflections from different surfaces {\em behind} the IM. However, Eve might manipulate the phase and delay of each consecutive pulse such that the reflections from each of those surfaces arrive at IM in-phase at the same time \cite{chaiwongkhot2024,a-etsi2021}. Those pulses will interfere constructively, resulting in the total photon number passing through the IM being much higher than a mere sum of individual reflections. This effective reflectance depends on the number of reflective surfaces Eve could exploit. Although measuring individual reflections that are widely spaced apart is possible \cite{gisin2006,jain2014,lucamarini2015,sushchev2021}, a general characterisation technique that takes into account closely spaced reflections is complex and not yet proven \cite{a-etsi2021}. Also the individual reflections might be wavelength-dependent, which further adds to the challenge. It is much easier and safer to adopt a conservative assumption that all the photons behind the IM are reflected back \cite{chaiwongkhot2024,a-etsi2021}. Thus all the losses behind the IM are neglected.

\item The variable optical attenuator (VOA) can be set anywhere in the range $0.5$--$30~\deci\bel$. It might be used at the lower-attenuation end of the range during QKD, according to QRate. We thus assume here the worst case with the minimum attenuation of $0.5~\deci\bel$.
	
\item We neglect the loss in FC/PC connectors, which can typically be $0.3~\deci\bel$ per connection.
	
\item Eve can attempt to change the attenuation characteristics of the last isolator (Iso2) by the laser-damage attack (see \cref{sec:laser-damage}). Here we do not consider this.

\end{enumerate}

\begin{figure}
	\includegraphics{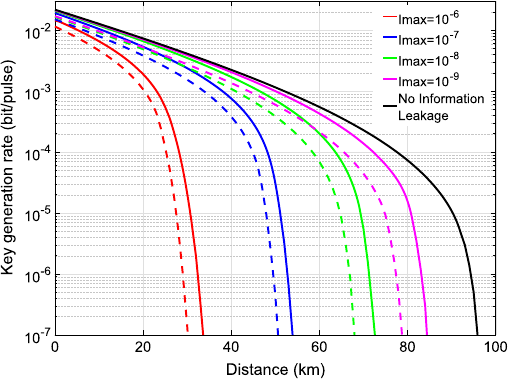}
	\caption{Secret key rates for different leaked intensities, for a typical QKD system (not QRate's), reprinted from \cite{wang2018a}. The total number of transmitted pulses $N=10^{12}$.}
	\label{fig:WangKeyRates}
\end{figure}

Combining \cref{eq:Imax,eq:IinValue,eq:AliceTrojanLoss} with the data from \cref{tab:component-loss} we obtain $\loss{A} \approx 172~\deci\bel$, $I_\text{max} \approx 1.5 \times 10^{-5}$ photons per pulse. We can quickly estimate an expected key rate by looking at the plots from \cite{wang2018a} calculated for a typical QKD system with slightly different parameters in the finite-key-size regime (\cref{fig:WangKeyRates}). Our high value of $I_\text{max}$ is not in the plots and leads to zero key rate at most distances. We can roughly estimate that an additional isolation of more than $40~\deci\bel$ is needed to approach the ideal case with no information leakage (i.e.,\ $I_\text{max} \lesssim 10^{-9}$).

\riskev{\Lr}{0 loophole is unlikely to be present taking into account the very conservative assumptions in our calculation, 0 requires research and possibly future technology to exploit because a complete low-reflectance Trojan-horse attack on the source has not been demonstrated, 0 probably gives Eve low key information}

\suggestion{First, it is necessary to characterise Alice's optical components in the wide spectral range (\cref{sec:spectral-testing}), to determine the minimum total loss $\loss{A}$ over the entire wavelength range accessible to Eve. Then, a more accurate calculation of the key rate \cite{wang2018a} should be done with our actual system parameters and the finite key size. Given our strategy of reducing the information leakage under each individual imperfection to a negligible value (\cref{sec:unified-proof}), an acceptable key rate reduction threshold should be set arbitrarily (e.g.,\ no less than 0.9 of the ideal-case rate) and additional isolators and, possibly, spectral filters should be installed in Alice to guarantee it. Note that the key rate curves always diverge near the maximum transmission distance (\cref{fig:WangKeyRates}); this means that a restriction should be implemented in the system software to prevent operation close to the distance limit.}

\subsection{Laser-seeding attack}
\label{sec:source-laser-seeding}

From the quantum channel, Eve might be able to inject light into Alice's laser diode (L1) and modify its emission characteristics, e.g.,\ phase, intensity, and wavelength \cite{sun2015,huang2019,pang2020,lovic2023}. According to previous research, the injection power reaching the connector of Alice's laser should be in the milliwatt range (assuming the laser has a built-in optical isolator) \cite{huang2019} or nanowatt range (assuming the laser without the built-in isolator) \cite {lovic2023}. Similarly to \cref{sec:THA}, we assume the laser power entering Alice is $100~\watt$. The loss in Alice for the laser-seeding attack
\begin{align}
\label{eq:source-laser-seeding-loss}
	\loss{As} = &~ \loss{IM} + \loss{PM1} + \loss{BS} + \loss{DWDM1} + \loss{VOA1} + \loss{Att} \nonumber \\
							& + \loss{DWDM2} + \loss{Iso1rev} + \loss{Iso2rev} = 123.7~\deci\bel.
\end{align}
The continuous-wave power reaching L1 $W_\text{L1} = 10^{-\sfrac{\loss{As}}{10}} W_\text{in} \approx 40~\pico\watt$, which is already orders of magnitude lower than the power needed for the successful hacking. Note that additional isolation will be added to Alice to protect her against the Trojan-horse attack. With this large margin, we consider this vulnerability to be eliminated.

\emph{Risk evaluation:} Solved.

\suggestion{None.}

\subsection{Light injection into Alice's power meter}
\label{sec:AD}

The Alice's internal power meter (PwM, see \cref{fig:setup}) is used to maintain the working point of her intensity modulator (IM; iXblue MX-LN-10). This intensity modulator internally consists of a Mach-Zehnder interferometer with a fast-modulation section and bias section in its arms. The zero point of the interferometer drifts over time and requires compensation by applying a static voltage at the bias section. The power meter indirectly measures the deviation from the zero point, by measuring an average power of the mix of vacuum, decoy, and signal states emitted by Alice in the normal QKD operation. If the zero point drifts, this power deviates from a factory-preset value, which lies in the range of $2$--$5~\micro\watt$. The difference acts on the bias voltage via a slow negative-feedback loop implemented in the system software. Currently the PwM is implemented with Thorlabs PM101 power meter with S154C sensor. The company plans to replace it with a discrete photodiode and their own current measurement circuit.

Injecting additional light into the PwM externally would thus cause the IM's working point to be set improperly. This would change the intensities of vacuum, signal, and decoy states, as well as their ratios. Notably the intensity of the vacuum state would be increased. This may lower the actual secure key rate below that calculated by the system.

Let's roughly estimate how much power Eve might inject in the current system at its operating wavelength of $1548.51~\nano\meter$, similarly to \cref{sec:THA}. The loss in Alice
\begin{align}
	\label{eq:AliceADLoss}
	\loss{Ap} = &~ \loss{DWDM1} + \loss{VOA1} + \loss{Att} + \loss{DWDM2} \nonumber \\
							& + \loss{Iso1rev} + \loss{Iso2rev} = 98.5~\deci\bel.
\end{align}
Here we conservatively assume the injected light totally reflects at the BS (the actual reflection coefficient is tricky to calculate owing to possible interference effects Eve might exploit). The upper bound on the power reaching PwM $W_\text{PwM} = 10^{-\sfrac{\loss{Ap}}{10}} W_\text{in} \approx 14~\nano\watt$, which is a fraction of the power it measures in the normal operation. This leaves a small risk Eve might manage to tamper with the operation of PwM and the state intensities emitted.

\riskev{\Lr}{1 vulnerability is known to exist in principle, 0 requires significant research and possibly future technology to exploit, 0 probably gives Eve low key information} Note added in 2023: an explicit attack on measurement-device-independent QKD that exploits this vulnerability has been published \cite{lu2023}, slightly raising the risk.

\suggestion{Re-evaluate $W_\text{PwM}$ after additional isolation is added to Alice to protect her against the Trojan-horse attack. This will likely solve this vulnerability as well.}

\subsection{Induced-photorefraction attack}
\label{sec:photorefraction}

Recently, a new light-injection attack based on photorefractive effect in modulators has been proposed \cite{ye2023,lu2023,han2023}. A demonstration has been made of Eve's shifting the bias point of Bob's lithium-niobate device by illuminating it using $405~\nano\meter$ laser emission with power of just $3~\nano\watt$. This might open security vulnerabilities and, in particular, in the case of variable optical attenuators, enables Eve to steal a secret key being undetected by the legitimate users. It is also claimed that the photorefraction is effective over a wide range of wavelengths (from ultraviolet to even $1549~\nano\meter$ \cite{kostritskii2009}).

In the QRate system, lithium niobate devices in Alice, namely IM and PM1, prepare the quantum state. They both might be affected by this attack. In the case of the phase modulator, a shift of its working point can have effects similar to those considered in \cref{sec:state-preparation,sec:calibration}. In the case of the intensity modulator, the effect will be similar to that in \cref{sec:AD}. I.e.,\ the induced-photorefraction attack on the modulators is a potential vulnerability.

Similarly to the light-injection attacks considered in \cref{sec:THA,sec:source-laser-seeding,sec:AD}, we really need a wide spectral characterisation of the system components to treat this vulnerability. The photorefractive effect in lithium niobate modulators is most easily produced by short-wavelength illumination of blue to green color \cite{ye2023,lu2023,han2023}, thus we need to consider primarily the short-wavelength end of the spectrum. But, as the first step, let's calculate how much Eve's power at $1548.51~\nano\meter$ might reach Alice's modulators. The loss in Alice before the PM1 and IM is
\begin{align}
	\label{eq:Alice_Apm1_Loss}
	\loss{Apm1} = &~ \loss{BS} + \loss{DWDM1} + \loss{VOA1} + \loss{Att} + \loss{DWDM2} \nonumber \\
							& + \loss{Iso1rev} + \loss{Iso2rev} = 118.5~\deci\bel,\\
	\label{eq:Alice_Aim_Loss}
	\loss{Aim} = &~\loss{PM1} + \loss{Apm1} = 121~\deci\bel.
\end{align}
Assuming the laser power entering Alice is $100~\watt$, the power reaching the PM1 and IM is about $141$ and $79~\pico\watt$. Owing to the low efficiency of the photorefractive effect at the long wavelengths \cite{ye2023}, the existing isolation in the system will prevent this attack at the operating wavelength. However, we stress that this attack should be characterised in the ultra-wide spectral range. 

\riskev{\Mr}{0 vulnerability is not likely to exist, 1 is exploitable with today's technology, 1 potentially gives Eve high key information}

\suggestion{Test IM and PM1 for sensitivity to induced photorefraction at short wavelengths, similarly to \cite{ye2023,lu2023,han2023}. This will establish the isolation required. It is also necessary to characterise Alice's optical components in the wide spectral range (\cref{sec:spectral-testing}), to determine the minimum total loss $\loss{Apm1}$ and \loss{Aim} over the entire wavelength range accessible to Eve.}

\subsection{Laser damage}
\label{sec:laser-damage}

High-power laser radiation may cause temporary or permanent changes of properties of both absorbing media (for example, via heating and vaporisation) and transparent media (for example, via nonlinear effects \cite{friberg1988}). This potentially affects many optical and optoelectronic components. Laser-damage attacks have been demonstrated on various QKD systems by targeting optical attenuators \cite{huang2020,bugai2022}, isolators \cite{ponosova2022,ruzhitskaya2021}, a photodiode \cite{makarov2016}, and an avalanche single-photon detector \cite{bugge2014}. Let's consider the laser-damage attack on the QRate system's Alice and Bob.

\begin{enumerate}

\item In earlier QKD systems, the optical attenuator was the last component in Alice before the quantum channel. However, its attenuation might be significantly decreased during the laser-damage attack, what leads to an increased Alice's output mean photon number and thus leakage of the secret key \cite{huang2020,bugai2022}. To mitigate this known risk, QRate's system has two optical isolators in series at its output (\cref{fig:setup}). According to our more recent experimental results \cite{ponosova2022,ruzhitskaya2021}, placing an additional sacrificial fiber-optic isolator or circulator at Alice's exit might be required to complete the countermeasure against the laser-damage attack, at least by a $1550$-$\nano\meter$ continuous-wave laser.

We have tested three models of fiber-optic circulators and four models of fiber-optic isolators, including the isolator previously used in the QRate QKD system (QRate has recently replaced the exit isolator Iso2 with another model Opneti D-P-1550-900-1-0.3-FC-5.5x35 that we have not tested) \cite{ponosova2022,ruzhitskaya2021}. The samples tested exhibit a temporary reduction of isolation by about $15$--$35~\deci\bel$ achieved at a certain cw laser power specific to each sample. In the current system configuration with two isolators, this reduction of isolation may open loopholes for the Trojan-horse attack, laser-seeding attack, and power-meter-injection attack (\cref{sec:THA,sec:source-laser-seeding,sec:AD}). However, attempts to reduce the isolation further under a higher illumination power result in the sample's catastrophic failure. The latter manifests in an extremely large insertion loss and isolation, safely and permanently interrupting key generation.

Almost all the samples tested had a residual isolation (before the catastrophic failure) of more than $17~\deci\bel$. This is sufficient to protect the next isolator behind it and the remaining system components from the laser damage, because the residual power reaching them never exceeds their specified maximum operating power. The isolator previously used by QRate (Thorlabs IO-G-1550APC; ISO~PM~2 in \cite{ponosova2022}) exhibited maximum isolation reduction from $37~\deci\bel$ to about $17~\deci\bel$ residual value at $3.37~\watt$ laser power. Therefore, this isolator may itself be a good passive countermeasure, when an extra copy of it is added at the channel interface. We stress that the current system configuration with untested isolator models is already unsafe against the Trojan-horse attack because of insufficient isolation (\cref{sec:THA}) and might be further impaired by the laser-damage attack.

We have only tested the isolators under cw illumination at $1550~\nano\meter$. However, damage mechanisms depend on illumination regime and wavelength. Continuous lasers and pulsed lasers with pulse duration longer than $1~\nano\second$ typically cause damage via thermal effects; short and ultrashort laser pulses often strip electrons from the lattice structure of optical material before causing thermal damage \cite{wood2003}. The damage thresholds strongly depend on wavelength. It is thus important to test the front-end components against damage by a short-pulsed laser and lasers at different wavelengths \cite{ruzhitskaya2021}.

Furthermore, the isolation properties of fiber-optic isolators often strongly depend on the wavelength. For instance, one model of isolator (not the one in QRate's system) has the minimum of $11~\deci\bel$ isolation at $1150~\nano\meter$ (\cref{fig:isolation}). Therefore, the laser-damage attack at this wavelength might bypass the isolators with enough power to affect the subsequent components. We discuss this problem further in \cref{sec:spectral-testing}.

\begin{figure}
	\includegraphics{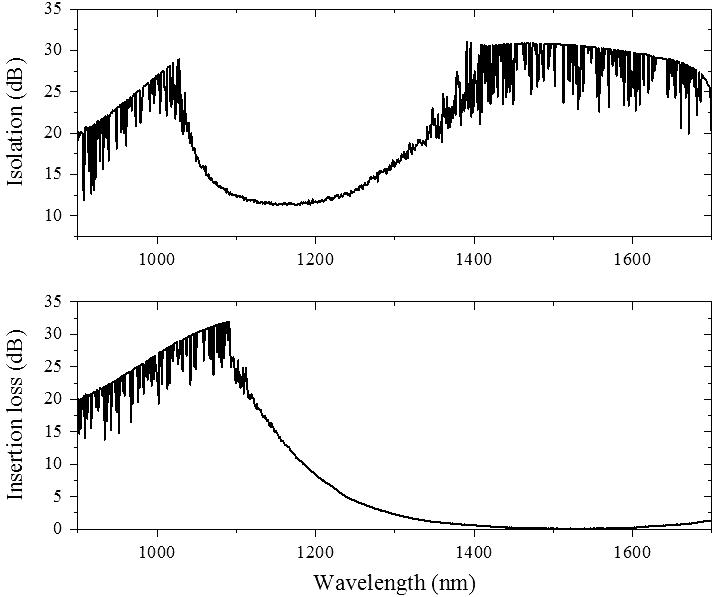}
	\caption{Typical isolation and insertion loss of a fiber-optic isolator (FOCI M-II-2-15-S-C-C-E-1-FC/FC; not the one in QRate's system) over a wider wavelength range. While the isolation is $\sim 60~\deci\bel$ at the operating wavelength of $1550~\nano\meter$ (measured separately with a narrowband source), it drops to $11~\deci\bel$ at $1150~\nano\meter$ (as the upper plot shows; note that the dynamic range of this wideband measurement is limited by the spectrometer's noise floor to $20$--$30~\deci\bel$).}
	\label{fig:isolation}
\end{figure}

\item Bob's setup is not protected against the laser damage. Theoretically, each component might be affected by the high-power laser. Let's consider them one by one.

SD: the laser-damage attack might reduce sensitivity of a photodiode \cite{makarov2016}. This does not compromise the security of QKD. However, SD shall not be employed as a countermeasure against the blinding attack.

DWDM: even if it is damaged by a high-power laser, this does not impact the QKD security. However, it shall not be employed as a security component against wavelength-dependent attacks.

PC and PM: their damage is unhelpful for Eve. We do not see how changing a detection basis setting could be exploited.

PBS: although changing the polarising beamsplitter's properties may assist the detector-efficiency-mismatch attacks (\cref{sec:efficiency-mismatch}), we consider this unlikely.

SPDs: Bob's detectors are unprotected against the high-power illumination. It is known that a single brief application of high power can cause a permanent switching of the single-photon detector into a linear regime, which is equivalent to its blinding \cite{bugge2014}. However, photocurrent monitoring should be an effective countermeasure against this.

\end{enumerate}

\riskev{\Mr}{in~Alice: 1 vulnerability exists, 0 requires research and possibly future technology to exploit similarly to the Trojan-horse attack in \cref{sec:THA}, 1 potentially gives Eve higher key information than the Trojan-horse attack without laser damage; in~Bob: 0 vulnerability is not likely to exist, 1 is exploitable with today's technology, 1 may give Eve high key information}

\suggestion{We recommend QRate to add the already tested isolator Thorlabs IO-G-1550APC as an additional sacrificial component at the exit of Alice, i.e., between the channel and the rest of Alice's setup. This component's only function is protecting the rest of the setup from damage, thus its own isolation should not be included into the isolation estimation of the source \cite{ponosova2022}. This isolator model should be further tested in a large range of laser powers and wavelengths, in continuous-wave and pulsed illumination regimes. We plan to test it under a $1064~\nano\meter$, sub-nanosecond pulsed laser \cite{ruzhitskaya2021}. Also, all the components in Alice including this isolator should be characterised in a wide spectral range (\cref{sec:spectral-testing}).}

\subsection{APD backflash}
\label{sec:detector-backflash}

It has been shown that avalanching APDs emit photons that are coupled back to the quantum channel \cite{meda2017,pinheiro2018,koehler-sidki2020}. This emission has a broad spectrum. Although the state of each photon might not be correlated to the photon that caused the detection, these backflash photons pass through optical components and carry information about the originating detector, thus leaking information about the key to Eve. In the QRate system, Bob's setup contains the PBS that splits the incoming photons into the two detectors. We assume this PBS encodes different polarisations into the backflash photons from different SPDs, allowing Eve to distinguish them and learn which detector has clicked, thus learning this bit of the raw key. The information leakage due to backflash is proportional to the probability of such events; the latter can be up to $10\%$ \cite{meda2017}. A modified formula for the key rate should be used in the system, reducing the key generation rate, considering the worst-case assumption that Eve can distinguish all the backflash photons and map them to the raw key of Alice and Bob. The secret key rate bound is estimated in \cite{pinheiro2018,koehler-sidki2020} but only for a perfect single-photon source. A modified decoy-state security analysis for a realistic photon source that takes into account several side channels, including the APD backflash, is attempted in \cite{molotkov2020}.

Given our strategy of reducing the information leakage under each individual imperfection to a negligible value (\cref{sec:unified-proof}), the emission probability from our SPDs needs to be measured and its transmission through Bob's scheme into the quantum channel calculated. The latter depends on the spectrum of the backflash and spectral properties of Bob's components, such as DWDM3. If necessary, additional spectral filters and isolators can then be added to Bob's scheme, to reduce the emission probability into the channel to a specified level.

While the spectral characterisation of Bob's passive components is straightforward (\cref{sec:spectral-testing}), the spectral measurement of the broadband backflash emission of the SPDs is more challenging, owing to the single-photon sensitivity required \cite{meda2017,shi2017,pinheiro2018,bogdanov2022}. The availability of single-photon detectors and their noise level restrict the wavelength range and spectral resolution of this measurement. We will probably need to make a relatively broadband integral measurement \cite{meda2017,pinheiro2018,bogdanov2022} and a crude bandpass measurement around the DWDM3's working wavelength \cite{meda2017,shi2017}, then make some reasonable assumptions about the SPD's true backflash spectrum to upper-bound the emission probability into the channel.

\riskev{\Mr}{1 vulnerability is known to exist, 1 exploitable with today's technology, 0 gives Eve low key information}

\suggestion{First, DWDM3 should be characterised in a wide spectral range in the reverse direction of light propagation; we may conservatively assume the other Bob's components to be transparent. Then, the probability of backflash emission from one of Bob's SPDs should be measured as outlined above. (Alternatively we may skip the latter measurement and assume the emission spectral density of our SPD to be equal to that of different devices measured in \cite{meda2017,koehler-sidki2020}.) Based on the results, additional spectral filters and isolators may need to be installed in Bob to reduce the backflash emission to a negligible level.}

\subsection{Intersymbol interference}
\label{sec:pattern-effect}

The security proof of the decoy-state BB84 protocol incorporates the assumption that all the intensity states (signal, weak and vacuum decoy) and the signal states are prepared independently of each other. However, a realistic frequency response of the intensity modulator that has a finite bandwidth might break this assumption and introduce correlations in the shape of the adjacent pulses \cite{yoshino2018}. The intensity of the pulse deviates from the set value, depending on the state of the preceding pulse, exhibiting a so-called pattern effect. The same study describes an experimental approach needed to quantify the intersymbol interference and suggests the additional post-processing procedures that effectively restore non-correlated pulses assumption. In a later study \cite{grunenfelder2020}, the correlation between adjacent pulses in both the intensity and phase modulators was quantified in a high-repetition rate polarisation-based QKD system.

At the same time, an intersymbol interference can occur even in the phase of the adjacent laser pulses. This effect becomes stronger as the QKD system's repetition rate increases. While, in principle, the decoy-state method relies on the assumption that the phase of the emitting weak coherent pulses is independently random in the range $[0, 2\pi)$, this might not be the case for most commercial QKD systems with gain-switching laser sources. The residual electromagnetic field after the emitted light pulse in the cavity of the laser can affect the next light pulse and their phases can be correlated \cite{kobayashi2014,grunenfelder2020}.

\riskev{\Lr}{1 vulnerability likely exists, 0 requires research and possibly future technology to exploit, 0 probably gives Eve low key information}

\suggestion{Our preliminary measurements on the QRate system have shown correlations between adjacent pulses in the electrical signals feeding the phase and intensity modulators \cite{trefilov2021}. This indirectly indicates the optical pulses have correlations as well. Optical measurements are planned to quantify the phase, intensity, and polarization correlations in the optical pulses. Once quantified, they can be incorporated in the security proof \cite{pereira2020,zapatero2021,pereira2023,curras-lorenzo2024,sixto2022}. A possible challenge here is that the available proof gives a zero or low key rate even for small correlations. In case our measurements result in an unsatisfactory key rate, we may additionally consider a software-based countermeasure in the post-processing \cite{yoshino2018} or replacement of the existing electro-optical modulators and/or their electronics with ones that have a higher bandwidth.}

\subsection{Imperfect individual state preparation}
\label{sec:state-preparation}

Most security proofs imply that the quantum states are prepared perfectly in any parameters like amplitude, relative phase, etc.,\ which is generally not the case in reality. Regardless of the above-mentioned effects of intersymbol interference in phase and intensity of the emitted qubits, deviations of these parameters from the ideal can be considered for both average and individual qubits. Such deviations have been studied experimentally and theoretically for a loss-tolerant protocol \cite{xu2015a,pereira2019,mizutani2019,curras-lorenzo2021}. To our knowledge, the analysis is yet to be developed for BB84.

\riskev{\Lr}{1 imperfection certainly exists, 0 research is required for exploitation, 0 probably gives Eve low key information}

\suggestion{The optical measurements of intersymbol correlations that we plan in our lab will also yield data on imperfect state preparation, including average deviation and its statistical distribution. Based on that, we may attempt to apply the loss-tolerant protocol to incorporate these flaws \cite{xu2015a,pereira2019,mizutani2019,curras-lorenzo2021}. Further theory development is needed for a full understanding of this imperfection.}

\subsection{Calibration performed via channel Alice--Bob}
\label{sec:calibration}

The current system implementation conducts several calibration routines through the channel Alice--Bob. During the calibration, Bob sends low-level commands to Alice via the classical channel, Alice transmits signals via the quantum channel, Bob receives them using his SPDs and collects photon click statistics. This calibration sets multiple vital parameters such as signal timing and working points of the modulators. It is always performed at the system power-up and repeated as necessary whenever the system fails to generate keys for a relatively long time (about $1~\hour$). The following parameters are determined by this calibration.
\begin{itemize}
	\item Precise timing of Bob's detector gate to maximise the count rate and correctly identify Alice's bit number (i.e.,\ not register clicks in an adjacent bit), separately for each of the two detectors. The scanning range is $6.4~\nano\second$, which spans two adjacent bit slots. The scanning is done with $100~\pico\second$ step.
	\item Zero point of Alice's IM, set by applying voltage at its bias section. While it is maintained by a feedback from PwM during QKD (\cref{sec:AD}), the initial setting is calibrated using Bob's SPDs.
	\item Precise timing of electrical signal applied at Bob's PM to correctly modulate the light pulse received from Alice. The scanning is done with $400~\pico\second$ step.
	\item Precise timing of electrical signal applied at Alice's PM to correctly modulate the laser pulse being sent. The scanning is done with $400~\pico\second$ step.
	\item Precise timing of electrical signal applied at Alice's IM to correctly modulate the laser pulse being sent. The scanning is done with $400~\pico\second$ step.
	\item Initial setting of Bob's PC to minimise QBER. (The PC is then adjusted in realtime during QKD to keep the QBER low.)
\end{itemize}
During QKD, three realtime adjustments are being performed continuously: Alice maintains her IM's zero point (\cref{sec:AD}), Bob adjusts his PC to maintain low QBER, and Bob's master clock generator is adjusted ten times per second.

Since the initial calibration is performed via the quantum channel, it is totally exposed to Eve's tampering \cite{jain2011,fei2018}. We have to assume, and it is likely the case, that Eve can set any values of the parameters being calibrated at her discretion. Additionally she may interfere with Bob's realtime clock adjustment and whatever timing parameters this affects. Additionally we have to assume Eve may issue low-level commands to Alice (unless this communication is strongly authenticated).

This has fairly horrible consequences. Several attacks become possible.
\begin{itemize}

	\item Eve can induce a large time-efficiency mismatch between Bob's detectors (\cref{sec:efficiency-mismatch}), which has been demonstrated experimentally in other QKD systems \cite{jain2011,fei2018}. An additional possibility is that, since the system scans both click acceptance window positions over the time range that spans more than one bit slot, Eve may diverge them. I.e.,\ she may set them such that a click resulting from a qubit detection at SPD1 is registered as one key bit while a click from the same qubit at SPD2 is registered as another key bit. This generally makes any security proof for BB84 inapplicable, because they all implicitly assume this situation is impossible. Besides, we can think of a practical attack that combines this diverged click registration with the simultaneous deadtime (\cref{sec:deadtime}) and allows Eve to suppress clicks in the detector that gets registered in the later bit slot and, possibly, exploit an asymmetric click discarding in one bit slot at the end of the deadtime.

	\item Eve can shift Alice's PM signal in time such that it modulates the light pulse at the transition between the modulation levels. The quantum states prepared are then less separated in phase and a phase-remapping attack may become possible \cite{xu2010}. Additionally, any careful characterisation of state preparation imperfections we perform (\cref{sec:pattern-effect,sec:state-preparation}) becomes meaningless. In particular, intersymbol correlations may be amplified.

If Eve can arbitrarily control Alice's and Bob's PM phase shifts, an extreme attack becomes possible. Eve sets Alice to use an identical pair of phase shifts in both her bases, and she tricks Bob to do the same as well. Then she performs a quantum intercept-resend attack in this one basis setting, while Alice and Bob think they are using different choices of bases. This attack does not increase the QBER and gives Eve the complete key. It is important to note that Alice and Bob can detect the presence of Eve in the discarded cases when their bases don't match since the measurement outcome for Bob will no longer be random. Nevertheless, there are no such verification procedures in the classical BB84 protocol and its version used in the QRate's system. The potential existence of this extreme attack hints that we have a problem that has to be treated by a security proof even in a milder case when Eve has a limited control over the PM settings. The most complete security proof for the BB84 protocol with imperfect state preparation \cite{pereira2023} cannot account for the most extreme case of this attack, where four quantum states in two bases in reality degenerate into two quantum states in one basis. 

	\item Eve can shift Alice's IM signal in time and/or make the IM operate with an incorrectly set zero point. This would have similar effects on the intensity states being prepared by Alice. The security proof of the decoy-state protocol \cite{gottesman2004} becomes inapplicable when the actual intensities are unknown and are chosen by Eve. Characterisation of state preparation imperfections becomes meaningless.

\end{itemize}

To summarise, the public exposure of the calibration routines presents multiple security issues. The analysis team did not know how to solve this and suspected that a significant redesign of the system hardware might be required. This is a decision that should be taken by the system manufacturer.

\riskev{\Hr}{1 vulnerabilities are likely exploitable, 1 with current technology, 1 might give Eve high key information}

\suggestion{An extensive discussion with QRate was needed. QRate has subsequently found solutions acceptable for the manufacturing process, see \cref{sec:countermeasures}. Regarding the four-state Bob countermeasure to the time mismatch vulnerability suggested in \cref{sec:efficiency-mismatch} and presently implemented by QRate, we are still not sure it is sufficient, given that Eve might be able to set all the time parameters and diverge them between bits.}

\subsection{Quantum random number generator}
\label{sec:QRNG}

We remind the manufacturer that, in order to be compliant with the security proof, a real quantum random number generator must provide all the state, basis, and intensity choices in Alice and Bob, as well as random bit-value assignment in the event of a double click and other random values needed in the protocol. A mathematical `random' number generator (used currently) or randomness expansion are, strictly speaking, insufficient.

\riskev{\Lr}{1 vulnerability is known to exist in principle, 0 requires significant research and possibly future technology to exploit, 0 probably gives Eve low key information}

\suggestion{The company should implement a full-bandwidth quantum random number generator, without resorting to the randomness expansion, and integrate it into the system.}

\subsection{Compromised chain of supply}
\label{sec:chain-of-supply}

Like most manufacturers of cryptographic hardware, QRate buys the constituent parts of its products from a multitude of external suppliers. It is the nature of cryptography that many of these parts may subvert the security of the product if the part's supplier, or a third party, modifies it in a malicious way before it is installed into the product. The modification may be a covert change of characteristics that enables an attack, a change of the part's behaviour that does the same, or a hidden transmitter (either radio-frequency or optical) that communicates the secret information outside the device. The modification will, of course, be difficult to detect: it will not reveal itself in the standard factory assembly and testing procedures, neither will it hinder the normal operation of the product. In the QKD system, the parts that may be compromised include optical, electrooptical and electronic components, third-party electronic modules, and even integrated circuits.

This problem is general to all cryptography hardware. We can also think of attacks and information leakage tactics specific to QKD that might be enabled in such a way.

A significant drawback of these attacks from Eve's point of view is the need to plan them well in advance. She must initiate them before the equipment is assembled and deployed for protection of the asset of interest.

\riskev{\Mr}{0 it is unlikely that any player will spend significant resources on preventively attacking a niche product that is not yet being deployed for protection of high-value information assets, 1 can certainly be done today, 1 can be arranged to leak the entire key}

\suggestion{The company should learn suitable mitigation strategies from the national cryptography licensing authority.}

\section{Summary of initial security analysis}
\label{sec:summary}

At the end of our initial security analysis concluded in January 2022, we identify more than ten potential implementation security issues in QRate $312.5~\mega\hertz$ QKD system and rank them by their practical risk (see \cref{tab:attacks}). The vulnerabilities in Bob's single-photon detection subsystem related to detector controllability and timing calibration are of a high concern (\cref{sec:detector-control,sec:efficiency-mismatch}). We are not sure if it is possible to construct sufficient countermeasures and stop all detector-related attacks that are implementable with today's technology. From this point of view, the measurement-device-independent and twin-field QKD schemes are an attractive alternative. The accessibility of the calibration routines for Eve's tampering (\cref{sec:calibration}) is another difficult problem that needs to be discussed.

Actions needed to address the remaining vulnerabilities are mostly clear. Most optical components in the scheme need to be spectrally characterised in a wide spectral range ($\sim 350$--$2400~\nano\meter$, see \cref{sec:spectral-testing}). Optical measurements of imperfections in the state preparation in Alice and light emission from Bob's APDs need to be performed. Several inexpensive additional passive components, such as isolators and spectral filters, should be added to the scheme. QRate should make minor improvements in the post-processing algorithms and update the key rate equation.

Finally, we ask QRate to provide the complete QKD system to our testing lab on a permanent basis. Further security analysis requires a level of familiarity with the system implementation that cannot be gained by reading technical documentation and can only be obtained via extensive hands-on experience during experiments. This sample of the system should be reserved for the hacking experiments and serve no other purposes.

\section{Addressing high-risk security issues}
\label{sec:countermeasures}

After the delivery of our initial analysis report, actions have taken place during the year 2022. The four high-risk security issues (marked \Hr in \cref{tab:attacks}) have been prioritised and QRate has implemented countermeasures to all of them. Meanwhile, we hope that a formal certification methodology that is being designed covers all, or most of, the security issues identified by us. QRate has also provided us the QKD system for testing.

The photocurrent-monitoring countermeasure against detector blinding (\cref{sec:detector-control}) has been implemented by QRate and tested in our lab \cite{acheva2023,kuzmin2023a}. It reliably protects against cw blinding. However, pulsed blinding and control remain possible, owing to the photocurrent measured being averaged over a relatively long time \cite{acheva2023}. A higher-bandwidth photocurrent registration scheme has subsequently been implemented in the sinusoidally-gated detector for this QKD system, in order to close this issue. Its testing on an automated testbench \cite{kuzmin2023a} is in progress. Also, testing this detector for the after-gate and falling-edge attacks is in progress \cite{kuzmin2023,zaitsev2024}.

The four-state Bob has been implemented by QRate as a countermeasure against the detector efficiency mismatch (\cref{sec:efficiency-mismatch}) and timing calibration vulnerability (\cref{sec:calibration}).

The software component of the simultaneous deadtime has been implemented, to complete the countermeasure against the deadtime attack (\cref{sec:deadtime}).

In order to address the calibration vulnerabilities (\cref{sec:calibration}), QRate has eliminated the calibration of Alice's IM and PM via the channel. The intensity modulator is now instead always calibrated via Alice's PwM. The phase modulator is now only calibrated at the factory once, then its settings remain fixed during the lifetime of the system. With these changes and the four-state Bob in place, we hope that the existing calibration of Bob's PM and timing of his detectors via the channel no longer constitute a vulnerability and may remain unchanged.

The above modifications to the system and additional tests planned should close all the high-risk vulnerabilities from \cref{tab:attacks}. This protects the system from the attacks known to be readily implementable today.

\section{Proposal for certification}
\label{sec:certification}

To perform a complete set of measurements and tests for certifying implementation security of the ``quantum'' part of the system (i.e.,\ to cover all the potential issues identified in this report), five testbenches are needed.
\begin{enumerate}
\item Wideband spectral characterisation of components, as detailed in \cref{sec:spectral-testing}.
\item Characterisation of detector controllability, deadtime, efficiency mismatch, and Bob's calibration routines. This includes testing the efficiency of any countermeasures to these issues. The testbench design is sketched in \cref{sec:test-of-detector-control} and \cite{acheva2023}.
\item Characterisation of state preparation imperfections in Alice. The testbench design can be based on \cite{grunenfelder2020,huang2023}.
\item Characterisation of light emission from the detectors, as detailed in \cite{meda2017}.
\item Laser damage, based on \cite{ponosova2022}. Although different lasers may be used by Eve, we propose to initially implement the basic testing under $1550~\nano\meter$ continuous-wave laser.
\end{enumerate}
A formal certification methodology for QKD is currently under development, in coordination with the Russian national cryptography licensing authority. This report is one of the inputs to this process. Traditionally, Russian national certification standards for cryptographic systems are classified. The actual domestic certification procedures being implemented thus cannot be disclosed.

\section{Conclusion}
\label{sec:conclusion}

We have performed security analysis of the commercial QKD prototype system from QRate. After several rounds of discussions, all the theoretical threats are eliminated and the system seems to be secure. This is to be verified by testing its final implementation during the certification. Since this system uses a fairly standard prepare-and-measure BB84 scheme, this analysis should be partially applicable to other systems of the same type. Out of several potential vulnerabilities identified (\cref{tab:attacks}), four are deemed high-risk (\Hr), because attacks exploiting them are likely implementable today. These four security issues are addressed first. QRate has implemented countermeasures to each of them (\cref{sec:countermeasures}). The remaining security issues might be addressed routinely in the course of the formal certification that is being developed (\cref{sec:certification}). We hope that this work contributes to the establishment of a Russian domestic certification lab and national certification standard for implementation imperfections in QKD. Since the recently published international standard \cite{iso23837-2023} prescribes similar evaluation and testing methods, our work also serves as an example of its application to a real QKD system.

\acknowledgments
This manuscript has been reviewed by QRate prior to its publication. We thank QRate's engineering division, theory group, and management for discussions and support. The measurement in \cref{fig:deadtime-test} was performed by Ilya Gerasimov and Nikita Rudavin. We also thank Marcos Curty, Anton Trushechkin, Aleksei Reutov, Daniil Menskoy, and Nikolay Borisov for comments and discussions.

\medskip

{\em Funding:} This work was funded by the Ministry of Education and Science of Russia (program NTI center for quantum communications) and Russian Science Foundation (grant 21-42-00040), which together contributed more than $95\%$ of the study costs. The remaining minor amounts of funding came from: P.C.\ acknowledges support from Thai DPST scholarship and NSRF via the Program Management Unit for Human Resources \& Institutional Development, Research and Innovation (grant B05F650024); A.H.\ acknowledges support from the National Natural Science Foundation of China (grants 62371459 and 62061136011) and Innovation Program for Quantum Science and Technology (2021ZD0300704); M.P.,\ D.T.\, and K.Z.\ acknowledge support from MICIN with funding from the European Union NextGenerationEU (PRTR-C17.I1) and the Galician Regional Government with own funding through the ``Planes Complementarios de I+D+I con las Comunidades Aut{\' o}nomas'' in Quantum Communication.

\medskip

{\em Author contributions:} All authors except A.K.F.,\ E.K.,\ and A.T.\ analysed the system documentation and different vulnerabilities, reviewed the general approach and conclusions. A.K.F.,\ E.K.,\ and A.T.\ described the post-processing and security proofs used in the system under analysis. All authors contributed to writing the manuscript.

\appendix

\section{Post-processing in QRate's system}
\label{sec:post-processing}

The concept of the QKD is that two legitimate users (Alice and Bob) generate ``long'' symmetric keys by using a classical and a quantum channels together with a ``short'' pre-shared key. The pre-shared key is used for authentication of a classical communication only and can be discarded, or even publicly announced, after the end of the first run (round) of the QKD protocol. In the next round, a piece of the previous quantum-generated key can be used for the authentication purposes. In this way, the QKD have to be considered as a {\em quantum key growing}.

The core of the QKD protocol is in preparing quantum states and encoding information on Alice's side, and measuring the states on Bob's side. In the BB84 protocol \cite{bennett1984}, Alice and Bob use four qubit states that form two orthogonal bases in two-dimensional Hilbert space, $Z:\{\ket{0_Z},\ket{1_Z}\}$ and $X:\{\ket{0_X},\ket{1_X}\}$, where 0 and 1 indicate a classical bit encoded by the corresponding basis vector. The basis vectors are related as
\begin{equation}
	\ket{0_X}=\frac{\ket{0_Z} + \ket{1_Z}}{\sqrt2}, \quad
	\ket{1_X}=\frac{\ket{0_Z} - \ket{1_Z}}{\sqrt2}.
	\label{eq:BB84_Xbasis}
\end{equation}
If the information is encoded into polarization of a single photon, then $\ket0_Z$ and $\ket1_Z$ can correspond to the horizontal and vertical polarizations. In this case, $\ket0_X$ and $\ket1_X$ represent two diagonal polarizations, rotated by $45\degree$ and $135\degree$ relative to the horizontal direction. This polarization encoding is used to illustrate the idea, but, in fact, there is no restriction on the method of information encoding. Formally, $\ket0_Z$, $\ket1_Z$, $\ket0_X$, and $\ket1_X$ are just vectors in the Hilbert space, and one can use any encoding scheme that fulfills \cref{eq:BB84_Xbasis}. The equivalence of the polarization and phase encodings is explained in detail in \cite{trushechkin2021}.

Importantly, as can be seen from \cref{eq:BB84_Xbasis}, when measuring a qubit in a basis different from the preparation one, the result is a completely random value. This is the consequence of the well-known fact that two non-orthogonal quantum states cannot be perfectly distinguished. On the contrary, if the preparation and measurement bases coincide, the result perfectly correlates with the initial qubit state (in the absence of errors in the channel, measuring devices, etc.). In this way, if Eve does not know the preparation basis, due to the no-cloning theorem \cite{dieks1982,wootters1982} she has to employ imperfect copying techniques that induce errors on Bob's side.

In practice, however, true single-photon states are very difficult to generate, and weak coherent states with a phase randomization are used instead.
In the case of the polarization encoding, the state preparation takes the form
\begin{equation} \label{eq:coherentstates}
	\begin{aligned}
		&\ket{0_Z} \rightarrow \rho_H(\alpha)\otimes \rho_V(0), \quad \ket{1_Z} \rightarrow \rho_H(0)\otimes \rho_V(\alpha),\\
		&\ket{0_X} \rightarrow \rho_D(\alpha)\otimes \rho_A(0), \quad \ket{1_X} \rightarrow \rho_D(0)\otimes \rho_A(\alpha),
	\end{aligned}
\end{equation}
where $\rho_M(\beta)$ stands for a phase-randomized coherent state in mode $M$ with mean photon number $\beta$:
\begin{equation}
	\rho_M(\beta) = \sum_{n=0}^{\infty} \frac{e^{-\beta}\beta^n}{n!} \ket{n}_M\bra{n},
	\label{eq:Poisson}
\end{equation}
$\ket{n}_M$ denotes an $n$-photon state in mode $M$, $H$ and $V$ ($D$ and $A$) indicate horizontal and vertical (diagonal and antidiagonal) modes with corresponding annihilation operators satisfying
\begin{equation}
	\hat{a}_D = \frac{1}{\sqrt{2}}(\hat{a}_H + \hat{a}_V), \quad	\hat{a}_A = \frac{1}{\sqrt{2}}(\hat{a}_H - \hat{a}_V).
\end{equation}
The chosen photon number $\alpha$ in \cref{eq:coherentstates} is specified by the protocol. The projection of considered states on a single-photon subspace results in four states
\begin{equation}
	\begin{aligned}
		&\ket{0_Z} \rightarrow \ket{1}_H\ket{0}_V, &\ket{0_X} \rightarrow \frac{1}{\sqrt{2}} (\ket{1}_H\ket{0}_V + \ket{0}_H\ket{1}_V),\\
		&\ket{1_Z} \rightarrow \ket{0}_H\ket{1}_V, &\ket{1_X} \rightarrow \frac{1}{\sqrt{2}} (\ket{1}_H\ket{0}_V - \ket{0}_H\ket{1}_V)~\\
	\end{aligned}
\end{equation}
that are suitable for the BB84 protocol [cf.~\cref{eq:BB84_Xbasis}]. Unfortunately, multiphoton components of states~\labelcref{eq:coherentstates} are vulnerable to a photon-number-splitting (PNS) attack and can not be used for secure key generation. Therefore, an estimation of the number of detections on Bob's side that resulted from single-photons states generated on Alice's side is required.

For the past decades, the BB84 protocol has been theoretically studied in detail. The first security proofs \cite{mayers1996,lo1999,shor2000} were made for ideal version of the protocol with perfect single-photon source, and then generalised for realistic photon source \cite{gottesman2004}. In order to eliminate the vulnerability against the PNS attack and increase the secure communication distance, the decoy-state technique was developed \cite{hwang2003} and combined with the entanglement distillation approach from~\cite{gottesman2004}. As a result, an improved secret key rate formula was obtained \cite{lo2005,ma2005}. The security against not only the PNS attack but all possible general attacks is usually considered by the community as ``obvious'', and until recently the complete mathematical proof has not been available in the literature. The formal security proof is summarised and presented in \cite{trushechkin2021}.

In the QKD system under evaluation, the practical realisation of the decoy-state BB84 protocol contains the following steps.
\begin{enumerate}

\item {\em State preparation and measurement.} Alice randomly with equal probabilities chooses a basis from the set $\{Z,X\}$ and an information bit from $\{0,1\}$.

In order to counteract the PNS attack, the widely applied decoy-state technique is used. Alice randomly chooses the laser pulse intensity [$\alpha$ in \cref{eq:coherentstates}] from the set $\{\mu,\nu_1,\nu_2\}$ with corresponding probabilities $\{p_\mu,p_{\nu_1},p_{\nu_2}\}$. Here $\mu$ corresponds to the signal-type state, $\nu_1$ and $\nu_2$ ($\nu_1+\nu_2<\mu$, $\nu_2<\nu_1$) correspond to the weak and vacuum decoy-type states respectively. The optimal intensities and probabilities are determined for a given communication distance and experimental setup from the numerical maximization of the simulated secret key rate.

Then, the photon pulse is prepared in the corresponding quantum state and is transmitted through the quantum channel. It is important that Alice's laser emits each pulse with a random phase, owing to it being internally seeded with spontaneous emission. Thus the photon number statistics of Alice's pulses is Poissonian [\cref{eq:Poisson}], as required in the decoy-state technique.

Bob randomly and independently of Alice chooses a measurement basis from $\{Z,X\}$ and measures the qubit state in the selected basis. In case of a double-click of Bob's detectors, he randomly chooses the bit value.

The above steps are repeated many times until a sufficient number of quantum states are detected. More specifically, Alice sends pulses in so-called ``trains'' of fixed size ($\sim10^6$ pulses per train). Owing to the time synchronisation with Alice, Bob knows the train number and the position of each detected pulse in the train.

\item {\em Sifting.} When Bob accumulates enough statistics ($\sim 1900$ clicks), he announces the train number and position of each registered pulse together with its measurement basis. Alice in turn compares her preparation basis with it and announces the positions with matching bases and their corresponding pulse types (signal, weak or vacuum decoy).

After that, Alice and Bob select the signal-type bits with matching bases and form two bit strings, called {\em sifted keys}. Ideally, they should be identical, but due to natural noise in the channel or adversary actions they do not match $100\%$. Moreover, Eve may have partial information about them.

\item {\em Statistics estimation.} For practical reasons, the sifted keys are assembled into post-processing blocks of equal fixed size. In order to minimise the effect of statistical fluctuations on the final secret key length and have a reasonable block generation time, it is chosen to be $\ell_\text{block}=1.36\times10^6$ bits. 

For each block, Alice counts the corresponding total numbers of transmitted ($N_\alpha$) and detected ($M_\alpha$) pulses of each intensity $\alpha\in\{\mu,\nu_1,\nu_2\}$ regardless of their preparation and measurement basis. Then Alice estimates a {\em gain} $Q_\alpha$ -- the probability that a pulse of intensity $\alpha$ is detected by Bob,
\begin{equation}
	\hat{Q}_\alpha = \frac{M_\alpha}{N_\alpha}, \quad \alpha\in\{\mu,\nu_1,\nu_2\},
\end{equation}
and sends all three sets $\{N_\alpha,\hat{Q}_\alpha\}$ to Bob. Here and below, $Q_\alpha$ denotes a true probability value of binomial distribution $M_\alpha\sim\text{Bi}(N_\alpha,Q_\alpha)$, while $\hat{Q}_\alpha$ denotes its statistical estimate (i.e.,\ a random variable). I.e.,\ Alice computes this statistics before sifting, in order to maximise the statistical sample size.

\item {\em Information reconciliation.} Alice's key is considered to be a reference one, while Bob attempts to eliminate the discrepancies between the keys caused by errors. In order to correct them, low-density parity-check (LDPC) codes are commonly used. Since $\ell_\text{block}$ is too large for high-speed and efficient LDPC-based algorithms, the block is split into 50 subblocks of length $\ell_\text{subblock}=27\,200$ bits and the error correction is performed on each subblock separately. If the correction of a subblock fails, it is discarded from the block by both sides. As a result, Alice and Bob obtain the corrected keys $K_\text{cor}^A$ and $K_\text{cor}^B$ of length $\ell_\text{cor}=n_\text{cor}\ell_\text{subblock}\leq\ell_\text{block}$, where $n_\text{cor}$ is the number of corrected subblocks. For a more detailed description of the symmetric blind information reconciliation scheme used, see \cite{kiktenko2017,borisov2022}.

For each successfully corrected subblock, Bob computes the signal QBER
\begin{equation}
	~ E_\mu^{(i)} = \frac{\text{number of errors in}~i^\text{th}~\text{subblock}}{\ell_\text{subblock}}.
\end{equation}

\item {\em Verification and parameter estimation.} The identity of obtained $K_\text{cor}^A$ and $K_\text{cor}^B$ is checked using an $\varepsilon$-universal polynomial hash function \texttt{PolyHash}, computed according to a modified PolyP32 algorithm \cite{krovetz2001,fedorov2018}. First, Alice generates a random number $k\in\{0,1,\dots,q-1\}$ where $q$ is a prime number, chosen to be $q=2^{50}-27$. Then she computes the hash-tag of her key and sends it together with $k$ to Bob to compare with his hash-tag. If $\texttt{PolyHash}(k,K_\text{cor}^A)=\texttt{PolyHash}(k,K_\text{cor}^B)$, the verification is considered successful and the protocol proceeds to the next step. Otherwise, Alice and Bob start comparing the hash-tags of every single subblock until all the corrupted subblocks are found and discarded. In this way, the legitimate users obtain identical {\em verified keys} $K_\text{ver}^A$ and $K_\text{ver}^B$ of length $\ell_\text{ver}=n_\text{ver}\ell_\text{subblock}\leq\ell_\text{cor}$, where $n_\text{ver}$ is the number of verified subblocks.

The probability of remaining errors in the verified keys can be estimated as
\begin{equation}
	\varepsilon_\text{ver} \leq \varepsilon_\text{col}(\ell_\text{cor})
\end{equation}
if $\texttt{PolyHash}(k,K_\text{cor}^A)=\texttt{PolyHash}(k,K_\text{cor}^B)$ or
\begin{equation}
	\varepsilon_\text{ver} \leq 1 - [1-\varepsilon_\text{col}(\ell_\text{subblock})]^{n_\text{ver}}
\end{equation}
otherwise. Here the probability of a hash collision, i.e.,\ $\texttt{PolyHash}(k,K^A)=\texttt{PolyHash}(k,K^B)$ when $K^A\neq K^B$, is evaluated as \cite{fedorov2018}
\begin{equation}
	\varepsilon_\text{col}(\ell_K) = \frac{\lceil \ell_K / \lfloor \log_2 q \rfloor \rceil - 1}{q}.
\end{equation}

At the end of this step Bob computes the overall average QBER
\begin{equation}
	E_\mu = \frac{1}{n_\text{ver}} \sum_{i \in \mathcal{V}} E_\mu^{(i)}, 
\end{equation}
where the summation is performed over the ensemble of successfully corrected and verified subblocks~$\mathcal{V}$.

\item {\em Estimation of the level of eavesdropping.} After the successful error correction and verification, Bob estimates the final secret key length \cite{trushechkin2017}
\begin{equation}
	~~~ \ell_\text{sec} = m_1^l \big[1 - h_2(E_1^u)\big] - \text{leak} - \log_2\varepsilon_\text{pa}^{-5},
	\label{eq:l_sec}
\end{equation}
where the first and last terms represent the privacy amplification step and are determined by $m_1^l$---the lower bound on the number of bits in the verified key, obtained from signal single-photon pulses, $E_1^u$---the upper bound on the single-photon QBER, and $\varepsilon_\text{pa}=10^{-12}$---the tolerable failure probability for the privacy amplification step. The $h_2$-function is the standard Shannon binary entropy. The second term in \cref{eq:l_sec} is the amount of information about the key leaked to Eve during the error correction and verification steps
\begin{equation}
	\text{leak} = \sum_{i \in \mathcal{V}} \big[ \ell_\text{synd} - p + d_i \big] + \xi \ell_\text{hash},
	\label{eq:leak}
\end{equation}
where $\ell_\text{synd}$ is the syndrome length, $p$ is the initial number of punctured bits, $d_i$ is the total number of disclosed punctured bits in additional rounds ($\ell_\text{synd}$, $p$ and $d_i$ depend on the LDPC code rate and {\em a priori} QBER estimation, see \cite{kiktenko2017,borisov2022}), $\ell_\text{hash}=\lceil\log_2q\rceil=50$ is the hash-tag length, $\xi=1$ if $\texttt{PolyHash}(k,K_\text{cor}^A)=\texttt{PolyHash}(k,K_\text{cor}^B)$ and $\xi=n_\text{cor}+1$ otherwise.

Using the decoy-state technique, the lower bound on the single-photon gain $Q_1$ is estimated as \cite{trushechkin2017,zhang2017}
\begin{equation}
	\begin{aligned}
		~~ Q_1^l &= \frac{\mu^2 e^{-\mu}}{(\nu_1 - \nu_2) (\mu - \nu_1 - \nu_2)} \bigg[ Q_{\nu_1}^l e^{\nu_1} \\
		 &~~~- Q_{\nu_2}^u e^{\nu_2} - \frac{\nu_1^2 - \nu_2^2}{\mu^2} \big( Q_\mu^u e^\mu - Y_0^l \big) \bigg],
	\end{aligned}
	\label{eq:Q1_l}
\end{equation}
\begin{equation}
	Y_0^l = \max \bigg\{ \frac{\nu_1 Q_{\nu_2}^l e^{\nu_2} - \nu_2 Q_{\nu_1}^u e^{\nu_1}}{\nu_1 - \nu_2} \,, 0 \bigg\} \,.
	\label{eq:Y0_l}
\end{equation}
The finite key effect and statistical fluctuations are taken into account in our analysis. According to the central limit theorem, the binomial distributions of $M_\alpha\sim\text{Bi}(N_\alpha,Q_\alpha)$ and $m_1\sim\text{Bi}(\ell_\text{ver},Q_1/Q_\mu)$ can be well approximated by the normal distribution. The upper and lower bounds on $Q_\alpha$ and $m_1$ are \cite{trushechkin2017}
\begin{equation}
	Q_\alpha^{u,l} = \hat{Q}_\alpha \pm z \sqrt \frac{\hat{Q}_\alpha (1 - \hat{Q}_\alpha)}{N_\alpha},
	\label{eq:Q_ul}
\end{equation}
\begin{equation}
	m_1^l = \ell_\text{ver} \frac{Q_1^l}{Q_\mu^u} - z \sqrt{\ell_\text{ver} \frac{Q_1^l}{Q_\mu^u} \bigg( 1 - \frac{Q_1^l}{Q_\mu^u} \bigg)},
	\label{eq:kappa1}
\end{equation}
where $z$ is the normal distribution quantile, and the bounds on the true value of $Q_\alpha$ are evaluated via the Wald confidence interval.

In general, one cannot use binomial distribution for QBER since if Eve performs a coherent attack, the errors in different positions in the key cannot be treated as independent events. Therefore, $E_1$ is estimated in a different way \cite{trushechkin2017}:
\begin{equation}
	E_1^u = \frac{\ell_\text{ver} E_\mu - \overline{m}_0^l}{m_1^l},
	\label{eq:E_1_u}
\end{equation}
where the lower bound on the number of bit errors in the verified key, obtained from the 0-photon pulses due to background events [$\overline{m}_0\sim\text{Bi}(N_\mu,e^{-\mu}Y_0/4)$], is given by
\begin{equation}
	~~~~~ \overline{m}_0^l = N_\mu \frac{e^{-\mu} Y_0^l}{4} - z \sqrt{N_\mu \frac{e^{-\mu} Y_0^l}{4} \bigg( 1 - \frac{e^{-\mu} Y_0^l}{4} \bigg)}.
	\label{eq:v}
\end{equation}

One can see that 7 confidence bounds in total are required to compute $\ell_\text{sec}$. Therefore, in order to have the estimation \cref{eq:l_sec} satisfied with probability not less than $1-\varepsilon_\text{decoy}$, one has to define the quantile as
\begin{equation}
	z = \Phi^{-1} \bigg( 1 - \frac{\varepsilon_\text{decoy}}{7} \bigg).
\end{equation}

If $\ell_\text{sec}\leq0$, Eve is assumed to have more information about Alice's string than Bob. Hence, the key block is considered insecure and is discarded by both sides. The protocol is aborted, and Alice and Bob proceed to the next accumulated sifted block.

\item {\em Privacy amplification.} If $\ell_\text{sec}>0$, Alice and Bob proceed to the privacy amplification step, aimed to shorten the verified key $K_\text{ver}$ even further and destroy Eve's potential knowledge about the key. This procedure is performed using a hash function from the Toeplitz family of 2-universal hash functions \cite{krawczyk1994,krawczyk1995}. Bob generates a random string $S$ of length $\ell_S=\ell_\text{ver}+\ell_\text{sec}-1$ and sends it to Alice \cite{kiktenko2016}. Alice computes $\ell_\text{sec}=\ell_S-\ell_\text{ver}+1$. Then both sides symmetrically generate a Toeplitz matrix $T_S$ of dimension $\ell_\text{sec}\times\ell_\text{ver}$ using $S$ and compute the final key $K_\text{sec}=T_SK_\text{ver}$. As a result, Alice and Bob obtain a common shorter {\em secret key} of length $\ell_\text{sec}$, Eve's information about which is now negligible. The security of privacy amplification is based on the leftover hash lemma \cite{tomamichel2011a}.

\end{enumerate}

One can notice that all the post-processing steps require Alice and Bob to communicate via the classical channel. In order to verify both data integrity and authenticity of each message, a hash-based message authentication code and a secret key taken from the common quantum key, is used. The message authentication code uses a Russian national standard hash function Streebog-512 (GOST~R 34.11-2012) \cite{dolmatov2013}. The authentication failure probability (i.e.,\ the probability that Eve will guess the secret key from the hash-tag of initial message) is considered to be much less than $10^{-12}$ and hence is neglected in \cref{eq:epsilon}. There is also an option to supply the system with a certified hardware authentication device (``Continent'' manufactured by the Russian company Security Code LLC) that replaces Streebog-512.

Theoretically the QKD security level is expressed in terms of the trace distance between the real classical-quantum state (in which the classical subsystem corresponds to the key, and the quantum one belongs to Eve) and the respective ideal state. The latter is characterised by a uniform distribution of the key and the absence of correlations between the key and Eve's quantum subsystem. If the trace distance does not exceed $\varepsilon$, the key is called $\varepsilon$-secure (see, e.g.,\ \cite{trushechkin2020}). This overall (in)security parameter of the entire QKD system has several contributions
\begin{equation}
	\begin{aligned}
		\varepsilon &= \varepsilon_\text{decoy} + \varepsilon_\text{ver} + \varepsilon_\text{pa} \\
		&= 10^{-12} + 2.5 \times 10^{-11} + 10^{-12} < 3 \times 10^{-11},
	\end{aligned}
	\label{eq:epsilon}
\end{equation}
where $\varepsilon_\text{decoy}$ is a failure probability of the single-photon gain and QBER estimation, $\varepsilon_\text{ver}$ and $\varepsilon_\text{pa}$ are the failure probabilities of key verification and privacy amplification. If the authentication at QKD round $r=2,3,\ldots$ is realized by using a part of a key generated at the $(r-1)^\text{th}$ round, then the security parameter for the $r^\text{th}$ round is given by $\varepsilon^{(r)}=r\varepsilon$.

\section{Implementation layers in a quantum communication system}
\label{sec:Q-reprint}

For convenience, we reprint the description of layers from \cite{sajeed2021} here:

\needspace{3\baselineskip} 
\begin{longtable}{>{\raggedright}p{25mm} >{\raggedright\arraybackslash}p{58mm}}
	\hhline{==}
	\bf{Layer} & \bf{Description} \\*
	\hline
	\hangindent=4.6ex Q7.~Installation and maintenance & Manual management procedures done by the manufacturer, network operator, and end users.\\
	\hangindent=4.6ex Q6.~Application interface & Handles the communication between the quantum communication protocol and the (classical) application that has asked for the service. For example, for QKD this layer may transfer the generated key to an encryption device or key distribution network.\\
	\hangindent=4.6ex Q5.~Post-processing & Handles the post-processing of the raw data. For QKD it involves preparation and storage of raw key data, sifting, error correction, privacy amplification, authentication, and the communication over a classical public channel involved in these steps.\\
	\hangindent=4.6ex Q4.~Operation cycle & State machine that decides when to run subsystems in different regimes, at any given time, alternating between qubit transmission, calibration and other service procedures.\\
	\hangindent=4.6ex Q3.~Driver and calibration algorithms & Firmware/software routines that control low-level operation of analog electronics and electro-optical devices in different regimes.\\
	\hangindent=4.6ex Q2.~Analog electronics interface & Electronic signal processing and conditioning between firmware/software and electro-optical devices. This includes for example current-to-voltage conversion, signal amplification, mixing, frequency filtering, limiting, sampling, timing-to-digital and analog-to-digital conversions.\\
	\hangindent=4.6ex Q1.~Optics & Generation, modulation, transmission and detection of optical signals, implemented with optical and electro-optical components. This includes both quantum states and service optical signals for synchronisation and calibration. For example, in a decoy-state BB84 QKD protocol this layer may include generation of weak coherent pulses with different polarisations and intensities, their transmission, polarisation splitting and detection.\\*
	\hhline{==}
\end{longtable}

\section{Attacks exploiting superlinear detector control}
\label{sec:superlinear-detector-control-survey}

Many commercially available gated SPDs exhibit superlinearity at the edge of the gate \cite{lydersen2011b,yuan2011a,qian2018}. This is an unwanted SPD behavior that creates a loophole in QKD security. It may be exploited, for instance, in the following intercept-resend attack on the BB84 \cite{bennett1984} family of protocols. Eve uses a random basis to measure quantum states sent by Alice and resends her measurement results as multiphoton pulses, which are split into four (with a passive basis choice) or two (with an active basis choice) detectors at Bob. If the basis and bit value of the detector coincides with Eve's basis and bit value, it will absorb twice as many photons as each detector in the opposite basis to Eve's. Due to the superlinearity of the SPD, the probability of detection for Bob in the basis matching Eve's is higher than in the opposite basis. This contradicts the assumptions on Bob's measurement in the BB84 security proof. If the superlinearity is strong enough, the quantum bit error rate (QBER) under attack falls below $11\%$. However, a constraint of this regime is that Bob doesn't always detect Eve's multiphoton pulse, even in the basis matching Eve's. She can compensate for this efficiency loss by making her intercept setup more efficient and placing it close to Alice (thus excluding line loss), which may make her attack successful depending on the setup parameters \cite{lydersen2011b}.

A step for Eve to improve her control of Bob would be to make his detection probability unity. If Bob uses gated detectors, she can achieve this by sending her multiphoton pulse in between the gates \cite{wiechers2011}. This is a so-called ``after-gate attack''. However Eve's pulse, typically of hundreds of $\femto\joule$ energy, creates afterpulses in Bob's SPDs in the following gates. They contribute to QBER, together with Bob's normal dark count rate.

The next step for Eve would be to take Bob's detectors under a complete control, by eliminating his dark counts. She can completely blind them to single photons and make the dark count rate zero. This is usually achieved by illuminating Bob with a continuous-wave laser of power ranging from $\nano\watt$ to $\watt$ depending on the type of SPD \cite{lydersen2010a,lydersen2010b,lydersen2011c}. The blinding is caused by either constant photocurrent through the avalanche photodiode \cite{lydersen2010a}, its raised temperature \cite{lydersen2010b}, or even its permanent damage from a brief one-time application of a high-power laser \cite{bugge2014}. There are versions of this attack that use pulsed blinding illumination \cite{makarov2009,lydersen2010b,sauge2011}. Eve causes Bob's blinded detectors to click controllably typically by adding a bright pulse with appropriate timing and energy ranging from hundreds $\atto\joule$ to dozens $\femto\joule$ \cite{lydersen2010a,lydersen2010b,lydersen2011c,sauge2011}, similarly to the after-gate attack. For some SPDs, she can make clicks by introducing gaps in her blinding illumination \cite{makarov2009,tanner2014}. The blinding attack often allows a total detector control, with unity probability and no artefacts like afterpulses or dark counts.

In summary, Eve has two passive ways to use superlinearity in Bob's detectors---find it at the edges of \cite{lydersen2011b,yuan2011a,qian2018} or between the gates \cite{wiechers2011}---and four active ways to induce it---influence electronics by constant light (creating photocurrent) \cite{lydersen2010a}, cause heating by constant light \cite{lydersen2010b}, influence electronics by blinding pulses \cite{lydersen2010b,sauge2011}, and change the properties of the SPD by laser damage \cite{bugge2014}. This gives her several ways to attack the SPD and makes it tricky to develop reliable countermeasures \cite{fedorov2019}. This is arguably the most difficult vulnerability in today's QKD.

Let's consider if the detector control attacks can be revealed by statistical means, e.g.,\ by analysing attack's signature and any possible artefacts in QBER, dark count rate, key rate, and other parameters. To begin discussing this we define two energy levels, $E_\text{never}$ and $E_\text{always}$ \cite{lydersen2010a}. The former is the highest pulse energy that the SPD does not respond to with a click and the latter is the lowest energy that always causes a click. I.e.,\
\begin{equation}
	\label{eq:detector-never-always}
	\begin{aligned}
	P(E_\text{never})=0,\\
	P(E_\text{always})=1,
	\end{aligned}
\end{equation}
where $P(E)$ is the probability of the SPD to respond to the light pulse with energy $E$. If $E_\text{never}$ is much higher that the single-photon energy (which means SPD works as a classical power meter), Eve can send the pulse with energy $4 E_\text{never}$ for passive basis choice or $2 E_\text{never}$ for active basis choice. In the former case energy $2 E_\text{never}$ would always impinge on the detector decoding her state and energy $E_\text{never}$ would impinge on each of the two detectors in the opposite basis. In the latter case either energy $2 E_\text{never}$ would impinge on the appropriate detector (if Bob's and Eve's bases match) or this energy would be split equally with $E_\text{never}$ impinging on both detectors (if the bases don't match). While this control method doesn't introduce any QBER, $P(2 E_\text{never})$ can be much less than one, reducing the key generation rate. As discussed above, Eve needs to consider this constraint carefully \cite{lydersen2011b}. The situation becomes easier for Eve when $2 E_\text{never} \geq E_\text{always}$, thus $P(2 E_\text{never})=P(E_\text{always})=1$. Under such condition Eve can always get her resent state detected by Bob in case of the passive basis choice or half the time in case of the active basis choice \cite{lydersen2010a}. This is generally enough to maintain the same key rate as before the attack.

\textit{Countermeasures:} Most SPDs suffer from superlinearity. In the ten years following the discovery of this vulnerability, many countermeasures have been proposed. Let's group and review them.

\begin{enumerate}

\item ``Too good to be true''. Many detector-control attacks ironically improve the system performance: they decrease the QBER, decrease the dark count rate and increase the detection rate. However, monitoring for improved performance cannot be a reliable countermeasure, because Eve can always throttle the rate and intentionally introduce random errors in order to simulate the normal system performance \cite{gerhardt2011}.

\item ``Change the paradigm''. Measurement-device-independent (MDI) and twin-field QKD protocols \cite{lo2012,lucamarini2018,wang2018,wang2022,fan-yuan2022} exclude the detectors from the secure environment. This solution totally removes all the detector vulnerabilities. Unfortunately implementing one of these protocols in a commercial system requires a complete redesign of the system and makes it more expensive and slow. So far, only laboratory demonstrations and prototypes have been made (notably one by Toshiba \cite{woodward2021}), but no commercial product.

\item ``Observe the observer''.\label{item:Observe-the-observer} When SPDs are pushed into the superlinear regime, they manifest some artifacts unusual for normal workflow. The countermeasure can be watching the parameters of detectors. For the blinding attack, it could be measuring photocurrent \cite{yuan2011,gras2020,wu2020,bulavkin2022} (for further reading, see \cite{sidki2018}; for probabilistic blinding attack model and security proofs of the photocurrent measurement, see \cite{jun2015}). The after-gate attack can be caught by exact time measurement and observing afterpulse effects \cite{silva2012,koehler-sidki2019}. The thermal blinding can be observed by temperature measurement. Such countermeasures are usually very effective against the specific type of attack but close one loophole only and make the system more complex and expensive. However, they cannot eliminate attacks that cause small changes of physical parameters. For example, the attack at the falling edge of the gate uses a small amount of energy and small time delay \cite{qian2018}.

\item ``Add a watchdog''.\label{item:Add-a-watchdog} Adding a beamsplitter and a separate monitoring detector at the entrance of the receiver allows in principle to monitor for bright blinding light \cite{lydersen2010a,chistiakov2019}. However a hack-proof construction of this detector is a separate challenge and it may miss attacks that use a small amount of energy. Practical implementations of such monitoring detector have not been reported in the literature.

\item ``Check double clicks''. The basic detector-control attack produces too frequent double clicks in pairs of SPDs, which can be the basis of a countermeasure \cite{alhussein2019}. This countermeasure needs further experiments to check if an improved attack that circumvents it can be constructed.

\item ``Test the detectors''. Placing a calibrated light source inside the receiver and activating it at random times allows to test the detector response during a QKD session, e.g.,\ check that it is not blinded \cite{lydersen2011a}. When this countermeasure is integrated into a security proof, this imposes tight conditions on the equipment \cite{maroy2017}. It is not clear if these conditions are sufficiently practical to implement.

\item ``Detector decoy''. Ideas similar to the well-known decoy-state protocol \cite{ma2005} but implemented by varying the detector sensitivity between two levels were suggested as a countermeasure against the detector-control attacks. Distinguishing between weak and strong avalanches in a self-differencing detector allows to detect its blinding \cite{lee2016}. Another implementation places variable attenuators in front of each detector that randomly introduce $3~\deci\bel$ loss \cite{qian2019}. QBER and qubit rate for both $0$ and $3~\deci\bel$ loss settings are measured. Without the attack, QBER is expected to be below 11\% at both loss settings and the rate with $3~\deci\bel$ is expected to be half that with $0~\deci\bel$. This countermeasure is promising but needs further experiments to check its security.

\item ``Shake the box''. To catch unexpected superlinear regime Bob can decrease sensitivity or even turn off his SPDs for some time. So he wouldn't expect any qubits from Alice would be measured. Whatever is measured is either noise or Eve's attack. Bob can randomly turn off his gate (to catch the blinding or after-gate attacks \cite{lim2015}) or shift the gate time (to catch the after-gate attack and even attack at the falling edge \cite{silva2015}). This can be effective against the basic attacks \cite{lydersen2010a,lydersen2011c,lydersen2011b} but can be hacked by some modification of the basic attack \cite{huang2016}. Note that this countermeasure requires individual control over each detector gate, which may be impossible to implement in sinusoidally-gated and self-differencing detector schemes.

\item ``Kill the superlinearity''. Eliminating the superlinearity would achieve perfect security against detector control attacks. Optical limiters may be investigated for this purpose \cite{tutt1993,derosa2003,martincek2012,zhang2021}. Unfortunately, they start nonlinear behavior at power much higher (from dozens milliwatt) than used for blinding attacks (microwatts to milliwatts) and need a sufficient time to react (milliseconds) \cite{zhang2021}.

\end{enumerate}

\section{Test of QRate's detector for bright-light control}
\label{sec:test-of-detector-control}

\begin{figure}
	\includegraphics{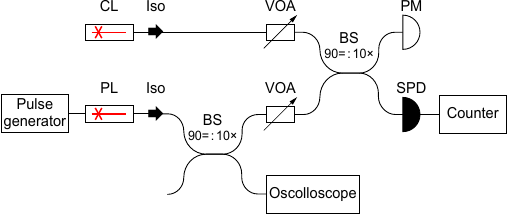}
	\caption{Setup for testing detector control by bright light. CL, cw laser ($1552~\nano\meter$, $40~\milli\watt$, Thorlabs SFL1550P); PL, pulsed laser ($1552~\nano\meter$, Gooch \& Housego AA1406); Iso, optical isolator; VOA, variable optical attenuator; BS, fiber beamsplitter; PM, power meter (Thorlabs PM400 with S155C head); SPD, single-photon detector under test. The pulse generator (Highland Technology P400) drives PL directly and can induce relaxation-limited short laser pulses. The counter (Stanford Research Systems SR620) was typically accumulating clicks over $100~\second$ for each data point. The oscilloscope (LeCroy 816Zi with OE555 optical-to-electrical converter) was used to observe the laser pulse shape.}
	\label{fig:blinding-scheme}
\end{figure}

Detector control by bright light was first proposed in \cite{makarov2009} and later demonstrated in a number of SPDs, making systems that use them insecure \cite{lydersen2010a,lydersen2010b,lydersen2011c,gerhardt2011,sauge2011,tanner2014,jogenfors2015,chistiakov2019}. We have subjected the SPD from QRate's system (detector serial number 20PD010013G ``Gleb'') to the same test, using a standard experimental setup shown in \cref{fig:blinding-scheme} \cite{lydersen2010a}. The scheme allows application of cw and pulsed light of controllable power to the SPD. We observe the SPD becoming blind (i.e.,\ stopping producing output pulses) at cw power of $3~\micro\watt$.

\begin{figure}
	\includegraphics{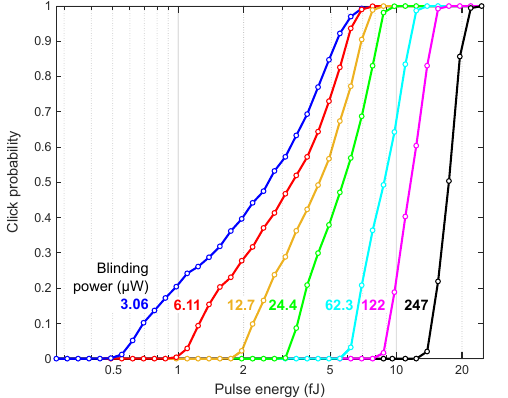}
	\caption{Detector control characteristics in the asynchronous regime at different cw blinding power levels. The trigger pulse FWHM was $0.4~\nano\second$.}
	\label{fig:blinding-plot}
\end{figure}

We then add bright trigger pulses that should produce a controlled click response. This SPD works in a sinusoidally-gated regime at $312.5~\mega\hertz$. In this test, we apply our trigger pulses at $100~\kilo\hertz$ and do not synchronise them to the detector gates. They thus impinge on the SPD at random times relative to the detector gate. One would expect the sensitivity of the blinded SPD to the trigger pulses to vary through the detector gate \cite{huang2016,qian2018}. Thus our asynchronous regime represents a worst-case condition for Eve. The measured control characteristics are shown in \cref{fig:blinding-plot}. While the pulse response at the minimum blinding power of $3~\micro\watt$ is somewhat unstable, from $6~\micro\watt$ on we observe a transition from 0 to exactly $100\%$ click probability. A close-to-perfect detector control, manifested in the click probability rising from 0 to $>99\%$ at $3~\deci\bel$ increase of the trigger pulse energy \cite{lydersen2010a}, is achieved in our SPD at blinding power $\geq 62~\micro\watt$.

This SPD is well controllable even in the asynchronous regime. The QKD system is thus certainly vulnerable and needs a countermeasure. A further treatment of this problem is given in \cite{acheva2023,kuzmin2023a}.

\section{Wide spectral testing}
\label{sec:spectral-testing}

Most of the known attacks and countermeasures for them traditionally considered Eve's access in Alice's and Bob's setups at around the QKD system operating wavelength ($\sim 1550~\nano\meter$). However, the transmission channel has much wider bandwidth (for quartz fiber it is $\sim 350$--$2400~\nano\meter$ \footnote{The quartz fiber transparency range is estimated approximately, since the boundaries of this range significantly depend on the fiber type and the technology of its production. The long-wavelength cutoff is determined by the radius of curvature of installed fiber, which is device-specific. We are not aware of reliable data in the literature.}) and gives Eve a potential to vary her light wavelength at which the attacks can be made. While the countermeasures implemented to protect from the attacks often work well at the QKD system operating wavelength, they may be completely unsafe in case of attacks in another spectral region \cite{jain2015}.

\begin{figure}
	\includegraphics{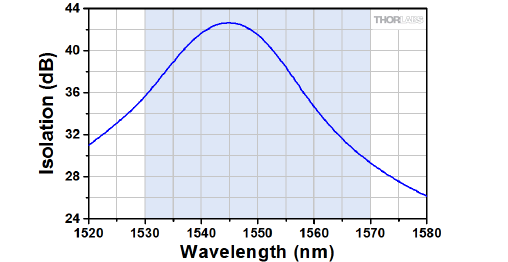}
	\caption{Isolation of Thorlabs IO-H-1550 fiber-optic isolator, from its specification sheet \cite{Thorlabs_IO-H-1550_spc_sheet}.}
	\label{fig:isolator-spec}
\end{figure}

\begin{figure}
	\includegraphics{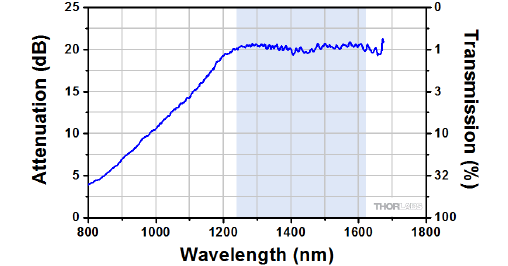}
	\caption{Attenuation of Thorlabs FA20T fiber-optic attenuator, from its specification sheet \cite{Thorlabs_FA20T_spc_sheet}.}
	\label{fig:attenuator-spec}
\end{figure}

As example, a standard approach to protect QKD systems from several attacks is their optical isolation with attenuators and optical isolators. Spectral characteristics of isolators (Iso1, Opneti IS-S-P-1550-900-1-0.3-FC-5.5x35; Iso2, Opneti IS-D-P-1550-900-1-0.3-FC-5.5x35) and attenuator (Att, Opneti FOA-P-1-20-FC), used for this purpose in QKD QRate system, are not provided by the manufacturer. But to illustrate the broadband attack principle let's consider similar devices from Thorlabs. Spectral attenuation of isolators (Thorlabs IO-H-1550) and attenuators (Thorlabs FA20T) are shown in \cref{fig:isolator-spec,fig:attenuator-spec} (from specification sheets). For IO-H-1550 isolation at the operating wavelength is about $43~\deci\bel$, but as the wavelength shift only to $1580~\nano\meter$ it becomes noticeably less, about $26~\deci\bel$. In the case of FA20T, its attenuation when shifting from the operating wavelength to $800~\nano\meter$ decreases from $20~\deci\bel$ to $4~\deci\bel$. Such a significant reduction of isolation may make the countermeasure ineffective. Note that the manufacturer's data shows their spectral properties only in a narrow range ($800$--$1700~\nano\meter$ and $1520$--$1580~\nano\meter$, respectively) while the quantum channel bandwidth is much wider ($\sim\:350$--$2400~\nano\meter$). It is possible that in the rest of the spectrum their attenuation is even less. 

In general, every QKD system has this problem. Most attacks can be optimised by Eve via varying the attack wavelength. Thus, the transmission and response to light of all the optical components involved in a particular attack must be characterised in the wide spectral range. Then, an optimum wavelength should be found at which each attack is the most efficient. The information leakage is quantified at this wavelength. To ease requirements on the dynamic range of the characterisation testbench, the components in the attack path are characterised individually, then their measured transmission characteristics are multiplied together.

We next list the attacks from this report that benefit from the wavelength-dependent component properties, and specify which components should be characterised. We then propose a testbench for the characterisation of fiber-optic components.

\subsection{Wavelength-dependent attacks}
\label{sec:wavelength-dependent-attacks}

\textbf{Superlinear detector control} (\cref{sec:detector-control}).
 
In the current implementation, the QRate QKD system is vulnerable to the detector control attacks at wavelengths over the entire sensitivity range of the SPD. In principle, SD (\cref{fig:setup}) might be used as a countermeasure to identify this attack's presence. But in this scheme SD is located after DWDM3 (Opneti DWDM-1-100-36-900-1-0.3-FC \cite{Opneti_fiber-optic_DWDM_specification_sheet_sheet}) in the $1554.94~\nano\meter$ port (channel 28). DWDM non-adjacent channel isolation is $>35~\deci\bel$ \cite{Opneti_fiber-optic_DWDM_specification_sheet_sheet}. This means that if Eve attacks outside the $1553.33$--$1556.55~\nano\meter$ range, the SD may not detect this. Thus, if a watchdog photodetector is used to monitor for the detector control attacks, it should be placed either in front of the DWDM3 or after it in the main signal path, via a beamsplitter not selective by wavelength. In order to exclude Eve's attempts to vary the attack wavelength, the watchdog photodetector's spectral sensitivity range should be wider than that of the SPDs. Otherwise Eve can choose a wavelength outside the photodetector's range, and blind and control Bob's SPDs unnoticed.

\textit{Components whose insertion loss or splitting ratio should be spectrally characterised:} DWDM3.

\textit{Components whose sensitivity should be spectrally characterised:} watchdog photodetector, SPD.

\textbf{Detector efficiency mismatch} (\cref{sec:efficiency-mismatch}).

Choosing the wavelength benefits Eve. The difference in the spectral and spatio-temporal properties of the beamsplitter (PBS) and photodetectors (SPD1, SPD2) mentioned above allows Eve to distinguish the photodetectors and activate them selectively, gaining the ability to steal the key. Eve can in addition select the attack wavelength at which the differences in the time and amplitude of the photodetector responses are maximized. To determine the optimal attack wavelength, one should measure the PBS splitting ratio over the wavelength, check each photodetector's spectral and time sensitivity separately, combine these measurements and determine the wavelength when the photodetectors' efficiency mismatch is maximized. However, if the four-state Bob is implemented as we suggest in \cref{sec:efficiency-mismatch}, this characterisation is not necessary.

\textit{Components whose insertion loss or splitting ratio should be spectrally characterised:} PBS.

\textit{Components whose sensitivity should be spectrally characterised:} SPD1, SPD2.

\textbf{Detector deadtime attack} (\cref{sec:deadtime}).

The deadtime loophole should be closed algorithmically as we suggest in \cref{sec:deadtime}. For completeness we remark that if it is not closed, the wavelength-dependent properties that affect the efficiency mismatch attack would also help Eve to select which of the two SPDs enters the deadtime.

\textbf{Trojan-horse attack} (\cref{sec:THA}).

As discussed in \cref{sec:THA}, if Eve uses $I_\text{in}=2.5\times10^{12}$ photons at the operating wavelength $\sim\!1550~\nano\meter$, the mean Trojan photon number $I_\text{max} \approx 1.5 \times 10^{-5}$ exiting Alice leads to significant information leakage, owing to $172~\deci\bel$ attenuation by Alice's components. 

We noted in \cref{sec:THA}, that to determine the maximum level of vulnerability of the QKD system by a Trojan-horse attack, one should find out the minimum total level of losses introduced by the entire system throughout whole range of quantum channel transparency and calculate corresponding maximum value of leaked signals $I_\text{max}$. Unfortunately, the manufacturer does not specify the spectral characteristics for QRate system components. These should be measured separately. Here, just to illustrate how crucial the choice of Trojan-horse attack wavelength is, we consider the spectral data of the Thorlabs devices discussed at the beginning of this section (\cref{sec:spectral-testing}), taking them as analogue to Iso2 and Att.

We roughly estimate Trojan-horse photon attenuation $\loss A$ (\cref{eq:AliceTrojanLoss}) in spectral region where Thorlabs elements (Iso2 and Att) are more transparent and determine the corresponding value of $I_\text{max}$. As mentioned above, manufacturer shows components' spectral data (\cref{fig:isolator-spec,fig:attenuator-spec}) only in a narrow band nearby operation wavelength, but even from these submitted short spectral range data it is obvious that, in principle, it is possible to select a spectral part in which attenuation becomes noticeably lower. From \cref{fig:isolator-spec,fig:attenuator-spec} we conservatively assume $\loss{Iso2rev}=26~\deci\bel$, $\loss{Att}=4~\deci\bel$. We assume that reverse loss of Iso1 decreases proportionally to that of Iso2 and $\loss{Iso1rev}=17~\deci\bel$. We guess that the attenuation of DWDM1 (Opneti DWDM-1-100-36-900-1-0.3-FC) and DWDM2 (Opneti DWDM-1-100-28-900-1-0.3-FC) for a non-adjacent channel is $>35~\deci\bel$ \cite{Opneti_fiber-optic_DWDM_specification_sheet_sheet}, but of course, losses outside the operating range certainly require experimental verification. We assume that the losses of VOA1, BS, PM1, and IM do not change significantly. From \cref{eq:AliceTrojanLoss} with these data we obtain $\loss A \approx 243~\deci\bel$ and $I_\text{max} \approx 1.25 \times 10^{-12}$. From \cite{wang2018a} and \cref{fig:WangKeyRates} we can estimate that in case of Eve's Trojan-horse attack at DWDM non-adjacent channel wavelength information leakage is low ($I_\text{max} \ll 10^{-9}$). It should be emphasised once again that the key value of DWDM loss outside its design spectral range requires experimental verification.

\textit{Components whose insertion loss or splitting ratio should be spectrally characterised:} Iso2, Iso1, DWDM2, Att, VOA1, DWDM1, BS, PM1, IM. 

\textbf{Laser-seeding attack} (\cref{sec:source-laser-seeding}).

As discussed in \cref{sec:source-laser-seeding} Eve might be able to inject light into Alice's L1 laser diode and modify its emission characteristics, e.g.,\ phase, intensity, and wavelength. But at operation wavelength attenuation inside Alice till laser L1 $\loss{As} \approx 124~\deci\bel$, making this attack unsuccessful. We estimate Alice's entry path losses in more transparent spectral region. With the assumptions made in \cref{sec:source-laser-seeding} and components' attenuation values discussed in the previous item, from \cref{eq:source-laser-seeding-loss} we get $\loss{As} \approx 142~\deci\bel$. The power reaching L1 is $W_\text{L1} = 10^{-\sfrac{\loss{As}}{10}} W_\text{in} \approx 0.63~\pico\watt$. This shows that even in the optimal wavelength case the seeding attack is still impossible \cite{huang2019}. Furthermore, L1 sensitivity for seeding outside the close vicinity of its emission wavelength is, in principle, significantly lower. This means that choosing the best wavelength is unlikely to give any benefit for this type of attack even if $W_\text{seed}$ reaches the values much higher than $0.63~\pico\watt$. Note that it is only a rough estimation and spectral minimum loss value in Alice requires additional broadband testing.

\textit{Components whose insertion loss or splitting ratio should be spectrally characterised:} Iso2, Iso1, DWDM2, Att, VOA1, DWDM1, BS, PM1, IM. 

\textit{Components whose sensitivity should be spectrally characterised:} L1.

\textbf{Light injection into power meter} (\cref{sec:AD}).

As explained in \cref{sec:AD}, attacking Alice's power meter PwM by injecting additional light into it, Eve tries to force Alice to disbalance her intensity modulator's zero point. This leads to a change in the intensities of the vacuum, decoy, and signal states, as well as their ratios. This would reduce the real secure key rate below that calculated by the system.

It was shown in \cref{sec:AD} that, if Eve attacks at the QKD operating wavelength, her additional power reaching PwM is less than $14~\nano\watt$, which is negligible in comparison with the power it measures in the normal operation. But, Eve can try to attack in the Alice's components maximum transparency spectral region. With all the \cref{sec:AD} assumptions and applying the above-mentioned component attenuation values in transparency region, by \cref{eq:AliceADLoss} we get the Alice's losses up to PwM $\loss{Ap}\approx 117~\deci\bel$ and seeding power reaching power meter $W_\text{PwM} = 10^{-\sfrac{\loss{Ap}}{10}} W_\text{in} \approx 0.2~\nano\watt$. This value is minor and even noticeably smaller that Eve can get with $1548.51~\nano\meter$ attack. Note that it is only a rough estimation and spectral minimum loss value in Alice requires additional broadband testing. 

\textit{Components whose insertion loss or splitting ratio should be spectrally characterised:} Iso2, Iso1, DWDM2, Att, VOA1, DWDM1.

\textbf{Induced-photorefraction attack} (\cref{sec:photorefraction}).

If Eve injects light into the QKD device this can change the photorefractive properties of Alice's and Bob's active lithium niobate elements and allow Eve to perform an induced photorefractive attack as described in \cref{sec:photorefraction}. Obviously, the magnitude of the photorefractive effect, and hence the effectiveness of the attack depends on radiation intensity reaching the active elements. Selection of injected light wavelength, in principle, can make it possible to use the spectral region with maximum optical channel transparency and increase the level of radiation power at PM and IM. We estimate radiation power at Alice's active elements in the case when the attack wavelength does not coincide with the QKD operating wavelength. Elements loss values are the same as previously assumed in this section. Using \cref{eq:Alice_Apm1_Loss,eq:Alice_Aim_Loss}, estimated power reaching PM1 and IM are as follows: $\loss{Apm1}\approx137.5~\deci\bel$, $W_\text{Apm1}\approx1.8~\pico\watt; \loss{Aim}\approx140~\deci\bel$, $W_\text{Aim}\approx1~\pico\watt.$ These power values turn out to be less than at the operating wavelength. This is due to the fact that two DWDMs introduce strong additional losses, which are essentially spectral filters that strictly pass only the operating wavelength. In this case, the photorefractive attack is completely ineffective.

It has already been discussed in this section, but needs to be emphasised again, that the DWDM insertion loss outside the working channel is conservatively assumed to be about $35~\deci\bel$, but we guess that outside the range of all working DWDM channels the transparency of this element may be significantly higher. This may cause much more power to reach the phase and amplitude modulators and the photorefractive attack may become possible. To evaluate this correctly, broadband spectral characterisation of the DWDM is needed.

\textit{Components whose insertion loss or splitting ratio should be spectrally characterised:} Iso2, Iso1, DWDM2, Att, VOA1, DWDM1, BS, PM1.

\textbf{Laser-damage attack} (\cref{sec:laser-damage}).

As has been shown above in this Appendix and in \cref{sec:laser-damage}, the Bob's side of QRate QKD is unprotected from all known types of receiver-side attacks (detector control, deadtime, mismatch). In this QKD implementation, there is no special isolation component at Bob's input to Eve's damage attack at the operating wavelength. In this way, Eve might try to reach inside Bob and implement the laser-damage attack on PBS, trying to change its polarisation splitting ratio. The DWDM3 installed at Bob's input limits Eve's ability to get inside Bob at other wavelengths to attack the PBS. But, if Eve applies a laser-damage attack to DWDM3, she might change its spectral properties and make DWDM3 transparent not only at a channel 36 wavelength, but also in other regions of the spectrum. After that, varying the attack wavelength Eve can choose the optimal one, at which changes in PBS under the laser light happen the most efficiently and as much as possible unbalance the PBS splitting ratio. Such PBS damaging may improve the detector deadtime and mismatch attacks.

Unlike Bob, Alice has protective components (Iso1, Iso2) that hamper Eve's attacks. \Cref{sec:laser-damage} recommends placing an additional isolator between Iso2 and the channel, as a countermeasure to the laser-damage attack. With such attack, under the action of laser radiation ($\sim\!1550~\nano\meter$, $\sim\!3.4~\watt$), the isolator either completely breaks and becomes practically opaque, or retains a residual isolation level of about $17~\deci\bel$. This limits the power of the attacking radiation passing through it and makes it impossible to defeat Iso2 and the other components beyond it.

However, Eve may benefit from the choice of wavelength. Due to different mechanisms of action, the impact of powerful attacking radiation with different wavelengths and illumination regimes may lead to different spectral changes in the losses introduced by the attacked isolator. It may be possible to choose such parameters of the attacking radiation that in some regions of the spectrum it will be possible to bleach the isolator and significantly reduce the losses introduced by it, which make it ineffective as a countermeasure and weaken protection against the other types of attacks. Furthermore, during the laser-damage attack the properties of optical components might not change uniformly in the entire spectral range. In principle, Eve might choose such a laser-damage regime when properties at the operating wavelength do not change significantly (i.e.,\ from the point of view of Alice and Bob, everything is okay), but change strongly in a different spectral range. This potentially gives Eve additional opportunities to implement ``invisible'' attacks. These properties of isolators under the laser-damage attack in different regimes require a separate detailed study.

\textit{Components whose insertion loss or splitting ratio should be spectrally characterised:} DWDM3, Iso2, Iso1, and PBS---all after the laser damage-attack.

\textbf{APD backflash} (\cref{sec:detector-backflash}).

Bob's components attenuate the spectrally broadband detector backflash. Their transmission in the backwards direction is needed for the calculation of emission probability into the channel, as explained in \cref{sec:detector-backflash}.

\textit{Components whose insertion loss or splitting ratio should be spectrally characterised:} DWDM3 in the reverse direction and eventually any additional components for Bob's setup.

\subsection{Ultra-wideband spectral testbench}
\label{sec:spectral-testing-testbench}

Since component manufacturers never provide the spectral characteristics in the full range we need ($\sim 350$--$2400~\nano\meter$), these need to be measured.

\begin{figure} 
	\includegraphics{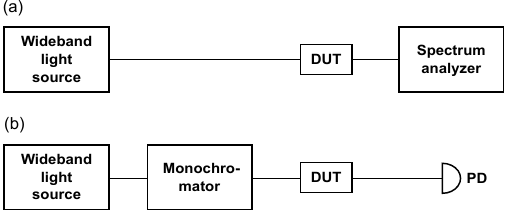}
	\caption{Measuring fiber-optic component attenuation spectra (a) with a spectrum analyser and (b) with a monochromator and photodetector. DUT, device under test; PD, photodetector.}
	\label{fig:WBS_setup}
\end{figure}

At least two different implementations for measuring fiber-optic elements attenuation spectra are possible. One uses a spectrum analyser [\cref{fig:WBS_setup}(a)], another a monochromator before device-under-test (DUT) and photodetector [\cref{fig:WBS_setup}(b)]. In both schemes, the spectrum is first scanned without the DUT (to characterise the instrument response), then with the DUT in place. The two spectral curves are then subtracted from one another, to obtain the DUT transmission curve. The testbench using a monochromator has the advantage that DUT is not exposed to high power, eliminating the possibility of heating it and potentially changing its characteristics. But, since the spectral range of measurements is wide and the required sensitivity is at least $60~\deci\bel$, as we need to characterize transmission of high-absorption fiber-optic components, in practice the setup with monochromator and photodetector is more difficult to implement instead of using a purchased spectrum analyser. The setup with spectrum analyser is also preferable due to its convenience of use and ergonomics. The difficulties encountered in the technical implementation of a setup using a monochromator are well illustrated in \cite{sushchev2021}, where a single-photon detector is employed as PD. They report $10~\nano\meter$ spectral resolution, spectral range of $1100$--$1800~\nano\meter$, and dynamic range of insertion loss measurement of about $70~\deci\bel$ (as visible in the plots in \cite{sushchev2021}). In comparison, our proposed testbench that uses an off-the-shelf spectrum analyser allows spectral resolution of $0.05~\nano\meter$, spectral range of $350$--$2400~\nano\meter$, and the dynamic range of measurements comparable to that demonstrated in \cite{sushchev2021}. We expect our testbench to be easier to align and operate. Next we will consider only measurements using the purchased spectrum analyser.

The key devices in the attenuation measurement spectrum setup are a spectrometer (spectrum analyser) and a light source. 

In today's optical instrumentation market there are spectrum analysers that completely cover the quantum channel transition wavelength range and have necessary sensitivity. Yokogawa here is the established leader. The advantage of these devices is high quality, user-friendly interface, a wide range of measured wavelengths, high sensitivity and dynamic range. The models line of Yokogawa spectrum analysers lacks a single device that completely covers the wavelength range necessary for our purposes. But a set of two devices does. These are the models AQ6374 ($350$--$1700~\nano\meter$) and AQ6375B ($1200$--$2400~\nano\meter$). Their spectral ranges overlap, allowing a more accurate ``stitching'' of data obtained in the two different wavelength ranges. 

Broadband light sources can be fundamentally divided into two large groups, according to their physical principle: incandescent lamps and laser (supercontinuum) sources. A practical disadvantage of incandescent-lamp-based sources is their low output power when coupled into a single-mode fiber. The supercontinuum (laser) sources are free from this shortcoming, their main disadvantage being a high cost. Supercontinuum sources that satisfy our requirements for power density and spectral range are commercially available.

The recognised world leader in the production of supercontinuum sources is NKT Photonics. This company offers a wide range of supercontinuum sources that differ in the range of generated wavelengths, average power and power spectral density in different parts of the spectrum. One of the optimal devices for our application is the SuperK Extreme/Fianium FIU-15 model. The range of light generation is from about $350$ to $2500~\nano\meter$. The integrated optical power is about $4.5~\watt$, the spectral power density varies through the wavelength range and averages about $3$--$4~\milli\watt\per\nano\meter$. The output radiation power can be manually and programmatically adjusted. The source is characterised by high stability of the output optical power in the entire range ($<\!0.5\%$). The polarisation of the output light is random.

Most of the fiber-optic components to be tested have standard single-mode fiber pigtails (9.5/125 $\micro\meter$) with FC type connectors. The advantage of Yokogawa spectrum analysers is that they have an input for a single-mode fiber with this type of connector. For connecting the supercontinuum source to the FC connectorised single-mode fiber, an accessory optical coupler from NKT Photonics has to be used. To cover our entire range, we can use two of them: $350$--$1200~\nano\meter$ model (SuperK Connect FD7) and $1200$--$2400~\nano\meter$ model (SuperK Connect FD6).

\begin{figure}
	\includegraphics{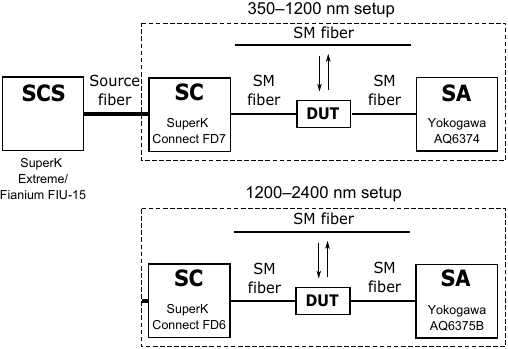}
	\caption{Setup implementation for measuring fiber-optic component transmission spectra with NKT Photonics and Yokogawa devices. SCS, supercontinuum source; SC, single-mode fiber coupler; SA, spectrum analyser.}
	\label{fig:scheme_WB_Measurement}
\end{figure}

One of the possible implementations of the testbench with these instruments is shown in \cref{fig:scheme_WB_Measurement}. The setup consists of three parts:
\begin{itemize}
	\item supercontinuum source SuperK Extreme/Fianium FIU-15;
	\item path for measurements in $350$--$1200~\nano\meter$ range: FD7 connector, DUT, AQ6374 spectrum analyser;
	\item path for measurements in $1200$--$2400~\nano\meter$ range: FD6 connector, DUT, AQ6375B spectrum analyser.
\end{itemize}
The device under test is measured in both paths and the results are combined to obtain its complete transmission spectrum.

To eliminate the possibility of influence on DUT by the source's radiation, which can potentially lead to its characteristics changing, we estimated the power impinging on DUT during the measurements. The maximum FIU-15 power density after single mode fiber coupler FD6 is approximately $1~\milli\watt\per\nano\meter$. We hope that significant changes to spectral characteristics of fiber-optic components under such power density is unlikely. However, this should be experimentally verified for every type of DUT. This can be done by carrying out several successive measurements with different radiation power of FIU-15 light source. If DUT's characteristics change under higher power, the results of these measurements will not match. Changing the integral output power of FIU-15 source is possible by varying its pulse repetition rate, which does not affect its spectrum.

We have assembled this testbench and are currently refining its usage methodology \cite{tan2024}. Meanwhile, the testbench with monochromator (an acousto-optic filter) and single-photon detector is implemented by SFB Lab in Moscow \cite{sushchev2021}.

As discussed in the previous section (\cref{sec:wavelength-dependent-attacks}), not only the fiber-optic components' spectral characteristics need to be measured, but the photodetectors' spectral sensitivity as well. These PD spectral characteristics are critical in the risk evaluation of several attacks: superlinear detector control, detector efficiency mismatch, and light injection into calibration photodetector.

The characterisation of photodetectors' spectral sensitivity is not as demanding task as for high-absorbing fiber-optic components, for the following reasons. Since QKD systems use sensitive photodetectors, there is no need to characterise them using a bright light source with high power spectral density. The spectral sensitivity range of photodetectors is limited (about $900$--$1700~\nano\meter$), which allows the use of non-ultra-wideband light sources. Thus, a simple to implement and relatively cheap setup can be used as a light source, shown in \cref{fig:WBS_setup}(b) (without the DUT), with the incandescent lamp (e.g.,\ Thorlabs SLS201L/M stabilised fiber coupled light source) and a narrow-band monochromator with a fiber-optic output (e.g.,\ Zolix Omni-$\lambda$305i).

\subsection{Consistency of broadband spectral properties}
\label{sec:Interpretation_spectral_test_results}

Even if all broadband spectral tests are performed properly for all the elements of a particular QKD system and its safety is fully proven, it is not possible to say with complete certainty, without additional assumptions, that another sample of the same QKD system is safe without performing exactly the same extensive tests. This can be due to a sample-to-sample variation of the system elements. Most manufacturers guarantee parameters of the elements only in a very narrow spectral region, close to a specific wavelength. In one manufacturing batch, the spectral properties of elements from the same manufacturer may coincide. But another batch of these elements may be made with a slight change of the manufacturing technology, maintaining the declared properties in the narrow spectral range. Such changes in technology can however lead to uncontrollable spectral transparency loopholes outside of the narrow range. To prevent this, three approaches are possible.
\begin{itemize}
\item Perform the spectral testing of all elements of the system, several samples for each of them. Assume that the properties of all the other samples will be the same as in those tested. There is a risk that this is not the case.
\item Spectrally test all elements for each particular QKD system. This is very expensive.
\item Install a filtering-and-isolation subsystem at Alice's output and Bob's input of each QKD system. This subsystem ensures a sufficient attenuation through the entire spectrum and is fully characterised in every manufacturing sample. The rest of the system is not characterised and is assumed to be unity-transmission for the purposes of the security proofs.
\end{itemize}

\def\bibsection{\medskip\begin{center}\rule{0.5\columnwidth}{.8pt}\end{center}\medskip} 


\begin{thebibliography}{165}%
\makeatletter
\providecommand \@ifxundefined [1]{%
 \@ifx{#1\undefined}
}%
\providecommand \@ifnum [1]{%
 \ifnum #1\expandafter \@firstoftwo
 \else \expandafter \@secondoftwo
 \fi
}%
\providecommand \@ifx [1]{%
 \ifx #1\expandafter \@firstoftwo
 \else \expandafter \@secondoftwo
 \fi
}%
\providecommand \natexlab [1]{#1}%
\providecommand \enquote  [1]{``#1''}%
\providecommand \bibnamefont  [1]{#1}%
\providecommand \bibfnamefont [1]{#1}%
\providecommand \citenamefont [1]{#1}%
\providecommand \href@noop [0]{\@secondoftwo}%
\providecommand \href [0]{\begingroup \@sanitize@url \@href}%
\providecommand \@href[1]{\@@startlink{#1}\@@href}%
\providecommand \@@href[1]{\endgroup#1\@@endlink}%
\providecommand \@sanitize@url [0]{\catcode `\\12\catcode `\$12\catcode
  `\&12\catcode `\#12\catcode `\^12\catcode `\_12\catcode `\%12\relax}%
\providecommand \@@startlink[1]{}%
\providecommand \@@endlink[0]{}%
\providecommand \url  [0]{\begingroup\@sanitize@url \@url }%
\providecommand \@url [1]{\endgroup\@href {#1}{\urlprefix }}%
\providecommand \urlprefix  [0]{URL }%
\providecommand \Eprint [0]{\href }%
\providecommand \doibase [0]{http://dx.doi.org/}%
\providecommand \selectlanguage [0]{\@gobble}%
\providecommand \bibinfo  [0]{\@secondoftwo}%
\providecommand \bibfield  [0]{\@secondoftwo}%
\providecommand \translation [1]{[#1]}%
\providecommand \BibitemOpen [0]{}%
\providecommand \bibitemStop [0]{}%
\providecommand \bibitemNoStop [0]{.\EOS\space}%
\providecommand \EOS [0]{\spacefactor3000\relax}%
\providecommand \BibitemShut  [1]{\csname bibitem#1\endcsname}%
\let\auto@bib@innerbib\@empty
\bibitem [{\citenamefont {Bennett}\ \emph {et~al.}(1992)\citenamefont
  {Bennett}, \citenamefont {Bessette}, \citenamefont {Salvail}, \citenamefont
  {Brassard},\ and\ \citenamefont {Smolin}}]{bennett1992b}%
  \BibitemOpen
  \bibfield  {author} {\bibinfo {author} {\bibfnamefont {C.~H.}\ \bibnamefont
  {Bennett}}, \bibinfo {author} {\bibfnamefont {F.}~\bibnamefont {Bessette}},
  \bibinfo {author} {\bibfnamefont {L.}~\bibnamefont {Salvail}}, \bibinfo
  {author} {\bibfnamefont {G.}~\bibnamefont {Brassard}}, \ and\ \bibinfo
  {author} {\bibfnamefont {J.}~\bibnamefont {Smolin}},\ }\bibfield  {title}
  {\enquote {\bibinfo {title} {Experimental quantum cryptography},}\ }\href
  {\doibase 10.1007/bf00191318} {\bibfield  {journal} {\bibinfo  {journal} {J.
  Cryptology}\ }\textbf {\bibinfo {volume} {5}},\ \bibinfo {pages} {3--28}
  (\bibinfo {year} {1992})}\BibitemShut {NoStop}%
\bibitem [{\citenamefont {Chen}\ \emph {et~al.}(2021)\citenamefont {Chen},
  \citenamefont {Zhang}, \citenamefont {Chen}, \citenamefont {Cai},
  \citenamefont {Liao}, \citenamefont {Zhang}, \citenamefont {Chen},
  \citenamefont {Yin}, \citenamefont {Ren}, \citenamefont {Chen}, \citenamefont
  {Han}, \citenamefont {Yu}, \citenamefont {Liang}, \citenamefont {Zhou},
  \citenamefont {Yuan}, \citenamefont {Zhao}, \citenamefont {Wang},
  \citenamefont {Jiang}, \citenamefont {Zhang}, \citenamefont {Liu},
  \citenamefont {Li}, \citenamefont {Shen}, \citenamefont {Cao}, \citenamefont
  {Lu}, \citenamefont {Shu}, \citenamefont {Wang}, \citenamefont {Li},
  \citenamefont {Liu}, \citenamefont {Xu}, \citenamefont {Wang}, \citenamefont
  {Peng},\ and\ \citenamefont {Pan}}]{chen2021}%
  \BibitemOpen
  \bibfield  {author} {\bibinfo {author} {\bibfnamefont {Yu-Ao}\ \bibnamefont
  {Chen}}, \bibinfo {author} {\bibfnamefont {Qiang}\ \bibnamefont {Zhang}},
  \bibinfo {author} {\bibfnamefont {Teng-Yun}\ \bibnamefont {Chen}}, \bibinfo
  {author} {\bibfnamefont {Wen-Qi}\ \bibnamefont {Cai}}, \bibinfo {author}
  {\bibfnamefont {Sheng-Kai}\ \bibnamefont {Liao}}, \bibinfo {author}
  {\bibfnamefont {Jun}\ \bibnamefont {Zhang}}, \bibinfo {author} {\bibfnamefont
  {Kai}\ \bibnamefont {Chen}}, \bibinfo {author} {\bibfnamefont {Juan}\
  \bibnamefont {Yin}}, \bibinfo {author} {\bibfnamefont {Ji-Gang}\ \bibnamefont
  {Ren}}, \bibinfo {author} {\bibfnamefont {Zhu}\ \bibnamefont {Chen}},
  \bibinfo {author} {\bibfnamefont {Sheng-Long}\ \bibnamefont {Han}}, \bibinfo
  {author} {\bibfnamefont {Qing}\ \bibnamefont {Yu}}, \bibinfo {author}
  {\bibfnamefont {Ken}\ \bibnamefont {Liang}}, \bibinfo {author} {\bibfnamefont
  {Fei}\ \bibnamefont {Zhou}}, \bibinfo {author} {\bibfnamefont {Xiao}\
  \bibnamefont {Yuan}}, \bibinfo {author} {\bibfnamefont {Mei-Sheng}\
  \bibnamefont {Zhao}}, \bibinfo {author} {\bibfnamefont {Tian-Yin}\
  \bibnamefont {Wang}}, \bibinfo {author} {\bibfnamefont {Xiao}\ \bibnamefont
  {Jiang}}, \bibinfo {author} {\bibfnamefont {Liang}\ \bibnamefont {Zhang}},
  \bibinfo {author} {\bibfnamefont {Wei-Yue}\ \bibnamefont {Liu}}, \bibinfo
  {author} {\bibfnamefont {Yang}\ \bibnamefont {Li}}, \bibinfo {author}
  {\bibfnamefont {Qi}~\bibnamefont {Shen}}, \bibinfo {author} {\bibfnamefont
  {Yuan}\ \bibnamefont {Cao}}, \bibinfo {author} {\bibfnamefont {Chao-Yang}\
  \bibnamefont {Lu}}, \bibinfo {author} {\bibfnamefont {Rong}\ \bibnamefont
  {Shu}}, \bibinfo {author} {\bibfnamefont {Jian-Yu}\ \bibnamefont {Wang}},
  \bibinfo {author} {\bibfnamefont {Li}~\bibnamefont {Li}}, \bibinfo {author}
  {\bibfnamefont {Nai-Le}\ \bibnamefont {Liu}}, \bibinfo {author}
  {\bibfnamefont {Feihu}\ \bibnamefont {Xu}}, \bibinfo {author} {\bibfnamefont
  {Xiang-Bin}\ \bibnamefont {Wang}}, \bibinfo {author} {\bibfnamefont
  {Cheng-Zhi}\ \bibnamefont {Peng}}, \ and\ \bibinfo {author} {\bibfnamefont
  {Jian-Wei}\ \bibnamefont {Pan}},\ }\bibfield  {title} {\enquote {\bibinfo
  {title} {An integrated space-to-ground quantum communication network over
  4,600 kilometres},}\ }\href {\doibase 10.1038/s41586-020-03093-8} {\bibfield
  {journal} {\bibinfo  {journal} {Nature}\ }\textbf {\bibinfo {volume} {589}},\
  \bibinfo {pages} {214} (\bibinfo {year} {2021})}\BibitemShut {NoStop}%
\bibitem [{\citenamefont {Vedovato}\ \emph {et~al.}()\citenamefont {Vedovato},
  \citenamefont {Villoresi}, \citenamefont {Vallone}, \citenamefont
  {Kutschera}, \citenamefont {Lopez}, \citenamefont {Martin}, \citenamefont
  {Bejarano}, \citenamefont {Piotr},\ and\ \citenamefont
  {Geitz}}]{vedovato2021}%
  \BibitemOpen
  \bibfield  {author} {\bibinfo {author} {\bibfnamefont {Francesco}\
  \bibnamefont {Vedovato}}, \bibinfo {author} {\bibfnamefont {Paolo}\
  \bibnamefont {Villoresi}}, \bibinfo {author} {\bibfnamefont {Giuseppe}\
  \bibnamefont {Vallone}}, \bibinfo {author} {\bibfnamefont {Florian}\
  \bibnamefont {Kutschera}}, \bibinfo {author} {\bibfnamefont {Victor}\
  \bibnamefont {Lopez}}, \bibinfo {author} {\bibfnamefont {Vicente}\
  \bibnamefont {Martin}}, \bibinfo {author} {\bibfnamefont {Jose Luis~Rosales}\
  \bibnamefont {Bejarano}}, \bibinfo {author} {\bibfnamefont {Rydlichowski}\
  \bibnamefont {Piotr}}, \ and\ \bibinfo {author} {\bibfnamefont {Marc}\
  \bibnamefont {Geitz}},\ }\href@noop {} {\enquote {\bibinfo {title} {{OPENQKD}
  deliverable {D8.3.} {R}eport on testbed replicability and performance},}\
  }\bibinfo {note}
  {\url{https://ec.europa.eu/research/participants/documents/downloadPublic?documentIds=080166e5da0754d3&appId=PPGMS},
  visited 13 Feb 2023}\BibitemShut {NoStop}%
\bibitem [{\citenamefont {Mehic}\ \emph {et~al.}(2020)\citenamefont {Mehic},
  \citenamefont {Niemiec}, \citenamefont {Rass}, \citenamefont {Ma},
  \citenamefont {Peev}, \citenamefont {Aguado}, \citenamefont {Martin},
  \citenamefont {Schauer}, \citenamefont {Poppe}, \citenamefont {Pacher},\ and\
  \citenamefont {Voznak}}]{mehic2020}%
  \BibitemOpen
  \bibfield  {author} {\bibinfo {author} {\bibfnamefont {Miralem}\ \bibnamefont
  {Mehic}}, \bibinfo {author} {\bibfnamefont {Marcin}\ \bibnamefont {Niemiec}},
  \bibinfo {author} {\bibfnamefont {Stefan}\ \bibnamefont {Rass}}, \bibinfo
  {author} {\bibfnamefont {Jiajun}\ \bibnamefont {Ma}}, \bibinfo {author}
  {\bibfnamefont {Momtchil}\ \bibnamefont {Peev}}, \bibinfo {author}
  {\bibfnamefont {Alejandro}\ \bibnamefont {Aguado}}, \bibinfo {author}
  {\bibfnamefont {Vicente}\ \bibnamefont {Martin}}, \bibinfo {author}
  {\bibfnamefont {Stefan}\ \bibnamefont {Schauer}}, \bibinfo {author}
  {\bibfnamefont {Andreas}\ \bibnamefont {Poppe}}, \bibinfo {author}
  {\bibfnamefont {Christoph}\ \bibnamefont {Pacher}}, \ and\ \bibinfo {author}
  {\bibfnamefont {Miroslav}\ \bibnamefont {Voznak}},\ }\bibfield  {title}
  {\enquote {\bibinfo {title} {Quantum key distribution: a networking
  perspective},}\ }\href {\doibase 10.1145/3402192} {\bibfield  {journal}
  {\bibinfo  {journal} {ACM Comp. Surv.}\ }\textbf {\bibinfo {volume} {53}},\
  \bibinfo {pages} {96} (\bibinfo {year} {2020})}\BibitemShut {NoStop}%
\bibitem [{\citenamefont {L{\" a}nger}\ and\ \citenamefont
  {Lenhart}(2009)}]{langer2009}%
  \BibitemOpen
  \bibfield  {author} {\bibinfo {author} {\bibfnamefont {Thomas}\ \bibnamefont
  {L{\" a}nger}}\ and\ \bibinfo {author} {\bibfnamefont {Gaby}\ \bibnamefont
  {Lenhart}},\ }\bibfield  {title} {\enquote {\bibinfo {title} {Standardization
  of quantum key distribution and the {ETSI} standardization initiative
  {ISG-QKD}},}\ }\href {\doibase 10.1088/1367-2630/11/5/055051} {\bibfield
  {journal} {\bibinfo  {journal} {New J. Phys.}\ }\textbf {\bibinfo {volume}
  {11}},\ \bibinfo {pages} {055051} (\bibinfo {year} {2009})}\BibitemShut
  {NoStop}%
\bibitem [{\citenamefont {All{\' e}aume}\ \emph {et~al.}(2014)\citenamefont
  {All{\' e}aume}, \citenamefont {Chapuran}, \citenamefont {Chunnilall},
  \citenamefont {Degiovanni}, \citenamefont {L{\" u}tkenhaus}, \citenamefont
  {Martin}, \citenamefont {Mink}, \citenamefont {Peev}, \citenamefont
  {Lucamarini}, \citenamefont {Ward},\ and\ \citenamefont
  {Shields}}]{alleaume2014}%
  \BibitemOpen
  \bibfield  {author} {\bibinfo {author} {\bibfnamefont {Romain}\ \bibnamefont
  {All{\' e}aume}}, \bibinfo {author} {\bibfnamefont {Thomas~E.}\ \bibnamefont
  {Chapuran}}, \bibinfo {author} {\bibfnamefont {Christopher~J.}\ \bibnamefont
  {Chunnilall}}, \bibinfo {author} {\bibfnamefont {Ivo~P.}\ \bibnamefont
  {Degiovanni}}, \bibinfo {author} {\bibfnamefont {Norbert}\ \bibnamefont {L{\"
  u}tkenhaus}}, \bibinfo {author} {\bibfnamefont {Vincente}\ \bibnamefont
  {Martin}}, \bibinfo {author} {\bibfnamefont {Alan}\ \bibnamefont {Mink}},
  \bibinfo {author} {\bibfnamefont {Momtchil}\ \bibnamefont {Peev}}, \bibinfo
  {author} {\bibfnamefont {Marco}\ \bibnamefont {Lucamarini}}, \bibinfo
  {author} {\bibfnamefont {Martin}\ \bibnamefont {Ward}}, \ and\ \bibinfo
  {author} {\bibfnamefont {Andrew}\ \bibnamefont {Shields}},\ }\bibfield
  {title} {\enquote {\bibinfo {title} {Worldwide standardization activity for
  quantum key distribution},}\ }in\ \href {\doibase
  10.1109/GLOCOMW.2014.7063507} {\emph {\bibinfo {booktitle} {Proc. IEEE
  Globecom Workshop 2014}}}\ (\bibinfo  {publisher} {IEEE Press},\ \bibinfo
  {year} {2014})\ pp.\ \bibinfo {pages} {656--661}\BibitemShut {NoStop}%
\bibitem [{\citenamefont {Lo}\ \emph {et~al.}(2014)\citenamefont {Lo},
  \citenamefont {Curty},\ and\ \citenamefont {Tamaki}}]{lo2014}%
  \BibitemOpen
  \bibfield  {author} {\bibinfo {author} {\bibfnamefont {Hoi-Kwong}\
  \bibnamefont {Lo}}, \bibinfo {author} {\bibfnamefont {Marcos}\ \bibnamefont
  {Curty}}, \ and\ \bibinfo {author} {\bibfnamefont {Kiyoshi}\ \bibnamefont
  {Tamaki}},\ }\bibfield  {title} {\enquote {\bibinfo {title} {Secure quantum
  key distribution},}\ }\href {\doibase 10.1038/nphoton.2014.149} {\bibfield
  {journal} {\bibinfo  {journal} {Nat. Photonics}\ }\textbf {\bibinfo {volume}
  {8}},\ \bibinfo {pages} {595--604} (\bibinfo {year} {2014})}\BibitemShut
  {NoStop}%
\bibitem [{\citenamefont {Dixon}\ \emph {et~al.}(2017)\citenamefont {Dixon},
  \citenamefont {Dynes}, \citenamefont {Lucamarini}, \citenamefont {Fr{\"
  o}hlich}, \citenamefont {Sharpe}, \citenamefont {Plews}, \citenamefont {Tam},
  \citenamefont {Yuan}, \citenamefont {Tanizawa}, \citenamefont {Sato},
  \citenamefont {Kawamura}, \citenamefont {Fujiwara}, \citenamefont {Sasaki},\
  and\ \citenamefont {Shields}}]{dixon2017}%
  \BibitemOpen
  \bibfield  {author} {\bibinfo {author} {\bibfnamefont {A.~R.}\ \bibnamefont
  {Dixon}}, \bibinfo {author} {\bibfnamefont {J.~F.}\ \bibnamefont {Dynes}},
  \bibinfo {author} {\bibfnamefont {M.}~\bibnamefont {Lucamarini}}, \bibinfo
  {author} {\bibfnamefont {B.}~\bibnamefont {Fr{\" o}hlich}}, \bibinfo {author}
  {\bibfnamefont {A.~W.}\ \bibnamefont {Sharpe}}, \bibinfo {author}
  {\bibfnamefont {A.}~\bibnamefont {Plews}}, \bibinfo {author} {\bibfnamefont
  {W.}~\bibnamefont {Tam}}, \bibinfo {author} {\bibfnamefont {Z.~L.}\
  \bibnamefont {Yuan}}, \bibinfo {author} {\bibfnamefont {Y.}~\bibnamefont
  {Tanizawa}}, \bibinfo {author} {\bibfnamefont {H.}~\bibnamefont {Sato}},
  \bibinfo {author} {\bibfnamefont {S.}~\bibnamefont {Kawamura}}, \bibinfo
  {author} {\bibfnamefont {M.}~\bibnamefont {Fujiwara}}, \bibinfo {author}
  {\bibfnamefont {M.}~\bibnamefont {Sasaki}}, \ and\ \bibinfo {author}
  {\bibfnamefont {A.~J.}\ \bibnamefont {Shields}},\ }\bibfield  {title}
  {\enquote {\bibinfo {title} {Quantum key distribution with hacking
  countermeasures and long term field trial},}\ }\href {\doibase
  10.1038/s41598-017-01884-0} {\bibfield  {journal} {\bibinfo  {journal} {Sci.
  Rep.}\ }\textbf {\bibinfo {volume} {7}},\ \bibinfo {pages} {1978} (\bibinfo
  {year} {2017})}\BibitemShut {NoStop}%
\bibitem [{ets()}]{etsi2018}%
  \BibitemOpen
  \href@noop {} {\enquote {\bibinfo {title} {{ETSI} white paper no.~27.
  {I}mplementation security of quantum cryptography},}\ }\bibinfo {note}
  {\url{https://www.etsi.org/images/files/ETSIWhitePapers/etsi_wp27_qkd_imp_sec_FINAL.pdf},
  visited 13 Feb 2023}\BibitemShut {NoStop}%
\bibitem [{\citenamefont {Tomita}(2019)}]{tomita2019}%
  \BibitemOpen
  \bibfield  {author} {\bibinfo {author} {\bibfnamefont {Akihisa}\ \bibnamefont
  {Tomita}},\ }\bibfield  {title} {\enquote {\bibinfo {title} {Implementation
  security certification of decoy-{BB84} quantum key distribution systems},}\
  }\href {\doibase 10.1002/qute.201900005} {\bibfield  {journal} {\bibinfo
  {journal} {Adv. Quantum Technol.}\ }\textbf {\bibinfo {volume} {2}},\
  \bibinfo {pages} {1900005} (\bibinfo {year} {2019})}\BibitemShut {NoStop}%
\bibitem [{\citenamefont {Xu}\ \emph {et~al.}(2020)\citenamefont {Xu},
  \citenamefont {Ma}, \citenamefont {Zhang}, \citenamefont {Lo},\ and\
  \citenamefont {Pan}}]{xu2020}%
  \BibitemOpen
  \bibfield  {author} {\bibinfo {author} {\bibfnamefont {Feihu}\ \bibnamefont
  {Xu}}, \bibinfo {author} {\bibfnamefont {Xiongfeng}\ \bibnamefont {Ma}},
  \bibinfo {author} {\bibfnamefont {Qiang}\ \bibnamefont {Zhang}}, \bibinfo
  {author} {\bibfnamefont {Hoi-Kwong}\ \bibnamefont {Lo}}, \ and\ \bibinfo
  {author} {\bibfnamefont {Jian-Wei}\ \bibnamefont {Pan}},\ }\bibfield  {title}
  {\enquote {\bibinfo {title} {Secure quantum key distribution with realistic
  devices},}\ }\href {\doibase 10.1103/RevModPhys.92.025002} {\bibfield
  {journal} {\bibinfo  {journal} {Rev. Mod. Phys.}\ }\textbf {\bibinfo {volume}
  {92}},\ \bibinfo {pages} {025002} (\bibinfo {year} {2020})}\BibitemShut
  {NoStop}%
\bibitem [{\citenamefont {Sun}\ and\ \citenamefont {Huang}(2022)}]{sun2022}%
  \BibitemOpen
  \bibfield  {author} {\bibinfo {author} {\bibfnamefont {Shihai}\ \bibnamefont
  {Sun}}\ and\ \bibinfo {author} {\bibfnamefont {Anqi}\ \bibnamefont {Huang}},\
  }\bibfield  {title} {\enquote {\bibinfo {title} {A review of security
  evaluation of practical quantum key distribution system},}\ }\href {\doibase
  10.3390/e24020260} {\bibfield  {journal} {\bibinfo  {journal} {Entropy}\
  }\textbf {\bibinfo {volume} {24}},\ \bibinfo {pages} {260} (\bibinfo {year}
  {2022})}\BibitemShut {NoStop}%
\bibitem [{\citenamefont {Marquardt}\ \emph {et~al.}()\citenamefont
  {Marquardt}, \citenamefont {Seyfarth}, \citenamefont {Bettendorf},
  \citenamefont {Bohmann}, \citenamefont {Buchner}, \citenamefont {Curty},
  \citenamefont {Elser}, \citenamefont {Eul}, \citenamefont {Gehring},
  \citenamefont {Jain}, \citenamefont {Klocke}, \citenamefont {Reinecke},
  \citenamefont {Sieber}, \citenamefont {Ursin}, \citenamefont {Wehling},\ and\
  \citenamefont {Weier}}]{marquardt2023}%
  \BibitemOpen
  \bibfield  {author} {\bibinfo {author} {\bibfnamefont {Christoph}\
  \bibnamefont {Marquardt}}, \bibinfo {author} {\bibfnamefont {Ulrich}\
  \bibnamefont {Seyfarth}}, \bibinfo {author} {\bibfnamefont {Sven}\
  \bibnamefont {Bettendorf}}, \bibinfo {author} {\bibfnamefont {Martin}\
  \bibnamefont {Bohmann}}, \bibinfo {author} {\bibfnamefont {Alexander}\
  \bibnamefont {Buchner}}, \bibinfo {author} {\bibfnamefont {Marcos}\
  \bibnamefont {Curty}}, \bibinfo {author} {\bibfnamefont {Dominique}\
  \bibnamefont {Elser}}, \bibinfo {author} {\bibfnamefont {Silas}\ \bibnamefont
  {Eul}}, \bibinfo {author} {\bibfnamefont {Tobias}\ \bibnamefont {Gehring}},
  \bibinfo {author} {\bibfnamefont {Nitin}\ \bibnamefont {Jain}}, \bibinfo
  {author} {\bibfnamefont {Thomas}\ \bibnamefont {Klocke}}, \bibinfo {author}
  {\bibfnamefont {Marie}\ \bibnamefont {Reinecke}}, \bibinfo {author}
  {\bibfnamefont {Nico}\ \bibnamefont {Sieber}}, \bibinfo {author}
  {\bibfnamefont {Rupert}\ \bibnamefont {Ursin}}, \bibinfo {author}
  {\bibfnamefont {Marc}\ \bibnamefont {Wehling}}, \ and\ \bibinfo {author}
  {\bibfnamefont {Henning}\ \bibnamefont {Weier}},\ }\href@noop {} {\enquote
  {\bibinfo {title} {Implementation attacks against {QKD} systems},}\ }\bibinfo
  {note} {{BSI} technical report,
  \url{https://www.bsi.bund.de/EN/Service-Navi/Publikationen/Studien/QKD-Systems/Implementation_Attacks_QKD_Systems_node.html},
  visited 14 Feb 2024}\BibitemShut {NoStop}%
\bibitem [{iso()}]{iso23837-2023}%
  \BibitemOpen
  \href@noop {} {\enquote {\bibinfo {title} {{ISO/IEC 23837-2:2023(en).}
  {I}nformation security --- {S}ecurity requirements, test and evaluation
  methods for quantum key distribution --- {P}art~2: {E}valuation and testing
  methods},}\ }\bibinfo {note}
  {\url{https://www.iso.org/obp/ui/en/\#iso:std:iso-iec:23837:-2:ed-1:v1:en},
  visited 22 Nov 2023}\BibitemShut {NoStop}%
\bibitem [{a-e()}]{a-etsi2021}%
  \BibitemOpen
  \href@noop {} {\enquote {\bibinfo {title} {Draft {ETSI GS QKD 010 V0.4.1}
  (2021-06). {Q}uantum key distribution {(QKD);} {I}mplementation security:
  protection against {T}rojan horse attacks},}\ }\bibinfo {note}
  {\url{https://docbox.etsi.org/ISG/QKD/Open/GS-QKD-0010_ISTrojan_v0.4.1_OpenArea.pdf},
  visited 13 Oct 2023}\BibitemShut {NoStop}%
\bibitem [{\citenamefont {Sajeed}\ \emph {et~al.}(2021)\citenamefont {Sajeed},
  \citenamefont {Chaiwongkhot}, \citenamefont {Huang}, \citenamefont {Qin},
  \citenamefont {Egorov}, \citenamefont {Kozubov}, \citenamefont {Gaidash},
  \citenamefont {Chistiakov}, \citenamefont {Vasiliev}, \citenamefont {Gleim},\
  and\ \citenamefont {Makarov}}]{sajeed2021}%
  \BibitemOpen
  \bibfield  {author} {\bibinfo {author} {\bibfnamefont {Shihan}\ \bibnamefont
  {Sajeed}}, \bibinfo {author} {\bibfnamefont {Poompong}\ \bibnamefont
  {Chaiwongkhot}}, \bibinfo {author} {\bibfnamefont {Anqi}\ \bibnamefont
  {Huang}}, \bibinfo {author} {\bibfnamefont {Hao}\ \bibnamefont {Qin}},
  \bibinfo {author} {\bibfnamefont {Vladimir}\ \bibnamefont {Egorov}}, \bibinfo
  {author} {\bibfnamefont {Anton}\ \bibnamefont {Kozubov}}, \bibinfo {author}
  {\bibfnamefont {Andrei}\ \bibnamefont {Gaidash}}, \bibinfo {author}
  {\bibfnamefont {Vladimir}\ \bibnamefont {Chistiakov}}, \bibinfo {author}
  {\bibfnamefont {Artur}\ \bibnamefont {Vasiliev}}, \bibinfo {author}
  {\bibfnamefont {Artur}\ \bibnamefont {Gleim}}, \ and\ \bibinfo {author}
  {\bibfnamefont {Vadim}\ \bibnamefont {Makarov}},\ }\bibfield  {title}
  {\enquote {\bibinfo {title} {An approach for security evaluation and
  certification of a complete quantum communication system},}\ }\href {\doibase
  10.1038/s41598-021-84139-3} {\bibfield  {journal} {\bibinfo  {journal} {Sci.
  Rep.}\ }\textbf {\bibinfo {volume} {11}},\ \bibinfo {pages} {5110} (\bibinfo
  {year} {2021})}\BibitemShut {NoStop}%
\bibitem [{\citenamefont {Gerhardt}\ \emph {et~al.}(2011)\citenamefont
  {Gerhardt}, \citenamefont {Liu}, \citenamefont {Lamas-Linares}, \citenamefont
  {Skaar}, \citenamefont {Kurtsiefer},\ and\ \citenamefont
  {Makarov}}]{gerhardt2011}%
  \BibitemOpen
  \bibfield  {author} {\bibinfo {author} {\bibfnamefont {I.}~\bibnamefont
  {Gerhardt}}, \bibinfo {author} {\bibfnamefont {Q.}~\bibnamefont {Liu}},
  \bibinfo {author} {\bibfnamefont {A.}~\bibnamefont {Lamas-Linares}}, \bibinfo
  {author} {\bibfnamefont {J.}~\bibnamefont {Skaar}}, \bibinfo {author}
  {\bibfnamefont {C.}~\bibnamefont {Kurtsiefer}}, \ and\ \bibinfo {author}
  {\bibfnamefont {V.}~\bibnamefont {Makarov}},\ }\bibfield  {title} {\enquote
  {\bibinfo {title} {Full-field implementation of a perfect eavesdropper on a
  quantum cryptography system},}\ }\href {\doibase 10.1038/ncomms1348}
  {\bibfield  {journal} {\bibinfo  {journal} {Nat. Commun.}\ }\textbf {\bibinfo
  {volume} {2}},\ \bibinfo {pages} {349} (\bibinfo {year} {2011})}\BibitemShut
  {NoStop}%
\bibitem [{\citenamefont {Fokou}\ \emph {et~al.}(2023)\citenamefont {Fokou},
  \citenamefont {Mousavi}, \citenamefont {Jasion}, \citenamefont {Richardson},\
  and\ \citenamefont {Poletti}}]{fokoua2023}%
  \BibitemOpen
  \bibfield  {author} {\bibinfo {author} {\bibfnamefont {Eric~Numkam}\
  \bibnamefont {Fokou}}, \bibinfo {author} {\bibfnamefont {Seyed~Abokhamis}\
  \bibnamefont {Mousavi}}, \bibinfo {author} {\bibfnamefont {Gregory~T.}\
  \bibnamefont {Jasion}}, \bibinfo {author} {\bibfnamefont {David~J.}\
  \bibnamefont {Richardson}}, \ and\ \bibinfo {author} {\bibfnamefont
  {Francesco}\ \bibnamefont {Poletti}},\ }\bibfield  {title} {\enquote
  {\bibinfo {title} {Loss in hollow-core optical fibers: mechanisms, scaling
  rules, and limits},}\ }\href {\doibase 10.1364/AOP.470592} {\bibfield
  {journal} {\bibinfo  {journal} {Adv. Opt. Photonics}\ }\textbf {\bibinfo
  {volume} {15}},\ \bibinfo {pages} {1} (\bibinfo {year} {2023})}\BibitemShut
  {NoStop}%
\bibitem [{\citenamefont {Tamaki}\ \emph {et~al.}(2014)\citenamefont {Tamaki},
  \citenamefont {Curty}, \citenamefont {Kato}, \citenamefont {Lo},\ and\
  \citenamefont {Azuma}}]{tamaki2014}%
  \BibitemOpen
  \bibfield  {author} {\bibinfo {author} {\bibfnamefont {Kiyoshi}\ \bibnamefont
  {Tamaki}}, \bibinfo {author} {\bibfnamefont {Marcos}\ \bibnamefont {Curty}},
  \bibinfo {author} {\bibfnamefont {Go}~\bibnamefont {Kato}}, \bibinfo {author}
  {\bibfnamefont {Hoi-Kwong}\ \bibnamefont {Lo}}, \ and\ \bibinfo {author}
  {\bibfnamefont {Koji}\ \bibnamefont {Azuma}},\ }\bibfield  {title} {\enquote
  {\bibinfo {title} {Loss-tolerant quantum cryptography with imperfect
  sources},}\ }\href {\doibase 10.1103/PhysRevA.90.052314} {\bibfield
  {journal} {\bibinfo  {journal} {Phys. Rev. A}\ }\textbf {\bibinfo {volume}
  {90}},\ \bibinfo {pages} {052314} (\bibinfo {year} {2014})}\BibitemShut
  {NoStop}%
\bibitem [{\citenamefont {Mizutani}\ \emph {et~al.}(2015)\citenamefont
  {Mizutani}, \citenamefont {Curty}, \citenamefont {Lim}, \citenamefont
  {Imoto},\ and\ \citenamefont {Tamaki}}]{mizutani2015}%
  \BibitemOpen
  \bibfield  {author} {\bibinfo {author} {\bibfnamefont {Akihiro}\ \bibnamefont
  {Mizutani}}, \bibinfo {author} {\bibfnamefont {Marcos}\ \bibnamefont
  {Curty}}, \bibinfo {author} {\bibfnamefont {Charles Ci~Wen}\ \bibnamefont
  {Lim}}, \bibinfo {author} {\bibfnamefont {Nobuyuki}\ \bibnamefont {Imoto}}, \
  and\ \bibinfo {author} {\bibfnamefont {Kiyoshi}\ \bibnamefont {Tamaki}},\
  }\bibfield  {title} {\enquote {\bibinfo {title} {Finite-key security analysis
  of quantum key distribution with imperfect light sources},}\ }\href {\doibase
  10.1088/1367-2630/17/9/093011} {\bibfield  {journal} {\bibinfo  {journal}
  {New J. Phys.}\ }\textbf {\bibinfo {volume} {17}},\ \bibinfo {pages} {093011}
  (\bibinfo {year} {2015})}\BibitemShut {NoStop}%
\bibitem [{\citenamefont {Mizutani}\ \emph {et~al.}(2019)\citenamefont
  {Mizutani}, \citenamefont {Kato}, \citenamefont {Azuma}, \citenamefont
  {Curty}, \citenamefont {Ikuta}, \citenamefont {Yamamoto}, \citenamefont
  {Imoto}, \citenamefont {Lo},\ and\ \citenamefont {Tamaki}}]{mizutani2019}%
  \BibitemOpen
  \bibfield  {author} {\bibinfo {author} {\bibfnamefont {Akihiro}\ \bibnamefont
  {Mizutani}}, \bibinfo {author} {\bibfnamefont {Go}~\bibnamefont {Kato}},
  \bibinfo {author} {\bibfnamefont {Koji}\ \bibnamefont {Azuma}}, \bibinfo
  {author} {\bibfnamefont {Marcos}\ \bibnamefont {Curty}}, \bibinfo {author}
  {\bibfnamefont {Rikizo}\ \bibnamefont {Ikuta}}, \bibinfo {author}
  {\bibfnamefont {Takashi}\ \bibnamefont {Yamamoto}}, \bibinfo {author}
  {\bibfnamefont {Nobuyuki}\ \bibnamefont {Imoto}}, \bibinfo {author}
  {\bibfnamefont {Hoi-Kwong}\ \bibnamefont {Lo}}, \ and\ \bibinfo {author}
  {\bibfnamefont {Kiyoshi}\ \bibnamefont {Tamaki}},\ }\bibfield  {title}
  {\enquote {\bibinfo {title} {Quantum key distribution with
  setting-choice-independently correlated light sources},}\ }\href {\doibase
  10.1038/s41534-018-0122-y} {\bibfield  {journal} {\bibinfo  {journal} {npj
  Quantum Inf.}\ }\textbf {\bibinfo {volume} {5}},\ \bibinfo {pages} {8}
  (\bibinfo {year} {2019})}\BibitemShut {NoStop}%
\bibitem [{\citenamefont {Pereira}\ \emph {et~al.}(2019)\citenamefont
  {Pereira}, \citenamefont {Curty},\ and\ \citenamefont
  {Tamaki}}]{pereira2019}%
  \BibitemOpen
  \bibfield  {author} {\bibinfo {author} {\bibfnamefont {Margarida}\
  \bibnamefont {Pereira}}, \bibinfo {author} {\bibfnamefont {Marcos}\
  \bibnamefont {Curty}}, \ and\ \bibinfo {author} {\bibfnamefont {Kiyoshi}\
  \bibnamefont {Tamaki}},\ }\bibfield  {title} {\enquote {\bibinfo {title}
  {Quantum key distribution with flawed and leaky sources},}\ }\href {\doibase
  10.1038/s41534-019-0180-9} {\bibfield  {journal} {\bibinfo  {journal} {npj
  Quantum Inf.}\ }\textbf {\bibinfo {volume} {5}},\ \bibinfo {pages} {62}
  (\bibinfo {year} {2019})}\BibitemShut {NoStop}%
\bibitem [{\citenamefont {Pereira}\ \emph {et~al.}(2020)\citenamefont
  {Pereira}, \citenamefont {Kato}, \citenamefont {Mizutani}, \citenamefont
  {Curty},\ and\ \citenamefont {Tamaki}}]{pereira2020}%
  \BibitemOpen
  \bibfield  {author} {\bibinfo {author} {\bibfnamefont {Margarida}\
  \bibnamefont {Pereira}}, \bibinfo {author} {\bibfnamefont {Go}~\bibnamefont
  {Kato}}, \bibinfo {author} {\bibfnamefont {Akihiro}\ \bibnamefont
  {Mizutani}}, \bibinfo {author} {\bibfnamefont {Marcos}\ \bibnamefont
  {Curty}}, \ and\ \bibinfo {author} {\bibfnamefont {Kiyoshi}\ \bibnamefont
  {Tamaki}},\ }\bibfield  {title} {\enquote {\bibinfo {title} {Quantum key
  distribution with correlated sources},}\ }\href {\doibase
  10.1126/sciadv.aaz4487} {\bibfield  {journal} {\bibinfo  {journal} {Sci.
  Adv.}\ }\textbf {\bibinfo {volume} {6}},\ \bibinfo {pages} {eaaz4487}
  (\bibinfo {year} {2020})}\BibitemShut {NoStop}%
\bibitem [{\citenamefont {Pereira}\ \emph {et~al.}(2023)\citenamefont
  {Pereira}, \citenamefont {Curr\'{a}s-Lorenzo}, \citenamefont {Navarrete},
  \citenamefont {Mizutani}, \citenamefont {Kato}, \citenamefont {Curty},\ and\
  \citenamefont {Tamaki}}]{pereira2023}%
  \BibitemOpen
  \bibfield  {author} {\bibinfo {author} {\bibfnamefont {Margarida}\
  \bibnamefont {Pereira}}, \bibinfo {author} {\bibfnamefont {Guillermo}\
  \bibnamefont {Curr\'{a}s-Lorenzo}}, \bibinfo {author} {\bibfnamefont
  {\'{A}lvaro}\ \bibnamefont {Navarrete}}, \bibinfo {author} {\bibfnamefont
  {Akihiro}\ \bibnamefont {Mizutani}}, \bibinfo {author} {\bibfnamefont
  {Go}~\bibnamefont {Kato}}, \bibinfo {author} {\bibfnamefont {Marcos}\
  \bibnamefont {Curty}}, \ and\ \bibinfo {author} {\bibfnamefont {Kiyoshi}\
  \bibnamefont {Tamaki}},\ }\bibfield  {title} {\enquote {\bibinfo {title}
  {Modified {BB84} quantum key distribution protocol robust to source
  imperfections},}\ }\href {\doibase 10.1103/PhysRevResearch.5.023065}
  {\bibfield  {journal} {\bibinfo  {journal} {Phys. Rev. Res.}\ }\textbf
  {\bibinfo {volume} {5}},\ \bibinfo {pages} {023065} (\bibinfo {year}
  {2023})}\BibitemShut {NoStop}%
\bibitem [{\citenamefont {Curr\'as-Lorenzo}\ \emph {et~al.}()\citenamefont
  {Curr\'as-Lorenzo}, \citenamefont {Pereira}, \citenamefont {Kato},
  \citenamefont {Curty},\ and\ \citenamefont {Tamaki}}]{curras-lorenzo2023}%
  \BibitemOpen
  \bibfield  {author} {\bibinfo {author} {\bibfnamefont {Guillermo}\
  \bibnamefont {Curr\'as-Lorenzo}}, \bibinfo {author} {\bibfnamefont
  {Margarida}\ \bibnamefont {Pereira}}, \bibinfo {author} {\bibfnamefont
  {Go}~\bibnamefont {Kato}}, \bibinfo {author} {\bibfnamefont {Marcos}\
  \bibnamefont {Curty}}, \ and\ \bibinfo {author} {\bibfnamefont {Kiyoshi}\
  \bibnamefont {Tamaki}},\ }\bibfield  {title} {\enquote {\bibinfo {title} {A
  security framework for quantum key distribution implementations},}\
  }\href@noop {} {\ }\Eprint {http://arxiv.org/abs/2305.05930}
  {arXiv:2305.05930 [quant-ph]} \BibitemShut {NoStop}%
\bibitem [{\citenamefont {Molotkov}(2020)}]{molotkov2020}%
  \BibitemOpen
  \bibfield  {author} {\bibinfo {author} {\bibfnamefont {S.N.}\ \bibnamefont
  {Molotkov}},\ }\bibfield  {title} {\enquote {\bibinfo {title} {{T}rojan horse
  attacks, decoy state method, and side channels of information leakage in
  quantum cryptography},}\ }\href {\doibase 10.1134/S1063776120050064}
  {\bibfield  {journal} {\bibinfo  {journal} {J. Exp. Theor. Phys.}\ }\textbf
  {\bibinfo {volume} {130}},\ \bibinfo {pages} {809--832} (\bibinfo {year}
  {2020})}\BibitemShut {NoStop}%
\bibitem [{\citenamefont {Molotkov}(2021)}]{molotkov2021}%
  \BibitemOpen
  \bibfield  {author} {\bibinfo {author} {\bibfnamefont {S.N.}\ \bibnamefont
  {Molotkov}},\ }\bibfield  {title} {\enquote {\bibinfo {title} {Side channels
  of information leakage in quantum cryptography: nonstrictly single-photon
  states, different quantum efficiencies of detectors, and finite transmitted
  sequences},}\ }\href {\doibase 10.1134/S1063776121080136} {\bibfield
  {journal} {\bibinfo  {journal} {J. Exp. Theor. Phys.}\ }\textbf {\bibinfo
  {volume} {133}},\ \bibinfo {pages} {272--304} (\bibinfo {year}
  {2021})}\BibitemShut {NoStop}%
\bibitem [{\citenamefont {Sun}\ and\ \citenamefont {Xu}(2021)}]{sun2021}%
  \BibitemOpen
  \bibfield  {author} {\bibinfo {author} {\bibfnamefont {Shihai}\ \bibnamefont
  {Sun}}\ and\ \bibinfo {author} {\bibfnamefont {Feihu}\ \bibnamefont {Xu}},\
  }\bibfield  {title} {\enquote {\bibinfo {title} {Security of quantum key
  distribution with source and detection imperfections},}\ }\href {\doibase
  10.1088/1367-2630/abdf9b} {\bibfield  {journal} {\bibinfo  {journal} {New J.
  Phys.}\ }\textbf {\bibinfo {volume} {23}},\ \bibinfo {pages} {023011}
  (\bibinfo {year} {2021})}\BibitemShut {NoStop}%
\bibitem [{\citenamefont {Duplinskiy}(2019)}]{duplinskiy2019}%
  \BibitemOpen
  \bibfield  {author} {\bibinfo {author} {\bibfnamefont {Alexander}\
  \bibnamefont {Duplinskiy}},\ }\emph {\bibinfo {title} {Quantum key
  distribution with high-rate polarization encoding}},\ \href@noop {} {Ph.D.
  thesis},\ \bibinfo  {school} {Moscow Institute of Physics and Technology}
  (\bibinfo {year} {2019}),\ \bibinfo {note} {in Russian,
  \url{https://mipt.ru/upload/medialibrary/17f/dissertatsiya-duplinskiy.pdf},
  visited 11 Feb 2024}\BibitemShut {NoStop}%
\bibitem [{\citenamefont {Duplinskiy}\ \emph {et~al.}(2017)\citenamefont
  {Duplinskiy}, \citenamefont {Ustimchik}, \citenamefont {Kanapin},
  \citenamefont {Kurochkin},\ and\ \citenamefont {Kurochkin}}]{duplinskiy2017}%
  \BibitemOpen
  \bibfield  {author} {\bibinfo {author} {\bibfnamefont {A.}~\bibnamefont
  {Duplinskiy}}, \bibinfo {author} {\bibfnamefont {V.}~\bibnamefont
  {Ustimchik}}, \bibinfo {author} {\bibfnamefont {A.}~\bibnamefont {Kanapin}},
  \bibinfo {author} {\bibfnamefont {V.}~\bibnamefont {Kurochkin}}, \ and\
  \bibinfo {author} {\bibfnamefont {Y.}~\bibnamefont {Kurochkin}},\ }\bibfield
  {title} {\enquote {\bibinfo {title} {Low loss {QKD} optical scheme for fast
  polarization encoding},}\ }\href {\doibase 10.1364/OE.25.028886} {\bibfield
  {journal} {\bibinfo  {journal} {Opt. Express}\ }\textbf {\bibinfo {volume}
  {25}},\ \bibinfo {pages} {28886} (\bibinfo {year} {2017})}\BibitemShut
  {NoStop}%
\bibitem [{\citenamefont {Duplinskiy}\ \emph {et~al.}(2018)\citenamefont
  {Duplinskiy}, \citenamefont {Kiktenko}, \citenamefont {Pozhar}, \citenamefont
  {Anufriev}, \citenamefont {Ermakov}, \citenamefont {Kotov}, \citenamefont
  {Brodskiy}, \citenamefont {Yunusov}, \citenamefont {Kurochkin}, \citenamefont
  {Fedorov},\ and\ \citenamefont {Kurochkin}}]{duplinskiy2018}%
  \BibitemOpen
  \bibfield  {author} {\bibinfo {author} {\bibfnamefont {A.~V.}\ \bibnamefont
  {Duplinskiy}}, \bibinfo {author} {\bibfnamefont {E.~O.}\ \bibnamefont
  {Kiktenko}}, \bibinfo {author} {\bibfnamefont {N.~O.}\ \bibnamefont
  {Pozhar}}, \bibinfo {author} {\bibfnamefont {M.~N.}\ \bibnamefont
  {Anufriev}}, \bibinfo {author} {\bibfnamefont {R.~P.}\ \bibnamefont
  {Ermakov}}, \bibinfo {author} {\bibfnamefont {A.~I.}\ \bibnamefont {Kotov}},
  \bibinfo {author} {\bibfnamefont {A.~V.}\ \bibnamefont {Brodskiy}}, \bibinfo
  {author} {\bibfnamefont {R.~R.}\ \bibnamefont {Yunusov}}, \bibinfo {author}
  {\bibfnamefont {V.~L.}\ \bibnamefont {Kurochkin}}, \bibinfo {author}
  {\bibfnamefont {A.~K.}\ \bibnamefont {Fedorov}}, \ and\ \bibinfo {author}
  {\bibfnamefont {Y.~V.}\ \bibnamefont {Kurochkin}},\ }\bibfield  {title}
  {\enquote {\bibinfo {title} {Quantum-secured data transmission in urban
  fiber-optics communication lines},}\ }\href {\doibase
  10.1007/s10946-018-9697-1} {\bibfield  {journal} {\bibinfo  {journal} {J.
  Russ. Laser Res.}\ }\textbf {\bibinfo {volume} {39}},\ \bibinfo {pages} {113}
  (\bibinfo {year} {2018})}\BibitemShut {NoStop}%
\bibitem [{\citenamefont {Kiktenko}\ \emph {et~al.}(2017)\citenamefont
  {Kiktenko}, \citenamefont {Trushechkin}, \citenamefont {Lim}, \citenamefont
  {Kurochkin},\ and\ \citenamefont {Fedorov}}]{kiktenko2017}%
  \BibitemOpen
  \bibfield  {author} {\bibinfo {author} {\bibfnamefont {E.~O.}\ \bibnamefont
  {Kiktenko}}, \bibinfo {author} {\bibfnamefont {A.~S.}\ \bibnamefont
  {Trushechkin}}, \bibinfo {author} {\bibfnamefont {C.~C.~W.}\ \bibnamefont
  {Lim}}, \bibinfo {author} {\bibfnamefont {Y.~V.}\ \bibnamefont {Kurochkin}},
  \ and\ \bibinfo {author} {\bibfnamefont {A.~K.}\ \bibnamefont {Fedorov}},\
  }\bibfield  {title} {\enquote {\bibinfo {title} {Symmetric blind information
  reconciliation for quantum key distribution},}\ }\href {\doibase
  10.1103/PhysRevApplied.8.044017} {\bibfield  {journal} {\bibinfo  {journal}
  {Phys. Rev. Appl.}\ }\textbf {\bibinfo {volume} {8}},\ \bibinfo {pages}
  {044017} (\bibinfo {year} {2017})}\BibitemShut {NoStop}%
\bibitem [{\citenamefont {Borisov}\ \emph {et~al.}(2023)\citenamefont
  {Borisov}, \citenamefont {Petrov},\ and\ \citenamefont
  {Tayduganov}}]{borisov2022}%
  \BibitemOpen
  \bibfield  {author} {\bibinfo {author} {\bibfnamefont {Nikolay}\ \bibnamefont
  {Borisov}}, \bibinfo {author} {\bibfnamefont {Ivan}\ \bibnamefont {Petrov}},
  \ and\ \bibinfo {author} {\bibfnamefont {Andrey}\ \bibnamefont
  {Tayduganov}},\ }\bibfield  {title} {\enquote {\bibinfo {title} {Asymmetric
  adaptive {LDPC}-based information reconciliation for industrial quantum key
  distribution},}\ }\href {\doibase 10.3390/e25010031} {\bibfield  {journal}
  {\bibinfo  {journal} {Entropy}\ }\textbf {\bibinfo {volume} {25}},\ \bibinfo
  {pages} {31} (\bibinfo {year} {2023})}\BibitemShut {NoStop}%
\bibitem [{\citenamefont {Krovetz}\ and\ \citenamefont
  {Rogaway}(2001)}]{krovetz2001}%
  \BibitemOpen
  \bibfield  {author} {\bibinfo {author} {\bibfnamefont {Ted}\ \bibnamefont
  {Krovetz}}\ and\ \bibinfo {author} {\bibfnamefont {Phillip}\ \bibnamefont
  {Rogaway}},\ }\bibfield  {title} {\enquote {\bibinfo {title} {Fast universal
  hashing with small keys and no preprocessing: The {PolyR} construction},}\
  }\href {\doibase 10.1007/3-540-45247-8_7} {\bibfield  {journal} {\bibinfo
  {journal} {Lect. Notes Comp. Sci.}\ }\textbf {\bibinfo {volume} {2015}},\
  \bibinfo {pages} {73--89} (\bibinfo {year} {2001})}\BibitemShut {NoStop}%
\bibitem [{\citenamefont {Fedorov}\ \emph {et~al.}(2018)\citenamefont
  {Fedorov}, \citenamefont {Kiktenko},\ and\ \citenamefont
  {Trushechkin}}]{fedorov2018}%
  \BibitemOpen
  \bibfield  {author} {\bibinfo {author} {\bibfnamefont {A.~K.}\ \bibnamefont
  {Fedorov}}, \bibinfo {author} {\bibfnamefont {E.~O.}\ \bibnamefont
  {Kiktenko}}, \ and\ \bibinfo {author} {\bibfnamefont {A.~S.}\ \bibnamefont
  {Trushechkin}},\ }\bibfield  {title} {\enquote {\bibinfo {title} {Symmetric
  blind information reconciliation and hash-function-based verification for
  quantum key distribution},}\ }\href {\doibase 10.1134/s1995080218070107}
  {\bibfield  {journal} {\bibinfo  {journal} {Lobachevskii J. Math.}\ }\textbf
  {\bibinfo {volume} {39}},\ \bibinfo {pages} {992--996} (\bibinfo {year}
  {2018})}\BibitemShut {NoStop}%
\bibitem [{\citenamefont {Trushechkin}\ \emph {et~al.}(2017)\citenamefont
  {Trushechkin}, \citenamefont {Kiktenko},\ and\ \citenamefont
  {Fedorov}}]{trushechkin2017}%
  \BibitemOpen
  \bibfield  {author} {\bibinfo {author} {\bibfnamefont {A.~S.}\ \bibnamefont
  {Trushechkin}}, \bibinfo {author} {\bibfnamefont {E.~O.}\ \bibnamefont
  {Kiktenko}}, \ and\ \bibinfo {author} {\bibfnamefont {A.~K.}\ \bibnamefont
  {Fedorov}},\ }\bibfield  {title} {\enquote {\bibinfo {title} {Practical
  issues in decoy-state quantum key distribution based on the central limit
  theorem},}\ }\href {\doibase 10.1103/PhysRevA.96.022316} {\bibfield
  {journal} {\bibinfo  {journal} {Phys. Rev. A}\ }\textbf {\bibinfo {volume}
  {96}},\ \bibinfo {pages} {022316} (\bibinfo {year} {2017})}\BibitemShut
  {NoStop}%
\bibitem [{\citenamefont {Krawczyk}(1994)}]{krawczyk1994}%
  \BibitemOpen
  \bibfield  {author} {\bibinfo {author} {\bibfnamefont {Hugo}\ \bibnamefont
  {Krawczyk}},\ }\bibfield  {title} {\enquote {\bibinfo {title} {{LFSR}-based
  hashing and authentication},}\ }\href {\doibase 10.1007/3-540-48658-5_15}
  {\bibfield  {journal} {\bibinfo  {journal} {Lect. Notes Comp. Sci.}\ }\textbf
  {\bibinfo {volume} {839}},\ \bibinfo {pages} {129--139} (\bibinfo {year}
  {1994})}\BibitemShut {NoStop}%
\bibitem [{\citenamefont {Krawczyk}(1995)}]{krawczyk1995}%
  \BibitemOpen
  \bibfield  {author} {\bibinfo {author} {\bibfnamefont {Hugo}\ \bibnamefont
  {Krawczyk}},\ }\bibfield  {title} {\enquote {\bibinfo {title} {New hash
  functions for message authentication},}\ }\href {\doibase
  10.1007/3-540-49264-X_24} {\bibfield  {journal} {\bibinfo  {journal} {Lect.
  Notes Comp. Sci.}\ }\textbf {\bibinfo {volume} {921}},\ \bibinfo {pages}
  {301--310} (\bibinfo {year} {1995})}\BibitemShut {NoStop}%
\bibitem [{\citenamefont {Ye}\ \emph {et~al.}(2023)\citenamefont {Ye},
  \citenamefont {Chen}, \citenamefont {Zhang}, \citenamefont {Lu},
  \citenamefont {Wang}, \citenamefont {Huang}, \citenamefont {Wang},
  \citenamefont {He}, \citenamefont {Yin}, \citenamefont {Guo},\ and\
  \citenamefont {Han}}]{ye2023}%
  \BibitemOpen
  \bibfield  {author} {\bibinfo {author} {\bibfnamefont {Peng}\ \bibnamefont
  {Ye}}, \bibinfo {author} {\bibfnamefont {Wei}\ \bibnamefont {Chen}}, \bibinfo
  {author} {\bibfnamefont {Guo-Wei}\ \bibnamefont {Zhang}}, \bibinfo {author}
  {\bibfnamefont {Feng-Yu}\ \bibnamefont {Lu}}, \bibinfo {author}
  {\bibfnamefont {Fang-Xiang}\ \bibnamefont {Wang}}, \bibinfo {author}
  {\bibfnamefont {Guan-Zhong}\ \bibnamefont {Huang}}, \bibinfo {author}
  {\bibfnamefont {Shuang}\ \bibnamefont {Wang}}, \bibinfo {author}
  {\bibfnamefont {De-Yong}\ \bibnamefont {He}}, \bibinfo {author}
  {\bibfnamefont {Zhen-Qiang}\ \bibnamefont {Yin}}, \bibinfo {author}
  {\bibfnamefont {Guang-Can}\ \bibnamefont {Guo}}, \ and\ \bibinfo {author}
  {\bibfnamefont {Zheng-Fu}\ \bibnamefont {Han}},\ }\bibfield  {title}
  {\enquote {\bibinfo {title} {Induced-photorefraction attack against quantum
  key distribution},}\ }\href {\doibase 10.1103/PhysRevApplied.19.054052}
  {\bibfield  {journal} {\bibinfo  {journal} {Phys. Rev. Appl.}\ }\textbf
  {\bibinfo {volume} {19}},\ \bibinfo {pages} {054052} (\bibinfo {year}
  {2023})}\BibitemShut {NoStop}%
\bibitem [{idq()}]{idq-Cerberis-XG}%
  \BibitemOpen
  \href@noop {} {\enquote {\bibinfo {title} {{C}erberis~{XG} {QKD} system},}\
  }\bibinfo {note}
  {\url{https://www.idquantique.com/quantum-safe-security/products/cerberis-xg-qkd-system/},
  visited 14 Feb 2023}\BibitemShut {NoStop}%
\bibitem [{\citenamefont {Walenta}\ \emph {et~al.}(2014)\citenamefont
  {Walenta}, \citenamefont {Burg}, \citenamefont {Caselunghe}, \citenamefont
  {Constantin}, \citenamefont {Gisin}, \citenamefont {Guinnard}, \citenamefont
  {Houlmann}, \citenamefont {Junod}, \citenamefont {Korzh}, \citenamefont
  {Kulesza}, \citenamefont {Legr{\' e}}, \citenamefont {Lim}, \citenamefont
  {Lunghi}, \citenamefont {Monat}, \citenamefont {Portmann}, \citenamefont
  {Soucarros}, \citenamefont {Thew}, \citenamefont {Trinkler}, \citenamefont
  {Trolliet}, \citenamefont {Vannel},\ and\ \citenamefont
  {Zbinden}}]{walenta2014}%
  \BibitemOpen
  \bibfield  {author} {\bibinfo {author} {\bibfnamefont {N.}~\bibnamefont
  {Walenta}}, \bibinfo {author} {\bibfnamefont {A.}~\bibnamefont {Burg}},
  \bibinfo {author} {\bibfnamefont {D.}~\bibnamefont {Caselunghe}}, \bibinfo
  {author} {\bibfnamefont {J.}~\bibnamefont {Constantin}}, \bibinfo {author}
  {\bibfnamefont {N.}~\bibnamefont {Gisin}}, \bibinfo {author} {\bibfnamefont
  {O.}~\bibnamefont {Guinnard}}, \bibinfo {author} {\bibfnamefont
  {R.}~\bibnamefont {Houlmann}}, \bibinfo {author} {\bibfnamefont
  {P.}~\bibnamefont {Junod}}, \bibinfo {author} {\bibfnamefont
  {B.}~\bibnamefont {Korzh}}, \bibinfo {author} {\bibfnamefont
  {N.}~\bibnamefont {Kulesza}}, \bibinfo {author} {\bibfnamefont
  {M.}~\bibnamefont {Legr{\' e}}}, \bibinfo {author} {\bibfnamefont {C.~W.}\
  \bibnamefont {Lim}}, \bibinfo {author} {\bibfnamefont {T.}~\bibnamefont
  {Lunghi}}, \bibinfo {author} {\bibfnamefont {L.}~\bibnamefont {Monat}},
  \bibinfo {author} {\bibfnamefont {C.}~\bibnamefont {Portmann}}, \bibinfo
  {author} {\bibfnamefont {M.}~\bibnamefont {Soucarros}}, \bibinfo {author}
  {\bibfnamefont {R.~T.}\ \bibnamefont {Thew}}, \bibinfo {author}
  {\bibfnamefont {P.}~\bibnamefont {Trinkler}}, \bibinfo {author}
  {\bibfnamefont {G.}~\bibnamefont {Trolliet}}, \bibinfo {author}
  {\bibfnamefont {F.}~\bibnamefont {Vannel}}, \ and\ \bibinfo {author}
  {\bibfnamefont {H.}~\bibnamefont {Zbinden}},\ }\bibfield  {title} {\enquote
  {\bibinfo {title} {A fast and versatile quantum key distribution system with
  hardware key distillation and wavelength multiplexing},}\ }\href {\doibase
  10.1088/1367-2630/16/1/013047} {\bibfield  {journal} {\bibinfo  {journal}
  {New J. Phys.}\ }\textbf {\bibinfo {volume} {16}},\ \bibinfo {pages} {013047}
  (\bibinfo {year} {2014})}\BibitemShut {NoStop}%
\bibitem [{\citenamefont {Stucki}\ \emph {et~al.}(2005)\citenamefont {Stucki},
  \citenamefont {Brunner}, \citenamefont {Gisin}, \citenamefont {Scarani},\
  and\ \citenamefont {Zbinden}}]{stucki2005}%
  \BibitemOpen
  \bibfield  {author} {\bibinfo {author} {\bibfnamefont {D.}~\bibnamefont
  {Stucki}}, \bibinfo {author} {\bibfnamefont {N.}~\bibnamefont {Brunner}},
  \bibinfo {author} {\bibfnamefont {N.}~\bibnamefont {Gisin}}, \bibinfo
  {author} {\bibfnamefont {V.}~\bibnamefont {Scarani}}, \ and\ \bibinfo
  {author} {\bibfnamefont {H.}~\bibnamefont {Zbinden}},\ }\bibfield  {title}
  {\enquote {\bibinfo {title} {Fast and simple one-way quantum key
  distribution},}\ }\href {\doibase 10.1063/1.2126792} {\bibfield  {journal}
  {\bibinfo  {journal} {Appl. Phys. Lett.}\ }\textbf {\bibinfo {volume} {87}},\
  \bibinfo {pages} {194108} (\bibinfo {year} {2005})}\BibitemShut {NoStop}%
\bibitem [{\citenamefont {Tr{\' e}nyi}\ and\ \citenamefont
  {Curty}(2021)}]{trenyi2021}%
  \BibitemOpen
  \bibfield  {author} {\bibinfo {author} {\bibfnamefont {R{\' o}bert}\
  \bibnamefont {Tr{\' e}nyi}}\ and\ \bibinfo {author} {\bibfnamefont {Marcos}\
  \bibnamefont {Curty}},\ }\bibfield  {title} {\enquote {\bibinfo {title}
  {Zero-error attack against coherent-one-way quantum key distribution},}\
  }\href {\doibase 10.1088/1367-2630/ac1e41} {\bibfield  {journal} {\bibinfo
  {journal} {New J. Phys.}\ }\textbf {\bibinfo {volume} {23}},\ \bibinfo
  {pages} {093005} (\bibinfo {year} {2021})}\BibitemShut {NoStop}%
\bibitem [{\citenamefont {Gaidash}\ \emph {et~al.}(2022)\citenamefont
  {Gaidash}, \citenamefont {Miroshnichenko},\ and\ \citenamefont
  {Kozubov}}]{gaidash2022}%
  \BibitemOpen
  \bibfield  {author} {\bibinfo {author} {\bibfnamefont {Andrei}\ \bibnamefont
  {Gaidash}}, \bibinfo {author} {\bibfnamefont {George}\ \bibnamefont
  {Miroshnichenko}}, \ and\ \bibinfo {author} {\bibfnamefont {Anton}\
  \bibnamefont {Kozubov}},\ }\bibfield  {title} {\enquote {\bibinfo {title}
  {Subcarrier wave quantum key distribution with leaky and flawed devices},}\
  }\href {\doibase 10.1364/JOSAB.439776} {\bibfield  {journal} {\bibinfo
  {journal} {J. Opt. Soc. Am. B}\ }\textbf {\bibinfo {volume} {39}},\ \bibinfo
  {pages} {577} (\bibinfo {year} {2022})}\BibitemShut {NoStop}%
\bibitem [{\citenamefont {Kozubov}\ \emph {et~al.}()\citenamefont {Kozubov},
  \citenamefont {Gaidash},\ and\ \citenamefont {Miroshnichenko}}]{kozubov2019}%
  \BibitemOpen
  \bibfield  {author} {\bibinfo {author} {\bibfnamefont {Anton}\ \bibnamefont
  {Kozubov}}, \bibinfo {author} {\bibfnamefont {Andrei}\ \bibnamefont
  {Gaidash}}, \ and\ \bibinfo {author} {\bibfnamefont {George}\ \bibnamefont
  {Miroshnichenko}},\ }\bibfield  {title} {\enquote {\bibinfo {title}
  {Finite-key security for quantum key distribution systems utilizing weak
  coherent states},}\ }\href@noop {} {\ }\Eprint
  {http://arxiv.org/abs/1903.04371} {arXiv:1903.04371 [quant-ph]} \BibitemShut
  {NoStop}%
\bibitem [{\citenamefont {Kulik}\ \emph {et~al.}(2007)\citenamefont {Kulik},
  \citenamefont {Molotkov},\ and\ \citenamefont {Makkaveev}}]{kulik2007}%
  \BibitemOpen
  \bibfield  {author} {\bibinfo {author} {\bibfnamefont {S.~P.}\ \bibnamefont
  {Kulik}}, \bibinfo {author} {\bibfnamefont {S.~N.}\ \bibnamefont {Molotkov}},
  \ and\ \bibinfo {author} {\bibfnamefont {A.~P.}\ \bibnamefont {Makkaveev}},\
  }\bibfield  {title} {\enquote {\bibinfo {title} {Combined phase--time
  encoding method in quantum cryptography},}\ }\href {\doibase
  10.1134/S0021364007060070} {\bibfield  {journal} {\bibinfo  {journal} {JETP
  Lett.}\ }\textbf {\bibinfo {volume} {85}},\ \bibinfo {pages} {297} (\bibinfo
  {year} {2007})}\BibitemShut {NoStop}%
\bibitem [{inf()}]{infotecs-Quandor}%
  \BibitemOpen
  \href@noop {} {\enquote {\bibinfo {title} {{ViPNet} {Q}uandor},}\ }\bibinfo
  {note} {\url{https://quantum-crypto.ru/projects/vipnet-quandor/}, visited 15
  Feb 2023}\BibitemShut {NoStop}%
\bibitem [{\citenamefont {Kronberg}(2023)}]{kronberg2023}%
  \BibitemOpen
  \bibfield  {author} {\bibinfo {author} {\bibfnamefont {D.~A.}\ \bibnamefont
  {Kronberg}},\ }\bibfield  {title} {\enquote {\bibinfo {title} {Vulnerability
  of quantum cryptography with phase--time coding under attenuation
  conditions},}\ }\href {\doibase 10.1134/S0040577923010075} {\bibfield
  {journal} {\bibinfo  {journal} {Theor. Math. Phys.}\ }\textbf {\bibinfo
  {volume} {214}},\ \bibinfo {pages} {121} (\bibinfo {year}
  {2023})}\BibitemShut {NoStop}%
\bibitem [{\citenamefont {Lydersen}\ \emph
  {et~al.}(2011{\natexlab{a}})\citenamefont {Lydersen}, \citenamefont {Jain},
  \citenamefont {Wittmann}, \citenamefont {Mar{\o}y}, \citenamefont {Skaar},
  \citenamefont {Marquardt}, \citenamefont {Makarov},\ and\ \citenamefont
  {Leuchs}}]{lydersen2011b}%
  \BibitemOpen
  \bibfield  {author} {\bibinfo {author} {\bibfnamefont {L.}~\bibnamefont
  {Lydersen}}, \bibinfo {author} {\bibfnamefont {N.}~\bibnamefont {Jain}},
  \bibinfo {author} {\bibfnamefont {C.}~\bibnamefont {Wittmann}}, \bibinfo
  {author} {\bibfnamefont {{\O}.}~\bibnamefont {Mar{\o}y}}, \bibinfo {author}
  {\bibfnamefont {J.}~\bibnamefont {Skaar}}, \bibinfo {author} {\bibfnamefont
  {C.}~\bibnamefont {Marquardt}}, \bibinfo {author} {\bibfnamefont
  {V.}~\bibnamefont {Makarov}}, \ and\ \bibinfo {author} {\bibfnamefont
  {G.}~\bibnamefont {Leuchs}},\ }\bibfield  {title} {\enquote {\bibinfo {title}
  {Superlinear threshold detectors in quantum cryptography},}\ }\href {\doibase
  10.1103/PhysRevA.84.032320} {\bibfield  {journal} {\bibinfo  {journal} {Phys.
  Rev. A}\ }\textbf {\bibinfo {volume} {84}},\ \bibinfo {pages} {032320}
  (\bibinfo {year} {2011}{\natexlab{a}})}\BibitemShut {NoStop}%
\bibitem [{\citenamefont {Cova}\ \emph {et~al.}(2004)\citenamefont {Cova},
  \citenamefont {Ghioni}, \citenamefont {Lotito}, \citenamefont {Rech},\ and\
  \citenamefont {Zappa}}]{cova2004}%
  \BibitemOpen
  \bibfield  {author} {\bibinfo {author} {\bibfnamefont {S.}~\bibnamefont
  {Cova}}, \bibinfo {author} {\bibfnamefont {M.}~\bibnamefont {Ghioni}},
  \bibinfo {author} {\bibfnamefont {A.}~\bibnamefont {Lotito}}, \bibinfo
  {author} {\bibfnamefont {I.}~\bibnamefont {Rech}}, \ and\ \bibinfo {author}
  {\bibfnamefont {F.}~\bibnamefont {Zappa}},\ }\bibfield  {title} {\enquote
  {\bibinfo {title} {Evolution and prospects for single-photon avalanche diodes
  and quenching circuits},}\ }\href {\doibase 10.1080/09500340408235272}
  {\bibfield  {journal} {\bibinfo  {journal} {J. Mod. Opt.}\ }\textbf {\bibinfo
  {volume} {51}},\ \bibinfo {pages} {1267--1288} (\bibinfo {year}
  {2004})}\BibitemShut {NoStop}%
\bibitem [{\citenamefont {Wiechers}\ \emph {et~al.}(2011)\citenamefont
  {Wiechers}, \citenamefont {Lydersen}, \citenamefont {Wittmann}, \citenamefont
  {Elser}, \citenamefont {Skaar}, \citenamefont {Marquardt}, \citenamefont
  {Makarov},\ and\ \citenamefont {Leuchs}}]{wiechers2011}%
  \BibitemOpen
  \bibfield  {author} {\bibinfo {author} {\bibfnamefont {C.}~\bibnamefont
  {Wiechers}}, \bibinfo {author} {\bibfnamefont {L.}~\bibnamefont {Lydersen}},
  \bibinfo {author} {\bibfnamefont {C.}~\bibnamefont {Wittmann}}, \bibinfo
  {author} {\bibfnamefont {D.}~\bibnamefont {Elser}}, \bibinfo {author}
  {\bibfnamefont {J.}~\bibnamefont {Skaar}}, \bibinfo {author} {\bibfnamefont
  {C.}~\bibnamefont {Marquardt}}, \bibinfo {author} {\bibfnamefont
  {V.}~\bibnamefont {Makarov}}, \ and\ \bibinfo {author} {\bibfnamefont
  {G.}~\bibnamefont {Leuchs}},\ }\bibfield  {title} {\enquote {\bibinfo {title}
  {After-gate attack on a quantum cryptosystem},}\ }\href {\doibase
  10.1088/1367-2630/13/1/013043} {\bibfield  {journal} {\bibinfo  {journal}
  {New J. Phys.}\ }\textbf {\bibinfo {volume} {13}},\ \bibinfo {pages} {013043}
  (\bibinfo {year} {2011})}\BibitemShut {NoStop}%
\bibitem [{\citenamefont {Bugge}\ \emph {et~al.}(2014)\citenamefont {Bugge},
  \citenamefont {Sauge}, \citenamefont {Ghazali}, \citenamefont {Skaar},
  \citenamefont {Lydersen},\ and\ \citenamefont {Makarov}}]{bugge2014}%
  \BibitemOpen
  \bibfield  {author} {\bibinfo {author} {\bibfnamefont {Audun~Nystad}\
  \bibnamefont {Bugge}}, \bibinfo {author} {\bibfnamefont {Sebastien}\
  \bibnamefont {Sauge}}, \bibinfo {author} {\bibfnamefont {Aina Mardhiyah~M.}\
  \bibnamefont {Ghazali}}, \bibinfo {author} {\bibfnamefont {Johannes}\
  \bibnamefont {Skaar}}, \bibinfo {author} {\bibfnamefont {Lars}\ \bibnamefont
  {Lydersen}}, \ and\ \bibinfo {author} {\bibfnamefont {Vadim}\ \bibnamefont
  {Makarov}},\ }\bibfield  {title} {\enquote {\bibinfo {title} {Laser damage
  helps the eavesdropper in quantum cryptography},}\ }\href {\doibase
  10.1103/PhysRevLett.112.070503} {\bibfield  {journal} {\bibinfo  {journal}
  {Phys. Rev. Lett.}\ }\textbf {\bibinfo {volume} {112}},\ \bibinfo {pages}
  {070503} (\bibinfo {year} {2014})}\BibitemShut {NoStop}%
\bibitem [{\citenamefont {Makarov}\ \emph {et~al.}(2016)\citenamefont
  {Makarov}, \citenamefont {Bourgoin}, \citenamefont {Chaiwongkhot},
  \citenamefont {Gagn{\'e}}, \citenamefont {Jennewein}, \citenamefont {Kaiser},
  \citenamefont {Kashyap}, \citenamefont {Legr{\'e}}, \citenamefont
  {Minshull},\ and\ \citenamefont {Sajeed}}]{makarov2016}%
  \BibitemOpen
  \bibfield  {author} {\bibinfo {author} {\bibfnamefont {Vadim}\ \bibnamefont
  {Makarov}}, \bibinfo {author} {\bibfnamefont {Jean-Philippe}\ \bibnamefont
  {Bourgoin}}, \bibinfo {author} {\bibfnamefont {Poompong}\ \bibnamefont
  {Chaiwongkhot}}, \bibinfo {author} {\bibfnamefont {Mathieu}\ \bibnamefont
  {Gagn{\'e}}}, \bibinfo {author} {\bibfnamefont {Thomas}\ \bibnamefont
  {Jennewein}}, \bibinfo {author} {\bibfnamefont {Sarah}\ \bibnamefont
  {Kaiser}}, \bibinfo {author} {\bibfnamefont {Raman}\ \bibnamefont {Kashyap}},
  \bibinfo {author} {\bibfnamefont {Matthieu}\ \bibnamefont {Legr{\'e}}},
  \bibinfo {author} {\bibfnamefont {Carter}\ \bibnamefont {Minshull}}, \ and\
  \bibinfo {author} {\bibfnamefont {Shihan}\ \bibnamefont {Sajeed}},\
  }\bibfield  {title} {\enquote {\bibinfo {title} {Creation of backdoors in
  quantum communications via laser damage},}\ }\href {\doibase
  10.1103/PhysRevA.94.030302} {\bibfield  {journal} {\bibinfo  {journal} {Phys.
  Rev. A}\ }\textbf {\bibinfo {volume} {94}},\ \bibinfo {pages} {030302}
  (\bibinfo {year} {2016})}\BibitemShut {NoStop}%
\bibitem [{\citenamefont {Gras}\ \emph {et~al.}(2020)\citenamefont {Gras},
  \citenamefont {Sultana}, \citenamefont {Huang}, \citenamefont {Jennewein},
  \citenamefont {Bussi{\` e}res}, \citenamefont {Makarov},\ and\ \citenamefont
  {Zbinden}}]{gras2020}%
  \BibitemOpen
  \bibfield  {author} {\bibinfo {author} {\bibfnamefont {Ga{\" e}tan}\
  \bibnamefont {Gras}}, \bibinfo {author} {\bibfnamefont {Nigar}\ \bibnamefont
  {Sultana}}, \bibinfo {author} {\bibfnamefont {Anqi}\ \bibnamefont {Huang}},
  \bibinfo {author} {\bibfnamefont {Thomas}\ \bibnamefont {Jennewein}},
  \bibinfo {author} {\bibfnamefont {F{\' e}lix}\ \bibnamefont {Bussi{\`
  e}res}}, \bibinfo {author} {\bibfnamefont {Vadim}\ \bibnamefont {Makarov}}, \
  and\ \bibinfo {author} {\bibfnamefont {Hugo}\ \bibnamefont {Zbinden}},\
  }\bibfield  {title} {\enquote {\bibinfo {title} {Optical control of
  single-photon negative-feedback avalanche diode detector},}\ }\href {\doibase
  10.1063/1.5140824} {\bibfield  {journal} {\bibinfo  {journal} {J. Appl.
  Phys.}\ }\textbf {\bibinfo {volume} {127}},\ \bibinfo {pages} {094502}
  (\bibinfo {year} {2020})}\BibitemShut {NoStop}%
\bibitem [{\citenamefont {Wu}\ \emph {et~al.}(2020)\citenamefont {Wu},
  \citenamefont {Huang}, \citenamefont {Chen}, \citenamefont {Sun},
  \citenamefont {Ding}, \citenamefont {Qiang}, \citenamefont {Fu},
  \citenamefont {Xu},\ and\ \citenamefont {Wu}}]{wu2020}%
  \BibitemOpen
  \bibfield  {author} {\bibinfo {author} {\bibfnamefont {Zhihao}\ \bibnamefont
  {Wu}}, \bibinfo {author} {\bibfnamefont {Anqi}\ \bibnamefont {Huang}},
  \bibinfo {author} {\bibfnamefont {Huan}\ \bibnamefont {Chen}}, \bibinfo
  {author} {\bibfnamefont {Shi-Hai}\ \bibnamefont {Sun}}, \bibinfo {author}
  {\bibfnamefont {Jiangfang}\ \bibnamefont {Ding}}, \bibinfo {author}
  {\bibfnamefont {Xiaogang}\ \bibnamefont {Qiang}}, \bibinfo {author}
  {\bibfnamefont {Xiang}\ \bibnamefont {Fu}}, \bibinfo {author} {\bibfnamefont
  {Ping}\ \bibnamefont {Xu}}, \ and\ \bibinfo {author} {\bibfnamefont {Junjie}\
  \bibnamefont {Wu}},\ }\bibfield  {title} {\enquote {\bibinfo {title} {Hacking
  single-photon avalanche detectors in quantum key distribution via pulse
  illumination},}\ }\href {\doibase 10.1364/OE.397962} {\bibfield  {journal}
  {\bibinfo  {journal} {Opt. Express}\ }\textbf {\bibinfo {volume} {28}},\
  \bibinfo {pages} {25574} (\bibinfo {year} {2020})}\BibitemShut {NoStop}%
\bibitem [{\citenamefont {Bulavkin}\ \emph {et~al.}(2022)\citenamefont
  {Bulavkin}, \citenamefont {Sushchev}, \citenamefont {Bugai}, \citenamefont
  {Bogdanov},\ and\ \citenamefont {Dvoretskiy}}]{bulavkin2022}%
  \BibitemOpen
  \bibfield  {author} {\bibinfo {author} {\bibfnamefont {D.~S.}\ \bibnamefont
  {Bulavkin}}, \bibinfo {author} {\bibfnamefont {I.~S.}\ \bibnamefont
  {Sushchev}}, \bibinfo {author} {\bibfnamefont {K.~E.}\ \bibnamefont {Bugai}},
  \bibinfo {author} {\bibfnamefont {S.~A.}\ \bibnamefont {Bogdanov}}, \ and\
  \bibinfo {author} {\bibfnamefont {D.~A.}\ \bibnamefont {Dvoretskiy}},\
  }\bibfield  {title} {\enquote {\bibinfo {title} {Study of a single-photon
  detector blinding attack with modulated bright light},}\ }\href {\doibase
  10.1117/12.2641984} {\bibfield  {journal} {\bibinfo  {journal} {Proc. SPIE}\
  }\textbf {\bibinfo {volume} {12323}},\ \bibinfo {pages} {123230E} (\bibinfo
  {year} {2022})}\BibitemShut {NoStop}%
\bibitem [{\citenamefont {Acheva}\ \emph {et~al.}(2023)\citenamefont {Acheva},
  \citenamefont {Zaitsev}, \citenamefont {Zavodilenko}, \citenamefont {Losev},
  \citenamefont {Huang},\ and\ \citenamefont {Makarov}}]{acheva2023}%
  \BibitemOpen
  \bibfield  {author} {\bibinfo {author} {\bibfnamefont {Polina}\ \bibnamefont
  {Acheva}}, \bibinfo {author} {\bibfnamefont {Konstantin}\ \bibnamefont
  {Zaitsev}}, \bibinfo {author} {\bibfnamefont {Vladimir}\ \bibnamefont
  {Zavodilenko}}, \bibinfo {author} {\bibfnamefont {Anton}\ \bibnamefont
  {Losev}}, \bibinfo {author} {\bibfnamefont {Anqi}\ \bibnamefont {Huang}}, \
  and\ \bibinfo {author} {\bibfnamefont {Vadim}\ \bibnamefont {Makarov}},\
  }\bibfield  {title} {\enquote {\bibinfo {title} {Automated verification of
  countermeasure against detector-control attack in quantum key
  distribution},}\ }\href {\doibase 10.1140/epjqt/s40507-023-00178-x}
  {\bibfield  {journal} {\bibinfo  {journal} {EPJ Quantum Technol.}\ }\textbf
  {\bibinfo {volume} {10}},\ \bibinfo {pages} {22} (\bibinfo {year}
  {2023})}\BibitemShut {NoStop}%
\bibitem [{\citenamefont {Kuzmin}(2023{\natexlab{a}})}]{kuzmin2023a}%
  \BibitemOpen
  \bibfield  {author} {\bibinfo {author} {\bibfnamefont {Mikhail}\ \bibnamefont
  {Kuzmin}},\ }\emph {\bibinfo {title} {Resistance of an avalanche photon
  detector to bright-light attacks in quantum key distribution}},\ \href@noop
  {} {Master's thesis},\ \bibinfo  {school} {Moscow Technical University of
  Communication and Informatics} (\bibinfo {year}
  {2023}{\natexlab{a}})\BibitemShut {NoStop}%
\bibitem [{\citenamefont {Lo}\ \emph {et~al.}(2012)\citenamefont {Lo},
  \citenamefont {Curty},\ and\ \citenamefont {Qi}}]{lo2012}%
  \BibitemOpen
  \bibfield  {author} {\bibinfo {author} {\bibfnamefont {H.-K.}\ \bibnamefont
  {Lo}}, \bibinfo {author} {\bibfnamefont {M.}~\bibnamefont {Curty}}, \ and\
  \bibinfo {author} {\bibfnamefont {B.}~\bibnamefont {Qi}},\ }\bibfield
  {title} {\enquote {\bibinfo {title} {Measurement-device-independent quantum
  key distribution},}\ }\href {\doibase 10.1103/PhysRevLett.108.130503}
  {\bibfield  {journal} {\bibinfo  {journal} {Phys. Rev. Lett.}\ }\textbf
  {\bibinfo {volume} {108}},\ \bibinfo {pages} {130503} (\bibinfo {year}
  {2012})}\BibitemShut {NoStop}%
\bibitem [{\citenamefont {Lucamarini}\ \emph {et~al.}(2018)\citenamefont
  {Lucamarini}, \citenamefont {Yuan}, \citenamefont {Dynes},\ and\
  \citenamefont {Shields}}]{lucamarini2018}%
  \BibitemOpen
  \bibfield  {author} {\bibinfo {author} {\bibfnamefont {M.}~\bibnamefont
  {Lucamarini}}, \bibinfo {author} {\bibfnamefont {Z.~L.}\ \bibnamefont
  {Yuan}}, \bibinfo {author} {\bibfnamefont {J.~F.}\ \bibnamefont {Dynes}}, \
  and\ \bibinfo {author} {\bibfnamefont {A.~J.}\ \bibnamefont {Shields}},\
  }\bibfield  {title} {\enquote {\bibinfo {title} {Overcoming the
  rate--distance limit of quantum key distribution without quantum
  repeaters},}\ }\href {\doibase 10.1038/s41586-018-0066-6} {\bibfield
  {journal} {\bibinfo  {journal} {Nature}\ }\textbf {\bibinfo {volume} {557}},\
  \bibinfo {pages} {400} (\bibinfo {year} {2018})}\BibitemShut {NoStop}%
\bibitem [{\citenamefont {Wang}\ \emph
  {et~al.}(2018{\natexlab{a}})\citenamefont {Wang}, \citenamefont {Yu},\ and\
  \citenamefont {Hu}}]{wang2018}%
  \BibitemOpen
  \bibfield  {author} {\bibinfo {author} {\bibfnamefont {Xiang-Bin}\
  \bibnamefont {Wang}}, \bibinfo {author} {\bibfnamefont {Zong-Wen}\
  \bibnamefont {Yu}}, \ and\ \bibinfo {author} {\bibfnamefont {Xiao-Long}\
  \bibnamefont {Hu}},\ }\bibfield  {title} {\enquote {\bibinfo {title}
  {Twin-field quantum key distribution with large misalignment error},}\ }\href
  {\doibase 10.1103/PhysRevA.98.062323} {\bibfield  {journal} {\bibinfo
  {journal} {Phys. Rev. A}\ }\textbf {\bibinfo {volume} {98}},\ \bibinfo
  {pages} {062323} (\bibinfo {year} {2018}{\natexlab{a}})}\BibitemShut
  {NoStop}%
\bibitem [{\citenamefont {Wang}\ \emph {et~al.}(2022)\citenamefont {Wang},
  \citenamefont {Yin}, \citenamefont {He}, \citenamefont {Chen}, \citenamefont
  {Wang}, \citenamefont {Ye}, \citenamefont {Zhou}, \citenamefont {Fan-Yuan},
  \citenamefont {Wang}, \citenamefont {Chen}, \citenamefont {Zhu},
  \citenamefont {Morozov}, \citenamefont {Divochiy}, \citenamefont {Zhou},
  \citenamefont {Guo},\ and\ \citenamefont {Han}}]{wang2022}%
  \BibitemOpen
  \bibfield  {author} {\bibinfo {author} {\bibfnamefont {Shuang}\ \bibnamefont
  {Wang}}, \bibinfo {author} {\bibfnamefont {Zhen-Qiang}\ \bibnamefont {Yin}},
  \bibinfo {author} {\bibfnamefont {De-Yong}\ \bibnamefont {He}}, \bibinfo
  {author} {\bibfnamefont {Wei}\ \bibnamefont {Chen}}, \bibinfo {author}
  {\bibfnamefont {Rui-Qiang}\ \bibnamefont {Wang}}, \bibinfo {author}
  {\bibfnamefont {Peng}\ \bibnamefont {Ye}}, \bibinfo {author} {\bibfnamefont
  {Yao}\ \bibnamefont {Zhou}}, \bibinfo {author} {\bibfnamefont {Guan-Jie}\
  \bibnamefont {Fan-Yuan}}, \bibinfo {author} {\bibfnamefont {Fang-Xiang}\
  \bibnamefont {Wang}}, \bibinfo {author} {\bibfnamefont {Wei}\ \bibnamefont
  {Chen}}, \bibinfo {author} {\bibfnamefont {Yong-Gang}\ \bibnamefont {Zhu}},
  \bibinfo {author} {\bibfnamefont {Pavel~V.}\ \bibnamefont {Morozov}},
  \bibinfo {author} {\bibfnamefont {Alexander~V.}\ \bibnamefont {Divochiy}},
  \bibinfo {author} {\bibfnamefont {Zheng}\ \bibnamefont {Zhou}}, \bibinfo
  {author} {\bibfnamefont {Guang-Can}\ \bibnamefont {Guo}}, \ and\ \bibinfo
  {author} {\bibfnamefont {Zheng-Fu}\ \bibnamefont {Han}},\ }\bibfield  {title}
  {\enquote {\bibinfo {title} {Twin-field quantum key distribution over 830-km
  fibre},}\ }\href {\doibase 10.1038/s41566-021-00928-2} {\bibfield  {journal}
  {\bibinfo  {journal} {Nat. Photonics}\ }\textbf {\bibinfo {volume} {16}},\
  \bibinfo {pages} {154} (\bibinfo {year} {2022})}\BibitemShut {NoStop}%
\bibitem [{\citenamefont {Fan-Yuan}\ \emph {et~al.}(2022)\citenamefont
  {Fan-Yuan}, \citenamefont {Lu}, \citenamefont {Wang}, \citenamefont {Yin},
  \citenamefont {He}, \citenamefont {Chen}, \citenamefont {Zhou}, \citenamefont
  {Wang}, \citenamefont {Teng}, \citenamefont {Guo},\ and\ \citenamefont
  {Han}}]{fan-yuan2022}%
  \BibitemOpen
  \bibfield  {author} {\bibinfo {author} {\bibfnamefont {Guan-Jie}\
  \bibnamefont {Fan-Yuan}}, \bibinfo {author} {\bibfnamefont {Feng-Yu}\
  \bibnamefont {Lu}}, \bibinfo {author} {\bibfnamefont {Shuang}\ \bibnamefont
  {Wang}}, \bibinfo {author} {\bibfnamefont {Zhen-Qiang}\ \bibnamefont {Yin}},
  \bibinfo {author} {\bibfnamefont {De-Yong}\ \bibnamefont {He}}, \bibinfo
  {author} {\bibfnamefont {Wei}\ \bibnamefont {Chen}}, \bibinfo {author}
  {\bibfnamefont {Zheng}\ \bibnamefont {Zhou}}, \bibinfo {author}
  {\bibfnamefont {Ze-Hao}\ \bibnamefont {Wang}}, \bibinfo {author}
  {\bibfnamefont {Jun}\ \bibnamefont {Teng}}, \bibinfo {author} {\bibfnamefont
  {Guang-Can}\ \bibnamefont {Guo}}, \ and\ \bibinfo {author} {\bibfnamefont
  {Zheng-Fu}\ \bibnamefont {Han}},\ }\bibfield  {title} {\enquote {\bibinfo
  {title} {Robust and adaptable quantum key distribution network without
  trusted nodes},}\ }\href {\doibase 10.1364/OPTICA.458937} {\bibfield
  {journal} {\bibinfo  {journal} {Optica}\ }\textbf {\bibinfo {volume} {9}},\
  \bibinfo {pages} {812} (\bibinfo {year} {2022})}\BibitemShut {NoStop}%
\bibitem [{\citenamefont {Rice}\ and\ \citenamefont {Harrington}()}]{rice2009}%
  \BibitemOpen
  \bibfield  {author} {\bibinfo {author} {\bibfnamefont {P.}~\bibnamefont
  {Rice}}\ and\ \bibinfo {author} {\bibfnamefont {J.}~\bibnamefont
  {Harrington}},\ }\bibfield  {title} {\enquote {\bibinfo {title} {Numerical
  analysis of decoy state quantum key distribution protocols},}\ }\href@noop {}
  {\ }\Eprint {http://arxiv.org/abs/0901.0013v2} {arXiv:0901.0013v2 [quant-ph]}
  \BibitemShut {NoStop}%
\bibitem [{\citenamefont {Bochkov}\ and\ \citenamefont
  {Trushechkin}(2019)}]{bochkov2019}%
  \BibitemOpen
  \bibfield  {author} {\bibinfo {author} {\bibfnamefont {M.~K.}\ \bibnamefont
  {Bochkov}}\ and\ \bibinfo {author} {\bibfnamefont {A.~S.}\ \bibnamefont
  {Trushechkin}},\ }\bibfield  {title} {\enquote {\bibinfo {title} {Security of
  quantum key distribution with detection-efficiency mismatch in the
  single-photon case: Tight bounds},}\ }\href {\doibase
  10.1103/PhysRevA.99.032308} {\bibfield  {journal} {\bibinfo  {journal} {Phys.
  Rev. A}\ }\textbf {\bibinfo {volume} {99}},\ \bibinfo {pages} {032308}
  (\bibinfo {year} {2019})}\BibitemShut {NoStop}%
\bibitem [{\citenamefont {Trushechkin}(2022)}]{trushechkin2022}%
  \BibitemOpen
  \bibfield  {author} {\bibinfo {author} {\bibfnamefont {Anton}\ \bibnamefont
  {Trushechkin}},\ }\bibfield  {title} {\enquote {\bibinfo {title} {Security of
  quantum key distribution with detection-efficiency mismatch in the
  multiphoton case},}\ }\href {\doibase 10.22331/q-2022-07-22-771} {\bibfield
  {journal} {\bibinfo  {journal} {Quantum}\ }\textbf {\bibinfo {volume} {6}},\
  \bibinfo {pages} {771} (\bibinfo {year} {2022})}\BibitemShut {NoStop}%
\bibitem [{\citenamefont {Qi}\ \emph {et~al.}(2007)\citenamefont {Qi},
  \citenamefont {Fung}, \citenamefont {Lo},\ and\ \citenamefont {Ma}}]{qi2007}%
  \BibitemOpen
  \bibfield  {author} {\bibinfo {author} {\bibfnamefont {Bing}\ \bibnamefont
  {Qi}}, \bibinfo {author} {\bibfnamefont {Chi-Hang~Fred}\ \bibnamefont
  {Fung}}, \bibinfo {author} {\bibfnamefont {Hoi-Kwong}\ \bibnamefont {Lo}}, \
  and\ \bibinfo {author} {\bibfnamefont {Xiongfeng}\ \bibnamefont {Ma}},\
  }\bibfield  {title} {\enquote {\bibinfo {title} {Time-shift attack in
  practical quantum cryptosystems},}\ }\href@noop {} {\bibfield  {journal}
  {\bibinfo  {journal} {Quantum Inf. Comput.}\ }\textbf {\bibinfo {volume}
  {7}},\ \bibinfo {pages} {73--82} (\bibinfo {year} {2007})}\BibitemShut
  {NoStop}%
\bibitem [{\citenamefont {Vakhitov}\ \emph {et~al.}(2001)\citenamefont
  {Vakhitov}, \citenamefont {Makarov},\ and\ \citenamefont
  {Hjelme}}]{vakhitov2001}%
  \BibitemOpen
  \bibfield  {author} {\bibinfo {author} {\bibfnamefont {A.}~\bibnamefont
  {Vakhitov}}, \bibinfo {author} {\bibfnamefont {V.}~\bibnamefont {Makarov}}, \
  and\ \bibinfo {author} {\bibfnamefont {D.~R.}\ \bibnamefont {Hjelme}},\
  }\bibfield  {title} {\enquote {\bibinfo {title} {Large pulse attack as a
  method of conventional optical eavesdropping in quantum cryptography},}\
  }\href {\doibase 10.1080/09500340108240904} {\bibfield  {journal} {\bibinfo
  {journal} {J. Mod. Opt.}\ }\textbf {\bibinfo {volume} {48}},\ \bibinfo
  {pages} {2023--2038} (\bibinfo {year} {2001})}\BibitemShut {NoStop}%
\bibitem [{\citenamefont {Jain}\ \emph {et~al.}(2014)\citenamefont {Jain},
  \citenamefont {Anisimova}, \citenamefont {Khan}, \citenamefont {Makarov},
  \citenamefont {Marquardt},\ and\ \citenamefont {Leuchs}}]{jain2014}%
  \BibitemOpen
  \bibfield  {author} {\bibinfo {author} {\bibfnamefont {Nitin}\ \bibnamefont
  {Jain}}, \bibinfo {author} {\bibfnamefont {Elena}\ \bibnamefont {Anisimova}},
  \bibinfo {author} {\bibfnamefont {Imran}\ \bibnamefont {Khan}}, \bibinfo
  {author} {\bibfnamefont {Vadim}\ \bibnamefont {Makarov}}, \bibinfo {author}
  {\bibfnamefont {Christoph}\ \bibnamefont {Marquardt}}, \ and\ \bibinfo
  {author} {\bibfnamefont {Gerd}\ \bibnamefont {Leuchs}},\ }\bibfield  {title}
  {\enquote {\bibinfo {title} {Trojan-horse attacks threaten the security of
  practical quantum cryptography},}\ }\href {\doibase
  10.1088/1367-2630/16/12/123030} {\bibfield  {journal} {\bibinfo  {journal}
  {New J. Phys.}\ }\textbf {\bibinfo {volume} {16}},\ \bibinfo {pages} {123030}
  (\bibinfo {year} {2014})}\BibitemShut {NoStop}%
\bibitem [{\citenamefont {Sajeed}\ \emph {et~al.}(2017)\citenamefont {Sajeed},
  \citenamefont {Minshull}, \citenamefont {Jain},\ and\ \citenamefont
  {Makarov}}]{sajeed2017}%
  \BibitemOpen
  \bibfield  {author} {\bibinfo {author} {\bibfnamefont {Shihan}\ \bibnamefont
  {Sajeed}}, \bibinfo {author} {\bibfnamefont {Carter}\ \bibnamefont
  {Minshull}}, \bibinfo {author} {\bibfnamefont {Nitin}\ \bibnamefont {Jain}},
  \ and\ \bibinfo {author} {\bibfnamefont {Vadim}\ \bibnamefont {Makarov}},\
  }\bibfield  {title} {\enquote {\bibinfo {title} {Invisible {T}rojan-horse
  attack},}\ }\href {\doibase 10.1038/s41598-017-08279-1} {\bibfield  {journal}
  {\bibinfo  {journal} {Sci. Rep.}\ }\textbf {\bibinfo {volume} {7}},\ \bibinfo
  {pages} {8403} (\bibinfo {year} {2017})}\BibitemShut {NoStop}%
\bibitem [{\citenamefont {Makarov}\ \emph {et~al.}(2006)\citenamefont
  {Makarov}, \citenamefont {Anisimov},\ and\ \citenamefont
  {Skaar}}]{makarov2006}%
  \BibitemOpen
  \bibfield  {author} {\bibinfo {author} {\bibfnamefont {V.}~\bibnamefont
  {Makarov}}, \bibinfo {author} {\bibfnamefont {A.}~\bibnamefont {Anisimov}}, \
  and\ \bibinfo {author} {\bibfnamefont {J.}~\bibnamefont {Skaar}},\ }\bibfield
   {title} {\enquote {\bibinfo {title} {Effects of detector efficiency mismatch
  on security of quantum cryptosystems},}\ }\href {\doibase
  10.1103/PhysRevA.74.022313} {\bibfield  {journal} {\bibinfo  {journal} {Phys.
  Rev. A}\ }\textbf {\bibinfo {volume} {74}},\ \bibinfo {pages} {022313}
  (\bibinfo {year} {2006})},\ \bibinfo {note} {erratum ibid. \textbf{78},
  019905 (2008)}\BibitemShut {NoStop}%
\bibitem [{\citenamefont {Zhao}\ \emph {et~al.}(2008)\citenamefont {Zhao},
  \citenamefont {Fung}, \citenamefont {Qi}, \citenamefont {Chen},\ and\
  \citenamefont {Lo}}]{zhao2008}%
  \BibitemOpen
  \bibfield  {author} {\bibinfo {author} {\bibfnamefont {Yi}~\bibnamefont
  {Zhao}}, \bibinfo {author} {\bibfnamefont {Chi-Hang~Fred}\ \bibnamefont
  {Fung}}, \bibinfo {author} {\bibfnamefont {Bing}\ \bibnamefont {Qi}},
  \bibinfo {author} {\bibfnamefont {Christine}\ \bibnamefont {Chen}}, \ and\
  \bibinfo {author} {\bibfnamefont {Hoi-Kwong}\ \bibnamefont {Lo}},\ }\bibfield
   {title} {\enquote {\bibinfo {title} {Quantum hacking: Experimental
  demonstration of time-shift attack against practical quantum-key-distribution
  systems},}\ }\href {\doibase 10.1103/PhysRevA.78.042333} {\bibfield
  {journal} {\bibinfo  {journal} {Phys. Rev. A}\ }\textbf {\bibinfo {volume}
  {78}},\ \bibinfo {pages} {042333} (\bibinfo {year} {2008})}\BibitemShut
  {NoStop}%
\bibitem [{\citenamefont {Li}\ \emph {et~al.}(2011)\citenamefont {Li},
  \citenamefont {Wang}, \citenamefont {Huang}, \citenamefont {Chen},
  \citenamefont {Yin}, \citenamefont {Li}, \citenamefont {Zhou}, \citenamefont
  {Liu}, \citenamefont {Zhang}, \citenamefont {Guo}, \citenamefont {Bao},\ and\
  \citenamefont {Han}}]{li2011a}%
  \BibitemOpen
  \bibfield  {author} {\bibinfo {author} {\bibfnamefont {Hong-Wei}\
  \bibnamefont {Li}}, \bibinfo {author} {\bibfnamefont {Shuang}\ \bibnamefont
  {Wang}}, \bibinfo {author} {\bibfnamefont {Jing-Zheng}\ \bibnamefont
  {Huang}}, \bibinfo {author} {\bibfnamefont {Wei}\ \bibnamefont {Chen}},
  \bibinfo {author} {\bibfnamefont {Zhen-Qiang}\ \bibnamefont {Yin}}, \bibinfo
  {author} {\bibfnamefont {Fang-Yi}\ \bibnamefont {Li}}, \bibinfo {author}
  {\bibfnamefont {Zheng}\ \bibnamefont {Zhou}}, \bibinfo {author}
  {\bibfnamefont {Dong}\ \bibnamefont {Liu}}, \bibinfo {author} {\bibfnamefont
  {Yang}\ \bibnamefont {Zhang}}, \bibinfo {author} {\bibfnamefont {Guang-Can}\
  \bibnamefont {Guo}}, \bibinfo {author} {\bibfnamefont {Wan-Su}\ \bibnamefont
  {Bao}}, \ and\ \bibinfo {author} {\bibfnamefont {Zheng-Fu}\ \bibnamefont
  {Han}},\ }\bibfield  {title} {\enquote {\bibinfo {title} {Attacking a
  practical quantum-key-distribution system with wavelength-dependent
  beam-splitter and multiwavelength sources},}\ }\href {\doibase
  10.1103/PhysRevA.84.062308} {\bibfield  {journal} {\bibinfo  {journal} {Phys.
  Rev. A}\ }\textbf {\bibinfo {volume} {84}},\ \bibinfo {pages} {062308}
  (\bibinfo {year} {2011})}\BibitemShut {NoStop}%
\bibitem [{\citenamefont {Huang}\ \emph {et~al.}(2013)\citenamefont {Huang},
  \citenamefont {Weedbrook}, \citenamefont {Yin}, \citenamefont {Wang},
  \citenamefont {Li}, \citenamefont {Chen}, \citenamefont {Guo},\ and\
  \citenamefont {Han}}]{huang2015}%
  \BibitemOpen
  \bibfield  {author} {\bibinfo {author} {\bibfnamefont {Jing-Zheng}\
  \bibnamefont {Huang}}, \bibinfo {author} {\bibfnamefont {Christian}\
  \bibnamefont {Weedbrook}}, \bibinfo {author} {\bibfnamefont {Zhen-Qiang}\
  \bibnamefont {Yin}}, \bibinfo {author} {\bibfnamefont {Shuang}\ \bibnamefont
  {Wang}}, \bibinfo {author} {\bibfnamefont {Hong-Wei}\ \bibnamefont {Li}},
  \bibinfo {author} {\bibfnamefont {Wei}\ \bibnamefont {Chen}}, \bibinfo
  {author} {\bibfnamefont {Guang-Can}\ \bibnamefont {Guo}}, \ and\ \bibinfo
  {author} {\bibfnamefont {Zheng-Fu}\ \bibnamefont {Han}},\ }\bibfield  {title}
  {\enquote {\bibinfo {title} {Quantum hacking of a continuous-variable
  quantum-key-distribution system using a wavelength attack},}\ }\href
  {\doibase 10.1103/PhysRevA.87.062329} {\bibfield  {journal} {\bibinfo
  {journal} {Phys. Rev. A}\ }\textbf {\bibinfo {volume} {87}},\ \bibinfo
  {pages} {062329} (\bibinfo {year} {2013})}\BibitemShut {NoStop}%
\bibitem [{\citenamefont {Weier}\ \emph {et~al.}(2011)\citenamefont {Weier},
  \citenamefont {Krauss}, \citenamefont {Rau}, \citenamefont {F{\"u}rst},
  \citenamefont {Nauerth},\ and\ \citenamefont {Weinfurter}}]{weier2011}%
  \BibitemOpen
  \bibfield  {author} {\bibinfo {author} {\bibfnamefont {H.}~\bibnamefont
  {Weier}}, \bibinfo {author} {\bibfnamefont {H.}~\bibnamefont {Krauss}},
  \bibinfo {author} {\bibfnamefont {M.}~\bibnamefont {Rau}}, \bibinfo {author}
  {\bibfnamefont {M.}~\bibnamefont {F{\"u}rst}}, \bibinfo {author}
  {\bibfnamefont {S.}~\bibnamefont {Nauerth}}, \ and\ \bibinfo {author}
  {\bibfnamefont {H.}~\bibnamefont {Weinfurter}},\ }\bibfield  {title}
  {\enquote {\bibinfo {title} {Quantum eavesdropping without interception: an
  attack exploiting the dead time of single-photon detectors},}\ }\href
  {\doibase 10.1088/1367-2630/13/7/073024} {\bibfield  {journal} {\bibinfo
  {journal} {New J. Phys.}\ }\textbf {\bibinfo {volume} {13}},\ \bibinfo
  {pages} {073024} (\bibinfo {year} {2011})}\BibitemShut {NoStop}%
\bibitem [{\citenamefont {Gisin}\ \emph {et~al.}(2006)\citenamefont {Gisin},
  \citenamefont {Fasel}, \citenamefont {Kraus}, \citenamefont {Zbinden},\ and\
  \citenamefont {Ribordy}}]{gisin2006}%
  \BibitemOpen
  \bibfield  {author} {\bibinfo {author} {\bibfnamefont {N.}~\bibnamefont
  {Gisin}}, \bibinfo {author} {\bibfnamefont {S.}~\bibnamefont {Fasel}},
  \bibinfo {author} {\bibfnamefont {B.}~\bibnamefont {Kraus}}, \bibinfo
  {author} {\bibfnamefont {H.}~\bibnamefont {Zbinden}}, \ and\ \bibinfo
  {author} {\bibfnamefont {G.}~\bibnamefont {Ribordy}},\ }\bibfield  {title}
  {\enquote {\bibinfo {title} {Trojan-horse attacks on quantum-key-distribution
  systems},}\ }\href {\doibase 10.1103/PhysRevA.73.022320} {\bibfield
  {journal} {\bibinfo  {journal} {Phys. Rev. A}\ }\textbf {\bibinfo {volume}
  {73}},\ \bibinfo {pages} {022320} (\bibinfo {year} {2006})}\BibitemShut
  {NoStop}%
\bibitem [{\citenamefont {Lucamarini}\ \emph {et~al.}(2015)\citenamefont
  {Lucamarini}, \citenamefont {Choi}, \citenamefont {Ward}, \citenamefont
  {Dynes}, \citenamefont {Yuan},\ and\ \citenamefont
  {Shields}}]{lucamarini2015}%
  \BibitemOpen
  \bibfield  {author} {\bibinfo {author} {\bibfnamefont {M.}~\bibnamefont
  {Lucamarini}}, \bibinfo {author} {\bibfnamefont {I.}~\bibnamefont {Choi}},
  \bibinfo {author} {\bibfnamefont {M.~B.}\ \bibnamefont {Ward}}, \bibinfo
  {author} {\bibfnamefont {J.~F.}\ \bibnamefont {Dynes}}, \bibinfo {author}
  {\bibfnamefont {Z.~L.}\ \bibnamefont {Yuan}}, \ and\ \bibinfo {author}
  {\bibfnamefont {A.~J.}\ \bibnamefont {Shields}},\ }\bibfield  {title}
  {\enquote {\bibinfo {title} {Practical security bounds against the
  {T}rojan-horse attack in quantum key distribution},}\ }\href {\doibase
  10.1103/PhysRevX.5.031030} {\bibfield  {journal} {\bibinfo  {journal} {Phys.
  Rev. X}\ }\textbf {\bibinfo {volume} {5}},\ \bibinfo {pages} {031030}
  (\bibinfo {year} {2015})}\BibitemShut {NoStop}%
\bibitem [{\citenamefont {Tamaki}\ \emph {et~al.}(2016)\citenamefont {Tamaki},
  \citenamefont {Curty},\ and\ \citenamefont {Lucamarini}}]{tamaki2016}%
  \BibitemOpen
  \bibfield  {author} {\bibinfo {author} {\bibfnamefont {Kiyoshi}\ \bibnamefont
  {Tamaki}}, \bibinfo {author} {\bibfnamefont {Marcos}\ \bibnamefont {Curty}},
  \ and\ \bibinfo {author} {\bibfnamefont {Marco}\ \bibnamefont {Lucamarini}},\
  }\bibfield  {title} {\enquote {\bibinfo {title} {Decoy-state quantum key
  distribution with a leaky source},}\ }\href {\doibase
  10.1088/1367-2630/18/6/065008} {\bibfield  {journal} {\bibinfo  {journal}
  {New J. Phys.}\ }\textbf {\bibinfo {volume} {18}},\ \bibinfo {pages} {065008}
  (\bibinfo {year} {2016})}\BibitemShut {NoStop}%
\bibitem [{\citenamefont {Wang}\ \emph
  {et~al.}(2018{\natexlab{b}})\citenamefont {Wang}, \citenamefont {Tamaki},\
  and\ \citenamefont {Curty}}]{wang2018a}%
  \BibitemOpen
  \bibfield  {author} {\bibinfo {author} {\bibfnamefont {Weilong}\ \bibnamefont
  {Wang}}, \bibinfo {author} {\bibfnamefont {Kiyoshi}\ \bibnamefont {Tamaki}},
  \ and\ \bibinfo {author} {\bibfnamefont {Marcos}\ \bibnamefont {Curty}},\
  }\bibfield  {title} {\enquote {\bibinfo {title} {Finite-key security analysis
  for quantum key distribution with leaky sources},}\ }\href {\doibase
  10.1088/1367-2630/aad839} {\bibfield  {journal} {\bibinfo  {journal} {New J.
  Phys.}\ }\textbf {\bibinfo {volume} {20}},\ \bibinfo {pages} {083027}
  (\bibinfo {year} {2018}{\natexlab{b}})}\BibitemShut {NoStop}%
\bibitem [{\citenamefont {Jain}\ \emph {et~al.}(2015)\citenamefont {Jain},
  \citenamefont {Stiller}, \citenamefont {Khan}, \citenamefont {Makarov},
  \citenamefont {Marquardt},\ and\ \citenamefont {Leuch}}]{jain2015}%
  \BibitemOpen
  \bibfield  {author} {\bibinfo {author} {\bibfnamefont {Nitin}\ \bibnamefont
  {Jain}}, \bibinfo {author} {\bibfnamefont {Birgit}\ \bibnamefont {Stiller}},
  \bibinfo {author} {\bibfnamefont {Imran}\ \bibnamefont {Khan}}, \bibinfo
  {author} {\bibfnamefont {Vadim}\ \bibnamefont {Makarov}}, \bibinfo {author}
  {\bibfnamefont {Christoph}\ \bibnamefont {Marquardt}}, \ and\ \bibinfo
  {author} {\bibfnamefont {Gerd}\ \bibnamefont {Leuch}},\ }\bibfield  {title}
  {\enquote {\bibinfo {title} {Risk analysis of {T}rojan-horse attacks on
  practical quantum key distribution systems},}\ }\href {\doibase
  10.1109/JSTQE.2014.2365585} {\bibfield  {journal} {\bibinfo  {journal} {IEEE
  J. Sel. Top. Quantum Electron.}\ }\textbf {\bibinfo {volume} {21}},\ \bibinfo
  {pages} {6600710} (\bibinfo {year} {2015})}\BibitemShut {NoStop}%
\bibitem [{\citenamefont {Chaiwongkhot}\ \emph {et~al.}(2024)\citenamefont
  {Chaiwongkhot}, \citenamefont {Qin},\ and\ \citenamefont
  {Makarov}}]{chaiwongkhot2024}%
  \BibitemOpen
  \bibfield  {author} {\bibinfo {author} {\bibfnamefont {Poompong}\
  \bibnamefont {Chaiwongkhot}}, \bibinfo {author} {\bibfnamefont {Hao}\
  \bibnamefont {Qin}}, \ and\ \bibinfo {author} {\bibfnamefont {Vadim}\
  \bibnamefont {Makarov}},\ }\href@noop {} {\enquote {\bibinfo {title} {The
  role of interference in {T}rojan-horse attack on quantum cryptography},}\ }
  (\bibinfo {year} {2024}),\ \bibinfo {note} {{u}npublished}\BibitemShut
  {NoStop}%
\bibitem [{\citenamefont {Sushchev}\ \emph {et~al.}(2021)\citenamefont
  {Sushchev}, \citenamefont {Guzairova}, \citenamefont {Klimov}, \citenamefont
  {Dvoretskiy}, \citenamefont {Bogdanov}, \citenamefont {Bondar},\ and\
  \citenamefont {Naumenko}}]{sushchev2021}%
  \BibitemOpen
  \bibfield  {author} {\bibinfo {author} {\bibfnamefont {Ivan~S.}\ \bibnamefont
  {Sushchev}}, \bibinfo {author} {\bibfnamefont {Diana~M.}\ \bibnamefont
  {Guzairova}}, \bibinfo {author} {\bibfnamefont {Andrey~N.}\ \bibnamefont
  {Klimov}}, \bibinfo {author} {\bibfnamefont {Dmitriy~A.}\ \bibnamefont
  {Dvoretskiy}}, \bibinfo {author} {\bibfnamefont {Sergey~A.}\ \bibnamefont
  {Bogdanov}}, \bibinfo {author} {\bibfnamefont {Klim~D.}\ \bibnamefont
  {Bondar}}, \ and\ \bibinfo {author} {\bibfnamefont {Anton~P.}\ \bibnamefont
  {Naumenko}},\ }\bibfield  {title} {\enquote {\bibinfo {title} {Practical
  security analysis against the {T}rojan-horse attacks on fiber-based
  phase-coding {QKD} system in the wide spectral range},}\ }\href {\doibase
  10.1117/12.2600126} {\bibfield  {journal} {\bibinfo  {journal} {Proc. SPIE}\
  }\textbf {\bibinfo {volume} {11868}},\ \bibinfo {pages} {118680H} (\bibinfo
  {year} {2021})}\BibitemShut {NoStop}%
\bibitem [{\citenamefont {Sun}\ \emph {et~al.}(2015)\citenamefont {Sun},
  \citenamefont {Xu}, \citenamefont {Jiang}, \citenamefont {Ma}, \citenamefont
  {Lo},\ and\ \citenamefont {Liang}}]{sun2015}%
  \BibitemOpen
  \bibfield  {author} {\bibinfo {author} {\bibfnamefont {Shi-Hai}\ \bibnamefont
  {Sun}}, \bibinfo {author} {\bibfnamefont {Feihu}\ \bibnamefont {Xu}},
  \bibinfo {author} {\bibfnamefont {Mu-Sheng}\ \bibnamefont {Jiang}}, \bibinfo
  {author} {\bibfnamefont {Xiang-Chun}\ \bibnamefont {Ma}}, \bibinfo {author}
  {\bibfnamefont {Hoi-Kwong}\ \bibnamefont {Lo}}, \ and\ \bibinfo {author}
  {\bibfnamefont {Lin-Mei}\ \bibnamefont {Liang}},\ }\bibfield  {title}
  {\enquote {\bibinfo {title} {Effect of source tampering in the security of
  quantum cryptography},}\ }\href {\doibase 10.1103/PhysRevA.92.022304}
  {\bibfield  {journal} {\bibinfo  {journal} {Phys. Rev. A}\ }\textbf {\bibinfo
  {volume} {92}},\ \bibinfo {pages} {022304} (\bibinfo {year}
  {2015})}\BibitemShut {NoStop}%
\bibitem [{\citenamefont {Huang}\ \emph {et~al.}(2019)\citenamefont {Huang},
  \citenamefont {Navarrete}, \citenamefont {Sun}, \citenamefont {Chaiwongkhot},
  \citenamefont {Curty},\ and\ \citenamefont {Makarov}}]{huang2019}%
  \BibitemOpen
  \bibfield  {author} {\bibinfo {author} {\bibfnamefont {Anqi}\ \bibnamefont
  {Huang}}, \bibinfo {author} {\bibfnamefont {{\'A}lvaro}\ \bibnamefont
  {Navarrete}}, \bibinfo {author} {\bibfnamefont {Shi-Hai}\ \bibnamefont
  {Sun}}, \bibinfo {author} {\bibfnamefont {Poompong}\ \bibnamefont
  {Chaiwongkhot}}, \bibinfo {author} {\bibfnamefont {Marcos}\ \bibnamefont
  {Curty}}, \ and\ \bibinfo {author} {\bibfnamefont {Vadim}\ \bibnamefont
  {Makarov}},\ }\bibfield  {title} {\enquote {\bibinfo {title} {Laser-seeding
  attack in quantum key distribution},}\ }\href {\doibase
  10.1103/PhysRevApplied.12.064043} {\bibfield  {journal} {\bibinfo  {journal}
  {Phys. Rev. Appl.}\ }\textbf {\bibinfo {volume} {12}},\ \bibinfo {pages}
  {064043} (\bibinfo {year} {2019})}\BibitemShut {NoStop}%
\bibitem [{\citenamefont {Pang}\ \emph {et~al.}(2020)\citenamefont {Pang},
  \citenamefont {Yang}, \citenamefont {Zhang}, \citenamefont {Dou},
  \citenamefont {Li}, \citenamefont {Gao},\ and\ \citenamefont
  {Jin}}]{pang2020}%
  \BibitemOpen
  \bibfield  {author} {\bibinfo {author} {\bibfnamefont {Xiao-Ling}\
  \bibnamefont {Pang}}, \bibinfo {author} {\bibfnamefont {Ai-Lin}\ \bibnamefont
  {Yang}}, \bibinfo {author} {\bibfnamefont {Chao-Ni}\ \bibnamefont {Zhang}},
  \bibinfo {author} {\bibfnamefont {Jian-Peng}\ \bibnamefont {Dou}}, \bibinfo
  {author} {\bibfnamefont {Hang}\ \bibnamefont {Li}}, \bibinfo {author}
  {\bibfnamefont {Jun}\ \bibnamefont {Gao}}, \ and\ \bibinfo {author}
  {\bibfnamefont {Xian-Min}\ \bibnamefont {Jin}},\ }\bibfield  {title}
  {\enquote {\bibinfo {title} {Hacking quantum key distribution via injection
  locking},}\ }\href {\doibase 10.1103/PhysRevApplied.13.034008} {\bibfield
  {journal} {\bibinfo  {journal} {Phys. Rev. Appl.}\ }\textbf {\bibinfo
  {volume} {13}},\ \bibinfo {pages} {034008} (\bibinfo {year}
  {2020})}\BibitemShut {NoStop}%
\bibitem [{\citenamefont {Lovic}\ \emph {et~al.}(2023)\citenamefont {Lovic},
  \citenamefont {Marangon}, \citenamefont {Smith}, \citenamefont {Woodward},\
  and\ \citenamefont {Shields}}]{lovic2023}%
  \BibitemOpen
  \bibfield  {author} {\bibinfo {author} {\bibfnamefont {V.}~\bibnamefont
  {Lovic}}, \bibinfo {author} {\bibfnamefont {D.G.}\ \bibnamefont {Marangon}},
  \bibinfo {author} {\bibfnamefont {P.R.}\ \bibnamefont {Smith}}, \bibinfo
  {author} {\bibfnamefont {R.I.}\ \bibnamefont {Woodward}}, \ and\ \bibinfo
  {author} {\bibfnamefont {A.J.}\ \bibnamefont {Shields}},\ }\bibfield  {title}
  {\enquote {\bibinfo {title} {Quantified effects of the laser-seeding attack
  in quantum key distribution},}\ }\href {\doibase
  10.1103/PhysRevApplied.20.044005} {\bibfield  {journal} {\bibinfo  {journal}
  {Phys. Rev. Appl.}\ }\textbf {\bibinfo {volume} {20}},\ \bibinfo {pages}
  {044005} (\bibinfo {year} {2023})}\BibitemShut {NoStop}%
\bibitem [{\citenamefont {Lu}\ \emph {et~al.}(2023)\citenamefont {Lu},
  \citenamefont {Ye}, \citenamefont {Wang}, \citenamefont {Wang}, \citenamefont
  {Yin}, \citenamefont {Wang}, \citenamefont {Huang}, \citenamefont {Chen},
  \citenamefont {He}, \citenamefont {Fan-Yuan}, \citenamefont {Guo},\ and\
  \citenamefont {Han}}]{lu2023}%
  \BibitemOpen
  \bibfield  {author} {\bibinfo {author} {\bibfnamefont {Feng-Yu}\ \bibnamefont
  {Lu}}, \bibinfo {author} {\bibfnamefont {Peng}\ \bibnamefont {Ye}}, \bibinfo
  {author} {\bibfnamefont {Ze-Hao}\ \bibnamefont {Wang}}, \bibinfo {author}
  {\bibfnamefont {Shuang}\ \bibnamefont {Wang}}, \bibinfo {author}
  {\bibfnamefont {Zhen-Qiang}\ \bibnamefont {Yin}}, \bibinfo {author}
  {\bibfnamefont {Rong}\ \bibnamefont {Wang}}, \bibinfo {author} {\bibfnamefont
  {Xiao-Juan}\ \bibnamefont {Huang}}, \bibinfo {author} {\bibfnamefont {Wei}\
  \bibnamefont {Chen}}, \bibinfo {author} {\bibfnamefont {De-Yong}\
  \bibnamefont {He}}, \bibinfo {author} {\bibfnamefont {Guan-Jie}\ \bibnamefont
  {Fan-Yuan}}, \bibinfo {author} {\bibfnamefont {Guang-Can}\ \bibnamefont
  {Guo}}, \ and\ \bibinfo {author} {\bibfnamefont {Zheng-Fu}\ \bibnamefont
  {Han}},\ }\bibfield  {title} {\enquote {\bibinfo {title} {Hacking
  measurement-device-independent quantum key distribution},}\ }\href {\doibase
  10.1364/OPTICA.485389} {\bibfield  {journal} {\bibinfo  {journal} {Optica}\
  }\textbf {\bibinfo {volume} {10}},\ \bibinfo {pages} {520} (\bibinfo {year}
  {2023})}\BibitemShut {NoStop}%
\bibitem [{\citenamefont {Han}\ \emph {et~al.}(2023)\citenamefont {Han},
  \citenamefont {Li}, \citenamefont {Tan}, \citenamefont {Zhang}, \citenamefont
  {Cai}, \citenamefont {Yin}, \citenamefont {Ren}, \citenamefont {Xu},
  \citenamefont {Liao},\ and\ \citenamefont {Peng}}]{han2023}%
  \BibitemOpen
  \bibfield  {author} {\bibinfo {author} {\bibfnamefont {Liying}\ \bibnamefont
  {Han}}, \bibinfo {author} {\bibfnamefont {Yang}\ \bibnamefont {Li}}, \bibinfo
  {author} {\bibfnamefont {Hao}\ \bibnamefont {Tan}}, \bibinfo {author}
  {\bibfnamefont {Weiyang}\ \bibnamefont {Zhang}}, \bibinfo {author}
  {\bibfnamefont {Wenqi}\ \bibnamefont {Cai}}, \bibinfo {author} {\bibfnamefont
  {Juan}\ \bibnamefont {Yin}}, \bibinfo {author} {\bibfnamefont {Jigang}\
  \bibnamefont {Ren}}, \bibinfo {author} {\bibfnamefont {Feihu}\ \bibnamefont
  {Xu}}, \bibinfo {author} {\bibfnamefont {Shengkai}\ \bibnamefont {Liao}}, \
  and\ \bibinfo {author} {\bibfnamefont {Chengzhi}\ \bibnamefont {Peng}},\
  }\bibfield  {title} {\enquote {\bibinfo {title} {Effect of light injection on
  the security of practical quantum key distribution},}\ }\href {\doibase
  10.1103/PhysRevApplied.20.044013} {\bibfield  {journal} {\bibinfo  {journal}
  {Phys. Rev. Appl.}\ }\textbf {\bibinfo {volume} {20}},\ \bibinfo {pages}
  {044013} (\bibinfo {year} {2023})}\BibitemShut {NoStop}%
\bibitem [{\citenamefont {Kostritskii}(2009)}]{kostritskii2009}%
  \BibitemOpen
  \bibfield  {author} {\bibinfo {author} {\bibfnamefont {S.M.}\ \bibnamefont
  {Kostritskii}},\ }\bibfield  {title} {\enquote {\bibinfo {title}
  {Photorefractive effect in {LiNbO\textsubscript{3}}-based integrated-optical
  circuits at wavelengths of third telecom window},}\ }\href {\doibase
  10.1007/s00340-009-3501-4} {\bibfield  {journal} {\bibinfo  {journal} {Appl.
  Phys. B}\ }\textbf {\bibinfo {volume} {95}},\ \bibinfo {pages} {421--428}
  (\bibinfo {year} {2009})}\BibitemShut {NoStop}%
\bibitem [{\citenamefont {Friberg}\ \emph {et~al.}(1988)\citenamefont
  {Friberg}, \citenamefont {Weiner}, \citenamefont {Silberberg}, \citenamefont
  {Sfez},\ and\ \citenamefont {Smith}}]{friberg1988}%
  \BibitemOpen
  \bibfield  {author} {\bibinfo {author} {\bibfnamefont {S.~R.}\ \bibnamefont
  {Friberg}}, \bibinfo {author} {\bibfnamefont {A.~M.}\ \bibnamefont {Weiner}},
  \bibinfo {author} {\bibfnamefont {Y.}~\bibnamefont {Silberberg}}, \bibinfo
  {author} {\bibfnamefont {B.~G.}\ \bibnamefont {Sfez}}, \ and\ \bibinfo
  {author} {\bibfnamefont {P.~S.}\ \bibnamefont {Smith}},\ }\bibfield  {title}
  {\enquote {\bibinfo {title} {Femotosecond switching in a dual-core-fiber
  nonlinear coupler},}\ }\href {\doibase 10.1364/OL.13.000904} {\bibfield
  {journal} {\bibinfo  {journal} {Opt. Lett.}\ }\textbf {\bibinfo {volume}
  {13}},\ \bibinfo {pages} {904--906} (\bibinfo {year} {1988})}\BibitemShut
  {NoStop}%
\bibitem [{\citenamefont {Huang}\ \emph {et~al.}(2020)\citenamefont {Huang},
  \citenamefont {Li}, \citenamefont {Egorov}, \citenamefont {Tchouragoulov},
  \citenamefont {Kumar},\ and\ \citenamefont {Makarov}}]{huang2020}%
  \BibitemOpen
  \bibfield  {author} {\bibinfo {author} {\bibfnamefont {Anqi}\ \bibnamefont
  {Huang}}, \bibinfo {author} {\bibfnamefont {Ruoping}\ \bibnamefont {Li}},
  \bibinfo {author} {\bibfnamefont {Vladimir}\ \bibnamefont {Egorov}}, \bibinfo
  {author} {\bibfnamefont {Serguei}\ \bibnamefont {Tchouragoulov}}, \bibinfo
  {author} {\bibfnamefont {Krtin}\ \bibnamefont {Kumar}}, \ and\ \bibinfo
  {author} {\bibfnamefont {Vadim}\ \bibnamefont {Makarov}},\ }\bibfield
  {title} {\enquote {\bibinfo {title} {Laser-damage attack against optical
  attenuators in quantum key distribution},}\ }\href {\doibase
  10.1103/PhysRevApplied.13.034017} {\bibfield  {journal} {\bibinfo  {journal}
  {Phys. Rev. Appl.}\ }\textbf {\bibinfo {volume} {13}},\ \bibinfo {pages}
  {034017} (\bibinfo {year} {2020})}\BibitemShut {NoStop}%
\bibitem [{\citenamefont {Bugai}\ \emph {et~al.}(2022)\citenamefont {Bugai},
  \citenamefont {Zyzykin}, \citenamefont {Bulavkin}, \citenamefont {Bogdanov},
  \citenamefont {Sushchev},\ and\ \citenamefont {Dvoretskiy}}]{bugai2022}%
  \BibitemOpen
  \bibfield  {author} {\bibinfo {author} {\bibfnamefont {K.~E.}\ \bibnamefont
  {Bugai}}, \bibinfo {author} {\bibfnamefont {A.~P.}\ \bibnamefont {Zyzykin}},
  \bibinfo {author} {\bibfnamefont {D.~S.}\ \bibnamefont {Bulavkin}}, \bibinfo
  {author} {\bibfnamefont {S.~A.}\ \bibnamefont {Bogdanov}}, \bibinfo {author}
  {\bibfnamefont {I.~S.}\ \bibnamefont {Sushchev}}, \ and\ \bibinfo {author}
  {\bibfnamefont {D.~A.}\ \bibnamefont {Dvoretskiy}},\ }\bibfield  {title}
  {\enquote {\bibinfo {title} {Laser damage attack on a simple optical
  attenuator widely used in fiber-based {QKD} systems},}\ }in\ \href {\doibase
  10.1109/ICLO54117.2022.9839749} {\emph {\bibinfo {booktitle} {Proc. 2022
  International Conference Laser Optics (ICLO)}}}\ (\bibinfo  {publisher}
  {IEEE},\ \bibinfo {year} {2022})\ p.\ \bibinfo {pages} {393}\BibitemShut
  {NoStop}%
\bibitem [{\citenamefont {Ponosova}\ \emph {et~al.}(2022)\citenamefont
  {Ponosova}, \citenamefont {Ruzhitskaya}, \citenamefont {Chaiwongkhot},
  \citenamefont {Egorov}, \citenamefont {Makarov},\ and\ \citenamefont
  {Huang}}]{ponosova2022}%
  \BibitemOpen
  \bibfield  {author} {\bibinfo {author} {\bibfnamefont {Anastasiya}\
  \bibnamefont {Ponosova}}, \bibinfo {author} {\bibfnamefont {Daria}\
  \bibnamefont {Ruzhitskaya}}, \bibinfo {author} {\bibfnamefont {Poompong}\
  \bibnamefont {Chaiwongkhot}}, \bibinfo {author} {\bibfnamefont {Vladimir}\
  \bibnamefont {Egorov}}, \bibinfo {author} {\bibfnamefont {Vadim}\
  \bibnamefont {Makarov}}, \ and\ \bibinfo {author} {\bibfnamefont {Anqi}\
  \bibnamefont {Huang}},\ }\bibfield  {title} {\enquote {\bibinfo {title}
  {Protecting fiber-optic quantum key distribution sources against
  light-injection attacks},}\ }\href {\doibase 10.1103/PRXQuantum.3.040307}
  {\bibfield  {journal} {\bibinfo  {journal} {PRX Quantum}\ }\textbf {\bibinfo
  {volume} {3}},\ \bibinfo {pages} {040307} (\bibinfo {year}
  {2022})}\BibitemShut {NoStop}%
\bibitem [{\citenamefont {Ruzhitskaya}\ \emph {et~al.}(2021)\citenamefont
  {Ruzhitskaya}, \citenamefont {Zhluktova}, \citenamefont {Petrov},
  \citenamefont {Zaitsev}, \citenamefont {Acheva}, \citenamefont {Zunikov},
  \citenamefont {Shilko}, \citenamefont {Akta}, \citenamefont {Johlinger},
  \citenamefont {Trefilov}, \citenamefont {Ponosova}, \citenamefont {Kamynin},\
  and\ \citenamefont {Makarov}}]{ruzhitskaya2021}%
  \BibitemOpen
  \bibfield  {author} {\bibinfo {author} {\bibfnamefont {Daria~D.}\
  \bibnamefont {Ruzhitskaya}}, \bibinfo {author} {\bibfnamefont {Irina~V.}\
  \bibnamefont {Zhluktova}}, \bibinfo {author} {\bibfnamefont {Mikhail~A.}\
  \bibnamefont {Petrov}}, \bibinfo {author} {\bibfnamefont {Konstantin~A.}\
  \bibnamefont {Zaitsev}}, \bibinfo {author} {\bibfnamefont {Polina~P.}\
  \bibnamefont {Acheva}}, \bibinfo {author} {\bibfnamefont {Nikolay~A.}\
  \bibnamefont {Zunikov}}, \bibinfo {author} {\bibfnamefont {Aleksei~V.}\
  \bibnamefont {Shilko}}, \bibinfo {author} {\bibfnamefont {Djeylan}\
  \bibnamefont {Akta}}, \bibinfo {author} {\bibfnamefont {Friederike}\
  \bibnamefont {Johlinger}}, \bibinfo {author} {\bibfnamefont {Daniil~O.}\
  \bibnamefont {Trefilov}}, \bibinfo {author} {\bibfnamefont {Anastasiya~A.}\
  \bibnamefont {Ponosova}}, \bibinfo {author} {\bibfnamefont {Vladimir~A.}\
  \bibnamefont {Kamynin}}, \ and\ \bibinfo {author} {\bibfnamefont {Vadim~V.}\
  \bibnamefont {Makarov}},\ }\bibfield  {title} {\enquote {\bibinfo {title}
  {Vulnerabilities in the quantum key distribution system induced under a
  pulsed laser attack},}\ }\href {\doibase
  10.17586/2226-1494-2021-21-6-837-847} {\bibfield  {journal} {\bibinfo
  {journal} {Sci. Tech. J. Inf. Technol. Mech. Opt.}\ }\textbf {\bibinfo
  {volume} {21}},\ \bibinfo {pages} {837--847} (\bibinfo {year} {2021})},\
  \bibinfo {note} {in~Russian}\BibitemShut {NoStop}%
\bibitem [{\citenamefont {Wood}(2003)}]{wood2003}%
  \BibitemOpen
  \bibfield  {author} {\bibinfo {author} {\bibfnamefont {Roger~M.}\
  \bibnamefont {Wood}},\ }\href@noop {} {\emph {\bibinfo {title} {Laser-Induced
  Damage of Optical Materials}}},\ \bibinfo {edition} {1st}\ ed.\ (\bibinfo
  {publisher} {CRC Press},\ \bibinfo {year} {2003})\BibitemShut {NoStop}%
\bibitem [{\citenamefont {Meda}\ \emph {et~al.}(2017)\citenamefont {Meda},
  \citenamefont {Degiovanni}, \citenamefont {Tosi}, \citenamefont {Yuan},
  \citenamefont {Brida},\ and\ \citenamefont {Genovese}}]{meda2017}%
  \BibitemOpen
  \bibfield  {author} {\bibinfo {author} {\bibfnamefont {Alice}\ \bibnamefont
  {Meda}}, \bibinfo {author} {\bibfnamefont {Ivo~Pietro}\ \bibnamefont
  {Degiovanni}}, \bibinfo {author} {\bibfnamefont {Alberto}\ \bibnamefont
  {Tosi}}, \bibinfo {author} {\bibfnamefont {Zhiliang}\ \bibnamefont {Yuan}},
  \bibinfo {author} {\bibfnamefont {Giorgio}\ \bibnamefont {Brida}}, \ and\
  \bibinfo {author} {\bibfnamefont {Marco}\ \bibnamefont {Genovese}},\
  }\bibfield  {title} {\enquote {\bibinfo {title} {Quantifying backflash
  radiation to prevent zero-error attacks in quantum key distribution},}\
  }\href {\doibase 10.1038/lsa.2016.261} {\bibfield  {journal} {\bibinfo
  {journal} {Light Sci. Appl.}\ }\textbf {\bibinfo {volume} {6}},\ \bibinfo
  {pages} {e16261} (\bibinfo {year} {2017})}\BibitemShut {NoStop}%
\bibitem [{\citenamefont {Pinheiro}\ \emph {et~al.}(2018)\citenamefont
  {Pinheiro}, \citenamefont {Chaiwongkhot}, \citenamefont {Sajeed},
  \citenamefont {Horn}, \citenamefont {Bourgoin}, \citenamefont {Jennewein},
  \citenamefont {L\"{u}tkenhaus},\ and\ \citenamefont
  {Makarov}}]{pinheiro2018}%
  \BibitemOpen
  \bibfield  {author} {\bibinfo {author} {\bibfnamefont {Paulo
  Vinicius~Pereira}\ \bibnamefont {Pinheiro}}, \bibinfo {author} {\bibfnamefont
  {Poompong}\ \bibnamefont {Chaiwongkhot}}, \bibinfo {author} {\bibfnamefont
  {Shihan}\ \bibnamefont {Sajeed}}, \bibinfo {author} {\bibfnamefont {Rolf~T.}\
  \bibnamefont {Horn}}, \bibinfo {author} {\bibfnamefont {Jean-Philippe}\
  \bibnamefont {Bourgoin}}, \bibinfo {author} {\bibfnamefont {Thomas}\
  \bibnamefont {Jennewein}}, \bibinfo {author} {\bibfnamefont {Norbert}\
  \bibnamefont {L\"{u}tkenhaus}}, \ and\ \bibinfo {author} {\bibfnamefont
  {Vadim}\ \bibnamefont {Makarov}},\ }\bibfield  {title} {\enquote {\bibinfo
  {title} {Eavesdropping and countermeasures for backflash side channel in
  quantum cryptography},}\ }\href {\doibase 10.1364/OE.26.021020} {\bibfield
  {journal} {\bibinfo  {journal} {Opt. Express}\ }\textbf {\bibinfo {volume}
  {26}},\ \bibinfo {pages} {21020--21032} (\bibinfo {year} {2018})}\BibitemShut
  {NoStop}%
\bibitem [{\citenamefont {Koehler-Sidki}\ \emph {et~al.}(2020)\citenamefont
  {Koehler-Sidki}, \citenamefont {Dynes}, \citenamefont {Para\"iso},
  \citenamefont {Lucamarini}, \citenamefont {Sharpe}, \citenamefont {Yuan},\
  and\ \citenamefont {Shields}}]{koehler-sidki2020}%
  \BibitemOpen
  \bibfield  {author} {\bibinfo {author} {\bibfnamefont {A.}~\bibnamefont
  {Koehler-Sidki}}, \bibinfo {author} {\bibfnamefont {J.F.}\ \bibnamefont
  {Dynes}}, \bibinfo {author} {\bibfnamefont {T.K.}\ \bibnamefont {Para\"iso}},
  \bibinfo {author} {\bibfnamefont {M.}~\bibnamefont {Lucamarini}}, \bibinfo
  {author} {\bibfnamefont {A.~W.}\ \bibnamefont {Sharpe}}, \bibinfo {author}
  {\bibfnamefont {Z.L.}\ \bibnamefont {Yuan}}, \ and\ \bibinfo {author}
  {\bibfnamefont {A.J.}\ \bibnamefont {Shields}},\ }\bibfield  {title}
  {\enquote {\bibinfo {title} {Backflashes from fast-gated avalanche
  photodiodes in quantum key distribution},}\ }\href {\doibase
  10.1063/1.5140548} {\bibfield  {journal} {\bibinfo  {journal} {Appl. Phys.
  Lett.}\ }\textbf {\bibinfo {volume} {116}},\ \bibinfo {pages} {154001}
  (\bibinfo {year} {2020})}\BibitemShut {NoStop}%
\bibitem [{\citenamefont {Shi}\ \emph {et~al.}(2017)\citenamefont {Shi},
  \citenamefont {Lim}, \citenamefont {Poh}, \citenamefont {Tan}, \citenamefont
  {Tan}, \citenamefont {Ling},\ and\ \citenamefont {Kurtsiefer}}]{shi2017}%
  \BibitemOpen
  \bibfield  {author} {\bibinfo {author} {\bibfnamefont {Yicheng}\ \bibnamefont
  {Shi}}, \bibinfo {author} {\bibfnamefont {Janet Zheng~Jie}\ \bibnamefont
  {Lim}}, \bibinfo {author} {\bibfnamefont {Hou~Shun}\ \bibnamefont {Poh}},
  \bibinfo {author} {\bibfnamefont {Peng~Kian}\ \bibnamefont {Tan}}, \bibinfo
  {author} {\bibfnamefont {Peiyu~Amelia}\ \bibnamefont {Tan}}, \bibinfo
  {author} {\bibfnamefont {Alexander}\ \bibnamefont {Ling}}, \ and\ \bibinfo
  {author} {\bibfnamefont {Christian}\ \bibnamefont {Kurtsiefer}},\ }\bibfield
  {title} {\enquote {\bibinfo {title} {Breakdown flash at telecom wavelengths
  in {InGaAs} avalanche photodiodes},}\ }\href {\doibase 10.1364/OE.25.030388}
  {\bibfield  {journal} {\bibinfo  {journal} {Opt. Express}\ }\textbf {\bibinfo
  {volume} {25}},\ \bibinfo {pages} {30388} (\bibinfo {year}
  {2017})}\BibitemShut {NoStop}%
\bibitem [{\citenamefont {Bogdanov}\ \emph {et~al.}(2022)\citenamefont
  {Bogdanov}, \citenamefont {Sushchev}, \citenamefont {Klimov}, \citenamefont
  {Bugai}, \citenamefont {Bulavkin},\ and\ \citenamefont
  {Dvoretskiy}}]{bogdanov2022}%
  \BibitemOpen
  \bibfield  {author} {\bibinfo {author} {\bibfnamefont {Sergey~A.}\
  \bibnamefont {Bogdanov}}, \bibinfo {author} {\bibfnamefont {Ivan~S.}\
  \bibnamefont {Sushchev}}, \bibinfo {author} {\bibfnamefont {Andrey~N.}\
  \bibnamefont {Klimov}}, \bibinfo {author} {\bibfnamefont {Kirill~E.}\
  \bibnamefont {Bugai}}, \bibinfo {author} {\bibfnamefont {Daniil~S.}\
  \bibnamefont {Bulavkin}}, \ and\ \bibinfo {author} {\bibfnamefont
  {Dmitriy~A.}\ \bibnamefont {Dvoretskiy}},\ }\bibfield  {title} {\enquote
  {\bibinfo {title} {Influence of {QKD} apparatus parameters on the
  ``backflash'' attack},}\ }\href {\doibase 10.1117/12.2622211} {\bibfield
  {journal} {\bibinfo  {journal} {Proc. SPIE}\ }\textbf {\bibinfo {volume}
  {12133}},\ \bibinfo {pages} {121330G} (\bibinfo {year} {2022})}\BibitemShut
  {NoStop}%
\bibitem [{\citenamefont {Yoshino}\ \emph {et~al.}(2018)\citenamefont
  {Yoshino}, \citenamefont {Fujiwara}, \citenamefont {Nakata}, \citenamefont
  {Sumiya}, \citenamefont {Sasaki}, \citenamefont {Takeoka}, \citenamefont
  {Sasaki}, \citenamefont {Tajima}, \citenamefont {Koashi},\ and\ \citenamefont
  {Tomita}}]{yoshino2018}%
  \BibitemOpen
  \bibfield  {author} {\bibinfo {author} {\bibfnamefont {Ken-ichiro}\
  \bibnamefont {Yoshino}}, \bibinfo {author} {\bibfnamefont {Mikio}\
  \bibnamefont {Fujiwara}}, \bibinfo {author} {\bibfnamefont {Kensuke}\
  \bibnamefont {Nakata}}, \bibinfo {author} {\bibfnamefont {Tatsuya}\
  \bibnamefont {Sumiya}}, \bibinfo {author} {\bibfnamefont {Toshihiko}\
  \bibnamefont {Sasaki}}, \bibinfo {author} {\bibfnamefont {Masahiro}\
  \bibnamefont {Takeoka}}, \bibinfo {author} {\bibfnamefont {Masahide}\
  \bibnamefont {Sasaki}}, \bibinfo {author} {\bibfnamefont {Akio}\ \bibnamefont
  {Tajima}}, \bibinfo {author} {\bibfnamefont {Masato}\ \bibnamefont {Koashi}},
  \ and\ \bibinfo {author} {\bibfnamefont {Akihisa}\ \bibnamefont {Tomita}},\
  }\bibfield  {title} {\enquote {\bibinfo {title} {Quantum key distribution
  with an efficient countermeasure against correlated intensity fluctuations in
  optical pulses},}\ }\href {\doibase 10.1038/s41534-017-0057-8} {\bibfield
  {journal} {\bibinfo  {journal} {npj Quantum Inf.}\ }\textbf {\bibinfo
  {volume} {4}},\ \bibinfo {pages} {8} (\bibinfo {year} {2018})}\BibitemShut
  {NoStop}%
\bibitem [{\citenamefont {Gr{\"u}nenfelder}\ \emph {et~al.}(2020)\citenamefont
  {Gr{\"u}nenfelder}, \citenamefont {Boaron}, \citenamefont {Rusca},
  \citenamefont {Martin},\ and\ \citenamefont {Zbinden}}]{grunenfelder2020}%
  \BibitemOpen
  \bibfield  {author} {\bibinfo {author} {\bibfnamefont {Fadri}\ \bibnamefont
  {Gr{\"u}nenfelder}}, \bibinfo {author} {\bibfnamefont {Alberto}\ \bibnamefont
  {Boaron}}, \bibinfo {author} {\bibfnamefont {Davide}\ \bibnamefont {Rusca}},
  \bibinfo {author} {\bibfnamefont {Anthony}\ \bibnamefont {Martin}}, \ and\
  \bibinfo {author} {\bibfnamefont {Hugo}\ \bibnamefont {Zbinden}},\ }\bibfield
   {title} {\enquote {\bibinfo {title} {Performance and security of 5~{GH}z
  repetition rate polarization-based quantum key distribution},}\ }\href
  {\doibase 10.1063/5.0021468} {\bibfield  {journal} {\bibinfo  {journal}
  {Appl. Phys. Lett.}\ }\textbf {\bibinfo {volume} {117}},\ \bibinfo {pages}
  {144003} (\bibinfo {year} {2020})}\BibitemShut {NoStop}%
\bibitem [{\citenamefont {Kobayashi}\ \emph {et~al.}(2014)\citenamefont
  {Kobayashi}, \citenamefont {Tomita},\ and\ \citenamefont
  {Okamoto}}]{kobayashi2014}%
  \BibitemOpen
  \bibfield  {author} {\bibinfo {author} {\bibfnamefont {Toshiya}\ \bibnamefont
  {Kobayashi}}, \bibinfo {author} {\bibfnamefont {Akihisa}\ \bibnamefont
  {Tomita}}, \ and\ \bibinfo {author} {\bibfnamefont {Atsushi}\ \bibnamefont
  {Okamoto}},\ }\bibfield  {title} {\enquote {\bibinfo {title} {Evaluation of
  the phase randomness of a light source in quantum-key-distribution systems
  with an attenuated laser},}\ }\href {\doibase 10.1103/PhysRevA.90.032320}
  {\bibfield  {journal} {\bibinfo  {journal} {Phys. Rev. A}\ }\textbf {\bibinfo
  {volume} {90}},\ \bibinfo {pages} {032320} (\bibinfo {year}
  {2014})}\BibitemShut {NoStop}%
\bibitem [{\citenamefont {Trefilov}(2021)}]{trefilov2021}%
  \BibitemOpen
  \bibfield  {author} {\bibinfo {author} {\bibfnamefont {Daniil}\ \bibnamefont
  {Trefilov}},\ }\emph {\bibinfo {title} {Imperfect state preparation in
  quantum key distribution}},\ \href@noop {} {Master's thesis},\ \bibinfo
  {school} {Higher School of Economics} (\bibinfo {year} {2021})\BibitemShut
  {NoStop}%
\bibitem [{\citenamefont {Zapatero}\ \emph {et~al.}(2021)\citenamefont
  {Zapatero}, \citenamefont {Navarrete}, \citenamefont {Tamaki},\ and\
  \citenamefont {Curty}}]{zapatero2021}%
  \BibitemOpen
  \bibfield  {author} {\bibinfo {author} {\bibfnamefont {V{\'\i}ctor}\
  \bibnamefont {Zapatero}}, \bibinfo {author} {\bibfnamefont {{\'A}lvaro}\
  \bibnamefont {Navarrete}}, \bibinfo {author} {\bibfnamefont {Kiyoshi}\
  \bibnamefont {Tamaki}}, \ and\ \bibinfo {author} {\bibfnamefont {Marcos}\
  \bibnamefont {Curty}},\ }\bibfield  {title} {\enquote {\bibinfo {title}
  {Security of quantum key distribution with intensity correlations},}\ }\href
  {\doibase 10.22331/q-2021-12-07-602} {\bibfield  {journal} {\bibinfo
  {journal} {Quantum}\ }\textbf {\bibinfo {volume} {5}},\ \bibinfo {pages}
  {602} (\bibinfo {year} {2021})}\BibitemShut {NoStop}%
\bibitem [{\citenamefont {Curr{\'a}s-Lorenzo}\ \emph
  {et~al.}(2024)\citenamefont {Curr{\'a}s-Lorenzo}, \citenamefont {Nahar},
  \citenamefont {L{\" u}tkenhaus}, \citenamefont {Tamaki},\ and\ \citenamefont
  {Curty}}]{curras-lorenzo2024}%
  \BibitemOpen
  \bibfield  {author} {\bibinfo {author} {\bibfnamefont {Guillermo}\
  \bibnamefont {Curr{\'a}s-Lorenzo}}, \bibinfo {author} {\bibfnamefont {Shlok}\
  \bibnamefont {Nahar}}, \bibinfo {author} {\bibfnamefont {Norbert}\
  \bibnamefont {L{\" u}tkenhaus}}, \bibinfo {author} {\bibfnamefont {Kiyoshi}\
  \bibnamefont {Tamaki}}, \ and\ \bibinfo {author} {\bibfnamefont {Marcos}\
  \bibnamefont {Curty}},\ }\bibfield  {title} {\enquote {\bibinfo {title}
  {Security of quantum key distribution with imperfect phase randomisation},}\
  }\href {\doibase 10.1088/2058-9565/ad141c} {\bibfield  {journal} {\bibinfo
  {journal} {New J. Phys.}\ }\textbf {\bibinfo {volume} {9}},\ \bibinfo {pages}
  {015025} (\bibinfo {year} {2024})}\BibitemShut {NoStop}%
\bibitem [{\citenamefont {Sixto}\ \emph {et~al.}(2022)\citenamefont {Sixto},
  \citenamefont {Zapatero},\ and\ \citenamefont {Curty}}]{sixto2022}%
  \BibitemOpen
  \bibfield  {author} {\bibinfo {author} {\bibfnamefont {Xoel}\ \bibnamefont
  {Sixto}}, \bibinfo {author} {\bibfnamefont {V\'{\i}ctor}\ \bibnamefont
  {Zapatero}}, \ and\ \bibinfo {author} {\bibfnamefont {Marcos}\ \bibnamefont
  {Curty}},\ }\bibfield  {title} {\enquote {\bibinfo {title} {Security of
  decoy-state quantum key distribution with correlated intensity
  fluctuations},}\ }\href {\doibase 10.1103/PhysRevApplied.18.044069}
  {\bibfield  {journal} {\bibinfo  {journal} {Phys. Rev. Appl.}\ }\textbf
  {\bibinfo {volume} {18}},\ \bibinfo {pages} {044069} (\bibinfo {year}
  {2022})}\BibitemShut {NoStop}%
\bibitem [{\citenamefont {Xu}\ \emph {et~al.}(2015)\citenamefont {Xu},
  \citenamefont {Wei}, \citenamefont {Sajeed}, \citenamefont {Kaiser},
  \citenamefont {Sun}, \citenamefont {Tang}, \citenamefont {Qian},
  \citenamefont {Makarov},\ and\ \citenamefont {Lo}}]{xu2015a}%
  \BibitemOpen
  \bibfield  {author} {\bibinfo {author} {\bibfnamefont {Feihu}\ \bibnamefont
  {Xu}}, \bibinfo {author} {\bibfnamefont {Kejin}\ \bibnamefont {Wei}},
  \bibinfo {author} {\bibfnamefont {Shihan}\ \bibnamefont {Sajeed}}, \bibinfo
  {author} {\bibfnamefont {Sarah}\ \bibnamefont {Kaiser}}, \bibinfo {author}
  {\bibfnamefont {Shihai}\ \bibnamefont {Sun}}, \bibinfo {author}
  {\bibfnamefont {Zhiyuan}\ \bibnamefont {Tang}}, \bibinfo {author}
  {\bibfnamefont {Li}~\bibnamefont {Qian}}, \bibinfo {author} {\bibfnamefont
  {Vadim}\ \bibnamefont {Makarov}}, \ and\ \bibinfo {author} {\bibfnamefont
  {Hoi-Kwong}\ \bibnamefont {Lo}},\ }\bibfield  {title} {\enquote {\bibinfo
  {title} {Experimental quantum key distribution with source flaws},}\ }\href
  {\doibase 10.1103/PhysRevA.92.032305} {\bibfield  {journal} {\bibinfo
  {journal} {Phys. Rev. A}\ }\textbf {\bibinfo {volume} {92}},\ \bibinfo
  {pages} {032305} (\bibinfo {year} {2015})}\BibitemShut {NoStop}%
\bibitem [{\citenamefont {Curr\'as-Lorenzo}\ \emph {et~al.}(2021)\citenamefont
  {Curr\'as-Lorenzo}, \citenamefont {Navarrete}, \citenamefont {Pereira},\ and\
  \citenamefont {Tamaki}}]{curras-lorenzo2021}%
  \BibitemOpen
  \bibfield  {author} {\bibinfo {author} {\bibfnamefont {Guillermo}\
  \bibnamefont {Curr\'as-Lorenzo}}, \bibinfo {author} {\bibfnamefont
  {\'Alvaro}\ \bibnamefont {Navarrete}}, \bibinfo {author} {\bibfnamefont
  {Margarida}\ \bibnamefont {Pereira}}, \ and\ \bibinfo {author} {\bibfnamefont
  {Kiyoshi}\ \bibnamefont {Tamaki}},\ }\bibfield  {title} {\enquote {\bibinfo
  {title} {Finite-key analysis of loss-tolerant quantum key distribution based
  on random sampling theory},}\ }\href {\doibase 10.1103/PhysRevA.104.012406}
  {\bibfield  {journal} {\bibinfo  {journal} {Phys. Rev. A}\ }\textbf {\bibinfo
  {volume} {104}},\ \bibinfo {pages} {012406} (\bibinfo {year}
  {2021})}\BibitemShut {NoStop}%
\bibitem [{\citenamefont {Jain}\ \emph {et~al.}(2011)\citenamefont {Jain},
  \citenamefont {Wittmann}, \citenamefont {Lydersen}, \citenamefont {Wiechers},
  \citenamefont {Elser}, \citenamefont {Marquardt}, \citenamefont {Makarov},\
  and\ \citenamefont {Leuchs}}]{jain2011}%
  \BibitemOpen
  \bibfield  {author} {\bibinfo {author} {\bibfnamefont {Nitin}\ \bibnamefont
  {Jain}}, \bibinfo {author} {\bibfnamefont {Christoffer}\ \bibnamefont
  {Wittmann}}, \bibinfo {author} {\bibfnamefont {Lars}\ \bibnamefont
  {Lydersen}}, \bibinfo {author} {\bibfnamefont {Carlos}\ \bibnamefont
  {Wiechers}}, \bibinfo {author} {\bibfnamefont {Dominique}\ \bibnamefont
  {Elser}}, \bibinfo {author} {\bibfnamefont {Christoph}\ \bibnamefont
  {Marquardt}}, \bibinfo {author} {\bibfnamefont {Vadim}\ \bibnamefont
  {Makarov}}, \ and\ \bibinfo {author} {\bibfnamefont {Gerd}\ \bibnamefont
  {Leuchs}},\ }\bibfield  {title} {\enquote {\bibinfo {title} {Device
  calibration impacts security of quantum key distribution},}\ }\href {\doibase
  10.1103/PhysRevLett.107.110501} {\bibfield  {journal} {\bibinfo  {journal}
  {Phys. Rev. Lett.}\ }\textbf {\bibinfo {volume} {107}},\ \bibinfo {pages}
  {110501} (\bibinfo {year} {2011})}\BibitemShut {NoStop}%
\bibitem [{\citenamefont {Fei}\ \emph {et~al.}(2018)\citenamefont {Fei},
  \citenamefont {Meng}, \citenamefont {Gao}, \citenamefont {Wang},\ and\
  \citenamefont {Ma}}]{fei2018}%
  \BibitemOpen
  \bibfield  {author} {\bibinfo {author} {\bibfnamefont {Yang-Yang}\
  \bibnamefont {Fei}}, \bibinfo {author} {\bibfnamefont {Xiang-Dong}\
  \bibnamefont {Meng}}, \bibinfo {author} {\bibfnamefont {Ming}\ \bibnamefont
  {Gao}}, \bibinfo {author} {\bibfnamefont {Hong}\ \bibnamefont {Wang}}, \ and\
  \bibinfo {author} {\bibfnamefont {Zhi}\ \bibnamefont {Ma}},\ }\bibfield
  {title} {\enquote {\bibinfo {title} {Quantum man-in-the-middle attack on the
  calibration process of quantum key distribution},}\ }\href {\doibase
  10.1038/s41598-018-22700-3} {\bibfield  {journal} {\bibinfo  {journal} {Sci.
  Rep.}\ }\textbf {\bibinfo {volume} {8}},\ \bibinfo {pages} {4283} (\bibinfo
  {year} {2018})}\BibitemShut {NoStop}%
\bibitem [{\citenamefont {Xu}\ \emph {et~al.}(2010)\citenamefont {Xu},
  \citenamefont {Qi},\ and\ \citenamefont {Lo}}]{xu2010}%
  \BibitemOpen
  \bibfield  {author} {\bibinfo {author} {\bibfnamefont {F.}~\bibnamefont
  {Xu}}, \bibinfo {author} {\bibfnamefont {B.}~\bibnamefont {Qi}}, \ and\
  \bibinfo {author} {\bibfnamefont {H.-K.}\ \bibnamefont {Lo}},\ }\bibfield
  {title} {\enquote {\bibinfo {title} {Experimental demonstration of
  phase-remapping attack in a practical quantum key distribution system},}\
  }\href {\doibase 10.1088/1367-2630/12/11/113026} {\bibfield  {journal}
  {\bibinfo  {journal} {New J. Phys.}\ }\textbf {\bibinfo {volume} {12}},\
  \bibinfo {pages} {113026} (\bibinfo {year} {2010})}\BibitemShut {NoStop}%
\bibitem [{\citenamefont {Gottesman}\ \emph {et~al.}(2004)\citenamefont
  {Gottesman}, \citenamefont {Lo}, \citenamefont {L{\" u}tkenhaus},\ and\
  \citenamefont {Preskill}}]{gottesman2004}%
  \BibitemOpen
  \bibfield  {author} {\bibinfo {author} {\bibfnamefont {D.}~\bibnamefont
  {Gottesman}}, \bibinfo {author} {\bibfnamefont {H.-K.}\ \bibnamefont {Lo}},
  \bibinfo {author} {\bibfnamefont {N.}~\bibnamefont {L{\" u}tkenhaus}}, \ and\
  \bibinfo {author} {\bibfnamefont {J.}~\bibnamefont {Preskill}},\ }\bibfield
  {title} {\enquote {\bibinfo {title} {Security of quantum key distribution
  with imperfect devices},}\ }\href@noop {} {\bibfield  {journal} {\bibinfo
  {journal} {Quantum Inf. Comput.}\ }\textbf {\bibinfo {volume} {4}},\ \bibinfo
  {pages} {325--360} (\bibinfo {year} {2004})}\BibitemShut {NoStop}%
\bibitem [{\citenamefont {Kuzmin}(2023{\natexlab{b}})}]{kuzmin2023}%
  \BibitemOpen
  \bibfield  {author} {\bibinfo {author} {\bibfnamefont {Dmitriy}\ \bibnamefont
  {Kuzmin}},\ }\emph {\bibinfo {title} {Resistance of a single-photon detector
  to after-gate attack in quantum key distribution}},\ \href@noop {} {Master's
  thesis},\ \bibinfo  {school} {Moscow Technical University of Communication
  and Informatics} (\bibinfo {year} {2023}{\natexlab{b}})\BibitemShut {NoStop}%
\bibitem [{\citenamefont {Zaitsev}\ \emph {et~al.}(2024)\citenamefont
  {Zaitsev}, \citenamefont {Kuzmin}, \citenamefont {Bizin},\ and\ \citenamefont
  {Makarov}}]{zaitsev2024}%
  \BibitemOpen
  \bibfield  {author} {\bibinfo {author} {\bibfnamefont {Konstantin}\
  \bibnamefont {Zaitsev}}, \bibinfo {author} {\bibfnamefont {Dmitriy}\
  \bibnamefont {Kuzmin}}, \bibinfo {author} {\bibfnamefont {Vladimir}\
  \bibnamefont {Bizin}}, \ and\ \bibinfo {author} {\bibfnamefont {Vadim}\
  \bibnamefont {Makarov}},\ }\href@noop {} {\enquote {\bibinfo {title}
  {Energy-time attack on detectors in quantum key distribution},}\ } (\bibinfo
  {year} {2024}),\ \bibinfo {note} {{u}npublished}\BibitemShut {NoStop}%
\bibitem [{\citenamefont {Huang}\ \emph {et~al.}(2023)\citenamefont {Huang},
  \citenamefont {Mizutani}, \citenamefont {Lo}, \citenamefont {Makarov},\ and\
  \citenamefont {Tamaki}}]{huang2023}%
  \BibitemOpen
  \bibfield  {author} {\bibinfo {author} {\bibfnamefont {Anqi}\ \bibnamefont
  {Huang}}, \bibinfo {author} {\bibfnamefont {Akihiro}\ \bibnamefont
  {Mizutani}}, \bibinfo {author} {\bibfnamefont {Hoi-Kwong}\ \bibnamefont
  {Lo}}, \bibinfo {author} {\bibfnamefont {Vadim}\ \bibnamefont {Makarov}}, \
  and\ \bibinfo {author} {\bibfnamefont {Kiyoshi}\ \bibnamefont {Tamaki}},\
  }\bibfield  {title} {\enquote {\bibinfo {title} {Characterization of
  state-preparation uncertainty in quantum key distribution},}\ }\href
  {\doibase 10.1103/PhysRevApplied.19.014048} {\bibfield  {journal} {\bibinfo
  {journal} {Phys. Rev. Appl.}\ }\textbf {\bibinfo {volume} {19}},\ \bibinfo
  {pages} {014048} (\bibinfo {year} {2023})}\BibitemShut {NoStop}%
\bibitem [{\citenamefont {Bennett}\ and\ \citenamefont
  {Brassard}(1984)}]{bennett1984}%
  \BibitemOpen
  \bibfield  {author} {\bibinfo {author} {\bibfnamefont {Charles~H.}\
  \bibnamefont {Bennett}}\ and\ \bibinfo {author} {\bibfnamefont {Gilles}\
  \bibnamefont {Brassard}},\ }\bibfield  {title} {\enquote {\bibinfo {title}
  {Quantum cryptography: public key distribution and coin tossing},}\ }in\
  \href@noop {} {\emph {\bibinfo {booktitle} {Proc. International Conference on
  Computers, Systems, and Signal Processing}}}\ (\bibinfo  {publisher} {IEEE
  Press, New York},\ \bibinfo {address} {Bangalore, India},\ \bibinfo {year}
  {1984})\ pp.\ \bibinfo {pages} {175--179}\BibitemShut {NoStop}%
\bibitem [{\citenamefont {Trushechkin}\ \emph {et~al.}(2021)\citenamefont
  {Trushechkin}, \citenamefont {Kiktenko}, \citenamefont {Kronberg},\ and\
  \citenamefont {Fedorov}}]{trushechkin2021}%
  \BibitemOpen
  \bibfield  {author} {\bibinfo {author} {\bibfnamefont {A.S.}\ \bibnamefont
  {Trushechkin}}, \bibinfo {author} {\bibfnamefont {E.O.}\ \bibnamefont
  {Kiktenko}}, \bibinfo {author} {\bibfnamefont {D.A.}\ \bibnamefont
  {Kronberg}}, \ and\ \bibinfo {author} {\bibfnamefont {A.K.}\ \bibnamefont
  {Fedorov}},\ }\bibfield  {title} {\enquote {\bibinfo {title} {Security of the
  decoy state method for quantum key distribution},}\ }\href {\doibase
  10.3367/UFNe.2020.11.038882} {\bibfield  {journal} {\bibinfo  {journal}
  {Phys. Uspekhi}\ }\textbf {\bibinfo {volume} {64}},\ \bibinfo {pages} {88}
  (\bibinfo {year} {2021})}\BibitemShut {NoStop}%
\bibitem [{\citenamefont {Dieks}(1982)}]{dieks1982}%
  \BibitemOpen
  \bibfield  {author} {\bibinfo {author} {\bibfnamefont {D.}~\bibnamefont
  {Dieks}},\ }\bibfield  {title} {\enquote {\bibinfo {title} {Communication by
  {EPR} devices},}\ }\href {\doibase 10.1016/0375-9601(82)90084-6} {\bibfield
  {journal} {\bibinfo  {journal} {Phys. Lett. A}\ }\textbf {\bibinfo {volume}
  {92}},\ \bibinfo {pages} {271--272} (\bibinfo {year} {1982})}\BibitemShut
  {NoStop}%
\bibitem [{\citenamefont {Wootters}\ and\ \citenamefont
  {Zurek}(1982)}]{wootters1982}%
  \BibitemOpen
  \bibfield  {author} {\bibinfo {author} {\bibfnamefont {W.~K.}\ \bibnamefont
  {Wootters}}\ and\ \bibinfo {author} {\bibfnamefont {W.~H.}\ \bibnamefont
  {Zurek}},\ }\bibfield  {title} {\enquote {\bibinfo {title} {A single quantum
  cannot be cloned},}\ }\href {\doibase 10.1038/299802a0} {\bibfield  {journal}
  {\bibinfo  {journal} {Nature}\ }\textbf {\bibinfo {volume} {299}},\ \bibinfo
  {pages} {802--803} (\bibinfo {year} {1982})}\BibitemShut {NoStop}%
\bibitem [{\citenamefont {Mayers}(1996)}]{mayers1996}%
  \BibitemOpen
  \bibfield  {author} {\bibinfo {author} {\bibfnamefont {Dominic}\ \bibnamefont
  {Mayers}},\ }\bibfield  {title} {\enquote {\bibinfo {title} {Quantum key
  distribution and string oblivious transfer in noisy channels},}\ }\href
  {\doibase 10.1007/3-540-68697-5_26} {\bibfield  {journal} {\bibinfo
  {journal} {Lect. Notes Comp. Sci.}\ }\textbf {\bibinfo {volume} {1109}},\
  \bibinfo {pages} {343--357} (\bibinfo {year} {1996})}\BibitemShut {NoStop}%
\bibitem [{\citenamefont {Lo}\ and\ \citenamefont {Chau}(1999)}]{lo1999}%
  \BibitemOpen
  \bibfield  {author} {\bibinfo {author} {\bibfnamefont {Hoi-Kwong}\
  \bibnamefont {Lo}}\ and\ \bibinfo {author} {\bibfnamefont {H.~F.}\
  \bibnamefont {Chau}},\ }\bibfield  {title} {\enquote {\bibinfo {title}
  {Unconditional security of quantum key distribution over arbitrarily long
  distances},}\ }\href {\doibase 10.1126/science.283.5410.2050} {\bibfield
  {journal} {\bibinfo  {journal} {Science}\ }\textbf {\bibinfo {volume}
  {283}},\ \bibinfo {pages} {2050--2056} (\bibinfo {year} {1999})}\BibitemShut
  {NoStop}%
\bibitem [{\citenamefont {Shor}\ and\ \citenamefont
  {Preskill}(2000)}]{shor2000}%
  \BibitemOpen
  \bibfield  {author} {\bibinfo {author} {\bibfnamefont {Peter~W.}\
  \bibnamefont {Shor}}\ and\ \bibinfo {author} {\bibfnamefont {John}\
  \bibnamefont {Preskill}},\ }\bibfield  {title} {\enquote {\bibinfo {title}
  {Simple proof of security of the {BB84} quantum key distribution protocol},}\
  }\href {\doibase 10.1103/PhysRevLett.85.441} {\bibfield  {journal} {\bibinfo
  {journal} {Phys. Rev. Lett.}\ }\textbf {\bibinfo {volume} {85}},\ \bibinfo
  {pages} {441--444} (\bibinfo {year} {2000})}\BibitemShut {NoStop}%
\bibitem [{\citenamefont {Hwang}(2003)}]{hwang2003}%
  \BibitemOpen
  \bibfield  {author} {\bibinfo {author} {\bibfnamefont {Won-Young}\
  \bibnamefont {Hwang}},\ }\bibfield  {title} {\enquote {\bibinfo {title}
  {Quantum key distribution with high loss: Toward global secure
  communication},}\ }\href {\doibase 10.1103/PhysRevLett.91.057901} {\bibfield
  {journal} {\bibinfo  {journal} {Phys. Rev. Lett.}\ }\textbf {\bibinfo
  {volume} {91}},\ \bibinfo {pages} {057901} (\bibinfo {year}
  {2003})}\BibitemShut {NoStop}%
\bibitem [{\citenamefont {Lo}\ \emph {et~al.}(2005)\citenamefont {Lo},
  \citenamefont {Ma},\ and\ \citenamefont {Chen}}]{lo2005}%
  \BibitemOpen
  \bibfield  {author} {\bibinfo {author} {\bibfnamefont {Hoi-Kwong}\
  \bibnamefont {Lo}}, \bibinfo {author} {\bibfnamefont {Xiongfeng}\
  \bibnamefont {Ma}}, \ and\ \bibinfo {author} {\bibfnamefont {Kai}\
  \bibnamefont {Chen}},\ }\bibfield  {title} {\enquote {\bibinfo {title} {Decoy
  state quantum key distribution},}\ }\href {\doibase
  10.1103/PhysRevLett.94.230504} {\bibfield  {journal} {\bibinfo  {journal}
  {Phys. Rev. Lett.}\ }\textbf {\bibinfo {volume} {94}},\ \bibinfo {pages}
  {230504} (\bibinfo {year} {2005})}\BibitemShut {NoStop}%
\bibitem [{\citenamefont {Ma}\ \emph {et~al.}(2005)\citenamefont {Ma},
  \citenamefont {Qi}, \citenamefont {Zhao},\ and\ \citenamefont {Lo}}]{ma2005}%
  \BibitemOpen
  \bibfield  {author} {\bibinfo {author} {\bibfnamefont {Xiongfeng}\
  \bibnamefont {Ma}}, \bibinfo {author} {\bibfnamefont {Bing}\ \bibnamefont
  {Qi}}, \bibinfo {author} {\bibfnamefont {Yi}~\bibnamefont {Zhao}}, \ and\
  \bibinfo {author} {\bibfnamefont {Hoi-Kwong}\ \bibnamefont {Lo}},\ }\bibfield
   {title} {\enquote {\bibinfo {title} {Practical decoy state for quantum key
  distribution},}\ }\href {\doibase 10.1103/PhysRevA.72.012326} {\bibfield
  {journal} {\bibinfo  {journal} {Phys. Rev. A}\ }\textbf {\bibinfo {volume}
  {72}},\ \bibinfo {pages} {012326} (\bibinfo {year} {2005})}\BibitemShut
  {NoStop}%
\bibitem [{\citenamefont {Zhang}\ \emph {et~al.}(2017)\citenamefont {Zhang},
  \citenamefont {Zhao}, \citenamefont {Razavi},\ and\ \citenamefont
  {Ma}}]{zhang2017}%
  \BibitemOpen
  \bibfield  {author} {\bibinfo {author} {\bibfnamefont {Zhen}\ \bibnamefont
  {Zhang}}, \bibinfo {author} {\bibfnamefont {Qi}~\bibnamefont {Zhao}},
  \bibinfo {author} {\bibfnamefont {Mohsen}\ \bibnamefont {Razavi}}, \ and\
  \bibinfo {author} {\bibfnamefont {Xiongfeng}\ \bibnamefont {Ma}},\ }\bibfield
   {title} {\enquote {\bibinfo {title} {Improved key-rate bounds for practical
  decoy-state quantum-key-distribution systems},}\ }\href {\doibase
  10.1103/PhysRevA.95.012333} {\bibfield  {journal} {\bibinfo  {journal} {Phys.
  Rev. A}\ }\textbf {\bibinfo {volume} {95}},\ \bibinfo {pages} {012333}
  (\bibinfo {year} {2017})}\BibitemShut {NoStop}%
\bibitem [{\citenamefont {Kiktenko}\ \emph {et~al.}(2016)\citenamefont
  {Kiktenko}, \citenamefont {Trushechkin}, \citenamefont {Kurochkin},\ and\
  \citenamefont {Fedorov}}]{kiktenko2016}%
  \BibitemOpen
  \bibfield  {author} {\bibinfo {author} {\bibfnamefont {Evgeny}\ \bibnamefont
  {Kiktenko}}, \bibinfo {author} {\bibfnamefont {Anton}\ \bibnamefont
  {Trushechkin}}, \bibinfo {author} {\bibfnamefont {Yury}\ \bibnamefont
  {Kurochkin}}, \ and\ \bibinfo {author} {\bibfnamefont {Aleksey}\ \bibnamefont
  {Fedorov}},\ }\bibfield  {title} {\enquote {\bibinfo {title} {Post-processing
  procedure for industrial quantum key distribution systems},}\ }\href
  {\doibase 10.1088/1742-6596/741/1/012081} {\bibfield  {journal} {\bibinfo
  {journal} {J. Phys. Conf. Ser.}\ }\textbf {\bibinfo {volume} {741}},\
  \bibinfo {pages} {012081} (\bibinfo {year} {2016})}\BibitemShut {NoStop}%
\bibitem [{\citenamefont {Tomamichel}\ \emph {et~al.}(2011)\citenamefont
  {Tomamichel}, \citenamefont {Schaffner}, \citenamefont {Smith},\ and\
  \citenamefont {Renner}}]{tomamichel2011a}%
  \BibitemOpen
  \bibfield  {author} {\bibinfo {author} {\bibfnamefont {Marco}\ \bibnamefont
  {Tomamichel}}, \bibinfo {author} {\bibfnamefont {Christian}\ \bibnamefont
  {Schaffner}}, \bibinfo {author} {\bibfnamefont {Adam}\ \bibnamefont {Smith}},
  \ and\ \bibinfo {author} {\bibfnamefont {Renato}\ \bibnamefont {Renner}},\
  }\bibfield  {title} {\enquote {\bibinfo {title} {Leftover hashing against
  quantum side information},}\ }\href {\doibase 10.1109/TIT.2011.2158473}
  {\bibfield  {journal} {\bibinfo  {journal} {IEEE Trans. Inf. Theory}\
  }\textbf {\bibinfo {volume} {57}},\ \bibinfo {pages} {5524--5535} (\bibinfo
  {year} {2011})}\BibitemShut {NoStop}%
\bibitem [{\citenamefont {Dolmatov{, Ed.}}\ and\ \citenamefont
  {Degtyarev}(2013)}]{dolmatov2013}%
  \BibitemOpen
  \bibfield  {author} {\bibinfo {author} {\bibfnamefont {V.}~\bibnamefont
  {Dolmatov{, Ed.}}}\ and\ \bibinfo {author} {\bibfnamefont {A.}~\bibnamefont
  {Degtyarev}},\ }\bibfield  {title} {\enquote {\bibinfo {title} {{GOST~R
  34.11-2012:} {H}ash function},}\ }\href
  {https://www.rfc-editor.org/rfc/rfc6986} {\bibfield  {journal} {\bibinfo
  {journal} {RFC}\ }\textbf {\bibinfo {volume} {6986}} (\bibinfo {year}
  {2013})}\BibitemShut {NoStop}%
\bibitem [{\citenamefont {Trushechkin}(2020)}]{trushechkin2020}%
  \BibitemOpen
  \bibfield  {author} {\bibinfo {author} {\bibfnamefont {Anton}\ \bibnamefont
  {Trushechkin}},\ }\bibfield  {title} {\enquote {\bibinfo {title} {On the
  operational meaning and practical aspects of using the security parameter in
  quantum key distribution},}\ }\href {\doibase 10.1070/qel17283} {\bibfield
  {journal} {\bibinfo  {journal} {Quantum Electron.}\ }\textbf {\bibinfo
  {volume} {50}},\ \bibinfo {pages} {426--439} (\bibinfo {year}
  {2020})}\BibitemShut {NoStop}%
\bibitem [{\citenamefont {Yuan}\ \emph
  {et~al.}(2011{\natexlab{a}})\citenamefont {Yuan}, \citenamefont {Dynes},\
  and\ \citenamefont {Shields}}]{yuan2011a}%
  \BibitemOpen
  \bibfield  {author} {\bibinfo {author} {\bibfnamefont {Z.~L.}\ \bibnamefont
  {Yuan}}, \bibinfo {author} {\bibfnamefont {J.~F.}\ \bibnamefont {Dynes}}, \
  and\ \bibinfo {author} {\bibfnamefont {A.~J.}\ \bibnamefont {Shields}},\
  }\bibfield  {title} {\enquote {\bibinfo {title} {Response to ``{C}omment on
  `{R}esilience of gated avalanche photodiodes against bright illumination
  attacks in quantum cryptography'{''} [{A}ppl.\ {P}hys.\ {L}ett. 99, 196101
  (2011)]},}\ }\href {\doibase 10.1063/1.3658807} {\bibfield  {journal}
  {\bibinfo  {journal} {Appl. Phys. Lett.}\ }\textbf {\bibinfo {volume} {99}},\
  \bibinfo {pages} {196102} (\bibinfo {year} {2011}{\natexlab{a}})}\BibitemShut
  {NoStop}%
\bibitem [{\citenamefont {Qian}\ \emph {et~al.}(2018)\citenamefont {Qian},
  \citenamefont {He}, \citenamefont {Wang}, \citenamefont {Chen}, \citenamefont
  {Yin}, \citenamefont {Guo},\ and\ \citenamefont {Han}}]{qian2018}%
  \BibitemOpen
  \bibfield  {author} {\bibinfo {author} {\bibfnamefont {Yong-Jun}\
  \bibnamefont {Qian}}, \bibinfo {author} {\bibfnamefont {De-Yong}\
  \bibnamefont {He}}, \bibinfo {author} {\bibfnamefont {Shuang}\ \bibnamefont
  {Wang}}, \bibinfo {author} {\bibfnamefont {Wei}\ \bibnamefont {Chen}},
  \bibinfo {author} {\bibfnamefont {Zhen-Qiang}\ \bibnamefont {Yin}}, \bibinfo
  {author} {\bibfnamefont {Guang-Can}\ \bibnamefont {Guo}}, \ and\ \bibinfo
  {author} {\bibfnamefont {Zheng-Fu}\ \bibnamefont {Han}},\ }\bibfield  {title}
  {\enquote {\bibinfo {title} {Hacking the quantum key distribution system by
  exploiting the avalanche-transition region of single-photon detectors},}\
  }\href {\doibase 10.1103/PhysRevApplied.10.064062} {\bibfield  {journal}
  {\bibinfo  {journal} {Phys. Rev. Appl.}\ }\textbf {\bibinfo {volume} {10}},\
  \bibinfo {pages} {064062} (\bibinfo {year} {2018})}\BibitemShut {NoStop}%
\bibitem [{\citenamefont {Lydersen}\ \emph
  {et~al.}(2010{\natexlab{a}})\citenamefont {Lydersen}, \citenamefont
  {Wiechers}, \citenamefont {Wittmann}, \citenamefont {Elser}, \citenamefont
  {Skaar},\ and\ \citenamefont {Makarov}}]{lydersen2010a}%
  \BibitemOpen
  \bibfield  {author} {\bibinfo {author} {\bibfnamefont {L.}~\bibnamefont
  {Lydersen}}, \bibinfo {author} {\bibfnamefont {C.}~\bibnamefont {Wiechers}},
  \bibinfo {author} {\bibfnamefont {C.}~\bibnamefont {Wittmann}}, \bibinfo
  {author} {\bibfnamefont {D.}~\bibnamefont {Elser}}, \bibinfo {author}
  {\bibfnamefont {J.}~\bibnamefont {Skaar}}, \ and\ \bibinfo {author}
  {\bibfnamefont {V.}~\bibnamefont {Makarov}},\ }\bibfield  {title} {\enquote
  {\bibinfo {title} {Hacking commercial quantum cryptography systems by
  tailored bright illumination},}\ }\href {\doibase 10.1038/nphoton.2010.214}
  {\bibfield  {journal} {\bibinfo  {journal} {Nat. Photonics}\ }\textbf
  {\bibinfo {volume} {4}},\ \bibinfo {pages} {686--689} (\bibinfo {year}
  {2010}{\natexlab{a}})}\BibitemShut {NoStop}%
\bibitem [{\citenamefont {Lydersen}\ \emph
  {et~al.}(2010{\natexlab{b}})\citenamefont {Lydersen}, \citenamefont
  {Wiechers}, \citenamefont {Wittmann}, \citenamefont {Elser}, \citenamefont
  {Skaar},\ and\ \citenamefont {Makarov}}]{lydersen2010b}%
  \BibitemOpen
  \bibfield  {author} {\bibinfo {author} {\bibfnamefont {L.}~\bibnamefont
  {Lydersen}}, \bibinfo {author} {\bibfnamefont {C.}~\bibnamefont {Wiechers}},
  \bibinfo {author} {\bibfnamefont {C.}~\bibnamefont {Wittmann}}, \bibinfo
  {author} {\bibfnamefont {D.}~\bibnamefont {Elser}}, \bibinfo {author}
  {\bibfnamefont {J.}~\bibnamefont {Skaar}}, \ and\ \bibinfo {author}
  {\bibfnamefont {V.}~\bibnamefont {Makarov}},\ }\bibfield  {title} {\enquote
  {\bibinfo {title} {Thermal blinding of gated detectors in quantum
  cryptography},}\ }\href {\doibase 10.1364/oe.18.027938} {\bibfield  {journal}
  {\bibinfo  {journal} {Opt. Express}\ }\textbf {\bibinfo {volume} {18}},\
  \bibinfo {pages} {27938--27954} (\bibinfo {year}
  {2010}{\natexlab{b}})}\BibitemShut {NoStop}%
\bibitem [{\citenamefont {Lydersen}\ \emph
  {et~al.}(2011{\natexlab{b}})\citenamefont {Lydersen}, \citenamefont
  {Akhlaghi}, \citenamefont {Majedi}, \citenamefont {Skaar},\ and\
  \citenamefont {Makarov}}]{lydersen2011c}%
  \BibitemOpen
  \bibfield  {author} {\bibinfo {author} {\bibfnamefont {L.}~\bibnamefont
  {Lydersen}}, \bibinfo {author} {\bibfnamefont {M.~K.}\ \bibnamefont
  {Akhlaghi}}, \bibinfo {author} {\bibfnamefont {A.~H.}\ \bibnamefont
  {Majedi}}, \bibinfo {author} {\bibfnamefont {J.}~\bibnamefont {Skaar}}, \
  and\ \bibinfo {author} {\bibfnamefont {V.}~\bibnamefont {Makarov}},\
  }\bibfield  {title} {\enquote {\bibinfo {title} {Controlling a
  superconducting nanowire single-photon detector using tailored bright
  illumination},}\ }\href {\doibase 10.1088/1367-2630/13/11/113042} {\bibfield
  {journal} {\bibinfo  {journal} {New J. Phys.}\ }\textbf {\bibinfo {volume}
  {13}},\ \bibinfo {pages} {113042} (\bibinfo {year}
  {2011}{\natexlab{b}})}\BibitemShut {NoStop}%
\bibitem [{\citenamefont {Makarov}(2009)}]{makarov2009}%
  \BibitemOpen
  \bibfield  {author} {\bibinfo {author} {\bibfnamefont {V.}~\bibnamefont
  {Makarov}},\ }\bibfield  {title} {\enquote {\bibinfo {title} {Controlling
  passively quenched single photon detectors by bright light},}\ }\href
  {\doibase 10.1088/1367-2630/11/6/065003} {\bibfield  {journal} {\bibinfo
  {journal} {New J. Phys.}\ }\textbf {\bibinfo {volume} {11}},\ \bibinfo
  {pages} {065003} (\bibinfo {year} {2009})}\BibitemShut {NoStop}%
\bibitem [{\citenamefont {Sauge}\ \emph {et~al.}(2011)\citenamefont {Sauge},
  \citenamefont {Lydersen}, \citenamefont {Anisimov}, \citenamefont {Skaar},\
  and\ \citenamefont {Makarov}}]{sauge2011}%
  \BibitemOpen
  \bibfield  {author} {\bibinfo {author} {\bibfnamefont {S.}~\bibnamefont
  {Sauge}}, \bibinfo {author} {\bibfnamefont {L.}~\bibnamefont {Lydersen}},
  \bibinfo {author} {\bibfnamefont {A.}~\bibnamefont {Anisimov}}, \bibinfo
  {author} {\bibfnamefont {J.}~\bibnamefont {Skaar}}, \ and\ \bibinfo {author}
  {\bibfnamefont {V.}~\bibnamefont {Makarov}},\ }\bibfield  {title} {\enquote
  {\bibinfo {title} {Controlling an actively-quenched single photon detector
  with bright light},}\ }\href {\doibase 10.1364/OE.19.023590} {\bibfield
  {journal} {\bibinfo  {journal} {Opt. Express}\ }\textbf {\bibinfo {volume}
  {19}},\ \bibinfo {pages} {23590--23600} (\bibinfo {year} {2011})}\BibitemShut
  {NoStop}%
\bibitem [{\citenamefont {Tanner}\ \emph {et~al.}(2014)\citenamefont {Tanner},
  \citenamefont {Makarov},\ and\ \citenamefont {Hadfield}}]{tanner2014}%
  \BibitemOpen
  \bibfield  {author} {\bibinfo {author} {\bibfnamefont {Michael~G.}\
  \bibnamefont {Tanner}}, \bibinfo {author} {\bibfnamefont {Vadim}\
  \bibnamefont {Makarov}}, \ and\ \bibinfo {author} {\bibfnamefont {Robert~H.}\
  \bibnamefont {Hadfield}},\ }\bibfield  {title} {\enquote {\bibinfo {title}
  {Optimised quantum hacking of superconducting nanowire single-photon
  detectors},}\ }\href {\doibase 10.1364/OE.22.006734} {\bibfield  {journal}
  {\bibinfo  {journal} {Opt. Express}\ }\textbf {\bibinfo {volume} {22}},\
  \bibinfo {pages} {6734--6748} (\bibinfo {year} {2014})}\BibitemShut {NoStop}%
\bibitem [{\citenamefont {Fedorov}\ \emph {et~al.}(2019)\citenamefont
  {Fedorov}, \citenamefont {Gerhardt}, \citenamefont {Huang}, \citenamefont
  {Jogenfors}, \citenamefont {Kurochkin}, \citenamefont {Lamas-Linares},
  \citenamefont {Larsson}, \citenamefont {Leuchs}, \citenamefont {Lydersen},
  \citenamefont {Makarov},\ and\ \citenamefont {Skaar}}]{fedorov2019}%
  \BibitemOpen
  \bibfield  {author} {\bibinfo {author} {\bibfnamefont {Aleksey}\ \bibnamefont
  {Fedorov}}, \bibinfo {author} {\bibfnamefont {Ilja}\ \bibnamefont
  {Gerhardt}}, \bibinfo {author} {\bibfnamefont {Anqi}\ \bibnamefont {Huang}},
  \bibinfo {author} {\bibfnamefont {Jonathan}\ \bibnamefont {Jogenfors}},
  \bibinfo {author} {\bibfnamefont {Yury}\ \bibnamefont {Kurochkin}}, \bibinfo
  {author} {\bibfnamefont {Ant{\' i}a}\ \bibnamefont {Lamas-Linares}}, \bibinfo
  {author} {\bibfnamefont {Jan-{\AA}ke}\ \bibnamefont {Larsson}}, \bibinfo
  {author} {\bibfnamefont {Gerd}\ \bibnamefont {Leuchs}}, \bibinfo {author}
  {\bibfnamefont {Lars}\ \bibnamefont {Lydersen}}, \bibinfo {author}
  {\bibfnamefont {Vadim}\ \bibnamefont {Makarov}}, \ and\ \bibinfo {author}
  {\bibfnamefont {Johannes}\ \bibnamefont {Skaar}},\ }\bibfield  {title}
  {\enquote {\bibinfo {title} {Comment on `{I}nherent security of phase coding
  quantum key distribution systems against detector blinding attacks' (2018
  {L}aser {P}hys.\ {L}ett.\ 15 095203)},}\ }\href {\doibase
  10.1088/1612-202X/aaf22d} {\bibfield  {journal} {\bibinfo  {journal} {Laser
  Phys. Lett.}\ }\textbf {\bibinfo {volume} {16}},\ \bibinfo {pages} {019401}
  (\bibinfo {year} {2019})}\BibitemShut {NoStop}%
\bibitem [{\citenamefont {Woodward}\ \emph {et~al.}(2021)\citenamefont
  {Woodward}, \citenamefont {Lo}, \citenamefont {Pittaluga}, \citenamefont
  {Minder}, \citenamefont {Para{\" i}so}, \citenamefont {Lucamarini},
  \citenamefont {Yuan},\ and\ \citenamefont {Shields}}]{woodward2021}%
  \BibitemOpen
  \bibfield  {author} {\bibinfo {author} {\bibfnamefont {R.~I.}\ \bibnamefont
  {Woodward}}, \bibinfo {author} {\bibfnamefont {Y.~S.}\ \bibnamefont {Lo}},
  \bibinfo {author} {\bibfnamefont {M.}~\bibnamefont {Pittaluga}}, \bibinfo
  {author} {\bibfnamefont {M.}~\bibnamefont {Minder}}, \bibinfo {author}
  {\bibfnamefont {T.~K.}\ \bibnamefont {Para{\" i}so}}, \bibinfo {author}
  {\bibfnamefont {M.}~\bibnamefont {Lucamarini}}, \bibinfo {author}
  {\bibfnamefont {Z.~L.}\ \bibnamefont {Yuan}}, \ and\ \bibinfo {author}
  {\bibfnamefont {A.~J.}\ \bibnamefont {Shields}},\ }\bibfield  {title}
  {\enquote {\bibinfo {title} {Gigahertz measurement-device-independent quantum
  key distribution using directly modulated lasers},}\ }\href {\doibase
  10.1038/s41534-021-00394-2} {\bibfield  {journal} {\bibinfo  {journal} {npj
  Quantum Inf.}\ }\textbf {\bibinfo {volume} {7}},\ \bibinfo {pages} {58}
  (\bibinfo {year} {2021})}\BibitemShut {NoStop}%
\bibitem [{\citenamefont {Yuan}\ \emph
  {et~al.}(2011{\natexlab{b}})\citenamefont {Yuan}, \citenamefont {Dynes},\
  and\ \citenamefont {Shields}}]{yuan2011}%
  \BibitemOpen
  \bibfield  {author} {\bibinfo {author} {\bibfnamefont {Z.~L.}\ \bibnamefont
  {Yuan}}, \bibinfo {author} {\bibfnamefont {J.~F.}\ \bibnamefont {Dynes}}, \
  and\ \bibinfo {author} {\bibfnamefont {A.~J.}\ \bibnamefont {Shields}},\
  }\bibfield  {title} {\enquote {\bibinfo {title} {Resilience of gated
  avalanche photodiodes against bright illumination attacks in quantum
  cryptography},}\ }\href {\doibase 10.1063/1.3597221} {\bibfield  {journal}
  {\bibinfo  {journal} {Appl. Phys. Lett.}\ }\textbf {\bibinfo {volume} {98}},\
  \bibinfo {eid} {231104} (\bibinfo {year} {2011}{\natexlab{b}})}\BibitemShut
  {NoStop}%
\bibitem [{\citenamefont {Koehler-Sidki}\ \emph {et~al.}(2018)\citenamefont
  {Koehler-Sidki}, \citenamefont {Dynes}, \citenamefont {Lucamarini},
  \citenamefont {Roberts}, \citenamefont {Sharpe}, \citenamefont {Yuan},\ and\
  \citenamefont {Shields}}]{sidki2018}%
  \BibitemOpen
  \bibfield  {author} {\bibinfo {author} {\bibfnamefont {A.}~\bibnamefont
  {Koehler-Sidki}}, \bibinfo {author} {\bibfnamefont {J.~F.}\ \bibnamefont
  {Dynes}}, \bibinfo {author} {\bibfnamefont {M.}~\bibnamefont {Lucamarini}},
  \bibinfo {author} {\bibfnamefont {G.~L.}\ \bibnamefont {Roberts}}, \bibinfo
  {author} {\bibfnamefont {A.~W.}\ \bibnamefont {Sharpe}}, \bibinfo {author}
  {\bibfnamefont {Z.~L.}\ \bibnamefont {Yuan}}, \ and\ \bibinfo {author}
  {\bibfnamefont {A.~J.}\ \bibnamefont {Shields}},\ }\bibfield  {title}
  {\enquote {\bibinfo {title} {Best-practice criteria for practical security of
  self-differencing avalanche photodiode detectors in quantum key
  distribution},}\ }\href {\doibase 10.1103/PhysRevApplied.9.044027} {\bibfield
   {journal} {\bibinfo  {journal} {Phys. Rev. Appl.}\ }\textbf {\bibinfo
  {volume} {9}},\ \bibinfo {pages} {044027} (\bibinfo {year}
  {2018})}\BibitemShut {NoStop}%
\bibitem [{\citenamefont {Qian}\ \emph {et~al.}(2015)\citenamefont {Qian},
  \citenamefont {Li}, \citenamefont {He}, \citenamefont {Yin}, \citenamefont
  {Zhang}, \citenamefont {Chen}, \citenamefont {Wang},\ and\ \citenamefont
  {Han}}]{jun2015}%
  \BibitemOpen
  \bibfield  {author} {\bibinfo {author} {\bibfnamefont {Yong-Jun}\
  \bibnamefont {Qian}}, \bibinfo {author} {\bibfnamefont {Hong-Wei}\
  \bibnamefont {Li}}, \bibinfo {author} {\bibfnamefont {De-Yong}\ \bibnamefont
  {He}}, \bibinfo {author} {\bibfnamefont {Zhen-Qiang}\ \bibnamefont {Yin}},
  \bibinfo {author} {\bibfnamefont {Chun-Mei}\ \bibnamefont {Zhang}}, \bibinfo
  {author} {\bibfnamefont {Wei}\ \bibnamefont {Chen}}, \bibinfo {author}
  {\bibfnamefont {Shuang}\ \bibnamefont {Wang}}, \ and\ \bibinfo {author}
  {\bibfnamefont {Zheng-Fu}\ \bibnamefont {Han}},\ }\bibfield  {title}
  {\enquote {\bibinfo {title} {Countermeasure against probabilistic blinding
  attack in practical quantum key distribution systems},}\ }\href {\doibase
  10.1088/1674-1056/24/9/090305} {\bibfield  {journal} {\bibinfo  {journal}
  {Chin. Phys. B}\ }\textbf {\bibinfo {volume} {24}},\ \bibinfo {pages}
  {090305} (\bibinfo {year} {2015})}\BibitemShut {NoStop}%
\bibitem [{\citenamefont {da~Silva}\ \emph {et~al.}(2012)\citenamefont
  {da~Silva}, \citenamefont {Xavier}, \citenamefont {Tempor\^{a}o},\ and\
  \citenamefont {von~der Weid}}]{silva2012}%
  \BibitemOpen
  \bibfield  {author} {\bibinfo {author} {\bibfnamefont {T.~F.}\ \bibnamefont
  {da~Silva}}, \bibinfo {author} {\bibfnamefont {G.~B.}\ \bibnamefont
  {Xavier}}, \bibinfo {author} {\bibfnamefont {G.~P.}\ \bibnamefont
  {Tempor\^{a}o}}, \ and\ \bibinfo {author} {\bibfnamefont {J.~P.}\
  \bibnamefont {von~der Weid}},\ }\bibfield  {title} {\enquote {\bibinfo
  {title} {Real-time monitoring of single-photon detectors against
  eavesdropping in quantum key distribution systems},}\ }\href {\doibase
  10.1364/OE.20.018911} {\bibfield  {journal} {\bibinfo  {journal} {Opt.
  Express}\ }\textbf {\bibinfo {volume} {20}},\ \bibinfo {pages} {18911--18924}
  (\bibinfo {year} {2012})}\BibitemShut {NoStop}%
\bibitem [{\citenamefont {Koehler-Sidki}\ \emph {et~al.}(2019)\citenamefont
  {Koehler-Sidki}, \citenamefont {Dynes}, \citenamefont {Martinez},
  \citenamefont {Lucamarini}, \citenamefont {Roberts}, \citenamefont {Sharpe},
  \citenamefont {Yuan},\ and\ \citenamefont {Shields}}]{koehler-sidki2019}%
  \BibitemOpen
  \bibfield  {author} {\bibinfo {author} {\bibfnamefont {A.}~\bibnamefont
  {Koehler-Sidki}}, \bibinfo {author} {\bibfnamefont {J.~F.}\ \bibnamefont
  {Dynes}}, \bibinfo {author} {\bibfnamefont {A.}~\bibnamefont {Martinez}},
  \bibinfo {author} {\bibfnamefont {M.}~\bibnamefont {Lucamarini}}, \bibinfo
  {author} {\bibfnamefont {G.~L.}\ \bibnamefont {Roberts}}, \bibinfo {author}
  {\bibfnamefont {A.~W.}\ \bibnamefont {Sharpe}}, \bibinfo {author}
  {\bibfnamefont {Z.~L.}\ \bibnamefont {Yuan}}, \ and\ \bibinfo {author}
  {\bibfnamefont {A.J.}\ \bibnamefont {Shields}},\ }\bibfield  {title}
  {\enquote {\bibinfo {title} {Intrinsic mitigation of the after-gate attack in
  quantum key distribution through fast-gated delayed detection},}\ }\href
  {\doibase 10.1103/PhysRevApplied.12.024050} {\bibfield  {journal} {\bibinfo
  {journal} {Phys. Rev. Appl.}\ }\textbf {\bibinfo {volume} {12}},\ \bibinfo
  {pages} {024050} (\bibinfo {year} {2019})}\BibitemShut {NoStop}%
\bibitem [{\citenamefont {Chistiakov}\ \emph {et~al.}(2019)\citenamefont
  {Chistiakov}, \citenamefont {Huang}, \citenamefont {Egorov},\ and\
  \citenamefont {Makarov}}]{chistiakov2019}%
  \BibitemOpen
  \bibfield  {author} {\bibinfo {author} {\bibfnamefont {Vladimir}\
  \bibnamefont {Chistiakov}}, \bibinfo {author} {\bibfnamefont {Anqi}\
  \bibnamefont {Huang}}, \bibinfo {author} {\bibfnamefont {Vladimir}\
  \bibnamefont {Egorov}}, \ and\ \bibinfo {author} {\bibfnamefont {Vadim}\
  \bibnamefont {Makarov}},\ }\bibfield  {title} {\enquote {\bibinfo {title}
  {Controlling single-photon detector {ID210} with bright light},}\ }\href
  {\doibase 10.1364/OE.27.032253} {\bibfield  {journal} {\bibinfo  {journal}
  {Opt. Express}\ }\textbf {\bibinfo {volume} {27}},\ \bibinfo {pages} {32253}
  (\bibinfo {year} {2019})}\BibitemShut {NoStop}%
\bibitem [{\citenamefont {Alhussein}\ \emph {et~al.}(2019)\citenamefont
  {Alhussein}, \citenamefont {Inoue},\ and\ \citenamefont
  {Honjo}}]{alhussein2019}%
  \BibitemOpen
  \bibfield  {author} {\bibinfo {author} {\bibfnamefont {Muataz}\ \bibnamefont
  {Alhussein}}, \bibinfo {author} {\bibfnamefont {Kyo}\ \bibnamefont {Inoue}},
  \ and\ \bibinfo {author} {\bibfnamefont {Toshimori}\ \bibnamefont {Honjo}},\
  }\bibfield  {title} {\enquote {\bibinfo {title} {Monitoring coincident clicks
  in differential-quadrature-phase shift {QKD} to reveal detector blinding and
  control attacks},}\ }\href {\doibase 10.7567/1347-4065/aaec1c} {\bibfield
  {journal} {\bibinfo  {journal} {Jpn. J. Appl. Phys.}\ }\textbf {\bibinfo
  {volume} {58}},\ \bibinfo {pages} {012006} (\bibinfo {year}
  {2019})}\BibitemShut {NoStop}%
\bibitem [{\citenamefont {Lydersen}\ \emph
  {et~al.}(2011{\natexlab{c}})\citenamefont {Lydersen}, \citenamefont
  {Makarov},\ and\ \citenamefont {Skaar}}]{lydersen2011a}%
  \BibitemOpen
  \bibfield  {author} {\bibinfo {author} {\bibfnamefont {L.}~\bibnamefont
  {Lydersen}}, \bibinfo {author} {\bibfnamefont {V.}~\bibnamefont {Makarov}}, \
  and\ \bibinfo {author} {\bibfnamefont {J.}~\bibnamefont {Skaar}},\ }\bibfield
   {title} {\enquote {\bibinfo {title} {Secure gated detection scheme for
  quantum cryptography},}\ }\href {\doibase 10.1103/PhysRevA.83.032306}
  {\bibfield  {journal} {\bibinfo  {journal} {Phys. Rev. A}\ }\textbf {\bibinfo
  {volume} {83}},\ \bibinfo {pages} {032306} (\bibinfo {year}
  {2011}{\natexlab{c}})}\BibitemShut {NoStop}%
\bibitem [{\citenamefont {Mar{\o}y}\ \emph {et~al.}(2017)\citenamefont
  {Mar{\o}y}, \citenamefont {Makarov},\ and\ \citenamefont
  {Skaar}}]{maroy2017}%
  \BibitemOpen
  \bibfield  {author} {\bibinfo {author} {\bibfnamefont {{\O}ystein}\
  \bibnamefont {Mar{\o}y}}, \bibinfo {author} {\bibfnamefont {Vadim}\
  \bibnamefont {Makarov}}, \ and\ \bibinfo {author} {\bibfnamefont {Johannes}\
  \bibnamefont {Skaar}},\ }\bibfield  {title} {\enquote {\bibinfo {title}
  {Secure detection in quantum key distribution by real-time calibration of
  receiver},}\ }\href {\doibase 10.1088/2058-9565/aa83c9} {\bibfield  {journal}
  {\bibinfo  {journal} {Quantum Sci. Technol.}\ }\textbf {\bibinfo {volume}
  {2}},\ \bibinfo {pages} {044013} (\bibinfo {year} {2017})}\BibitemShut
  {NoStop}%
\bibitem [{\citenamefont {Lee}\ \emph {et~al.}(2016)\citenamefont {Lee},
  \citenamefont {Park}, \citenamefont {Woo}, \citenamefont {Park},
  \citenamefont {Kim}, \citenamefont {Han},\ and\ \citenamefont
  {Moon}}]{lee2016}%
  \BibitemOpen
  \bibfield  {author} {\bibinfo {author} {\bibfnamefont {Min~Soo}\ \bibnamefont
  {Lee}}, \bibinfo {author} {\bibfnamefont {Byung~Kwon}\ \bibnamefont {Park}},
  \bibinfo {author} {\bibfnamefont {Min~Ki}\ \bibnamefont {Woo}}, \bibinfo
  {author} {\bibfnamefont {Chang~Hoon}\ \bibnamefont {Park}}, \bibinfo {author}
  {\bibfnamefont {Yong-Su}\ \bibnamefont {Kim}}, \bibinfo {author}
  {\bibfnamefont {Sang-Wook}\ \bibnamefont {Han}}, \ and\ \bibinfo {author}
  {\bibfnamefont {Sung}\ \bibnamefont {Moon}},\ }\bibfield  {title} {\enquote
  {\bibinfo {title} {Countermeasure against blinding attacks on low-noise
  detectors with a background-noise-cancellation scheme},}\ }\href {\doibase
  10.1103/PhysRevA.94.062321} {\bibfield  {journal} {\bibinfo  {journal} {Phys.
  Rev. A}\ }\textbf {\bibinfo {volume} {94}},\ \bibinfo {pages} {062321}
  (\bibinfo {year} {2016})}\BibitemShut {NoStop}%
\bibitem [{\citenamefont {Qian}\ \emph {et~al.}(2019)\citenamefont {Qian},
  \citenamefont {He}, \citenamefont {Wang}, \citenamefont {Chen}, \citenamefont
  {Yin}, \citenamefont {Guo},\ and\ \citenamefont {Han}}]{qian2019}%
  \BibitemOpen
  \bibfield  {author} {\bibinfo {author} {\bibfnamefont {Yong-Jun}\
  \bibnamefont {Qian}}, \bibinfo {author} {\bibfnamefont {De-Yong}\
  \bibnamefont {He}}, \bibinfo {author} {\bibfnamefont {Shuang}\ \bibnamefont
  {Wang}}, \bibinfo {author} {\bibfnamefont {Wei}\ \bibnamefont {Chen}},
  \bibinfo {author} {\bibfnamefont {Zhen-Qiang}\ \bibnamefont {Yin}}, \bibinfo
  {author} {\bibfnamefont {Guang-Can}\ \bibnamefont {Guo}}, \ and\ \bibinfo
  {author} {\bibfnamefont {Zheng-Fu}\ \bibnamefont {Han}},\ }\bibfield  {title}
  {\enquote {\bibinfo {title} {Robust countermeasure against detector control
  attack in a practical quantum key distribution system},}\ }\href {\doibase
  10.1364/OPTICA.6.001178} {\bibfield  {journal} {\bibinfo  {journal} {Optica}\
  }\textbf {\bibinfo {volume} {6}},\ \bibinfo {pages} {1178--1184} (\bibinfo
  {year} {2019})}\BibitemShut {NoStop}%
\bibitem [{\citenamefont {Lim}\ \emph {et~al.}(2015)\citenamefont {Lim},
  \citenamefont {Walenta}, \citenamefont {Legr\'e}, \citenamefont {Gisin},\
  and\ \citenamefont {Zbinden}}]{lim2015}%
  \BibitemOpen
  \bibfield  {author} {\bibinfo {author} {\bibfnamefont {Charles Ci~Wen}\
  \bibnamefont {Lim}}, \bibinfo {author} {\bibfnamefont {Nino}\ \bibnamefont
  {Walenta}}, \bibinfo {author} {\bibfnamefont {Matthieu}\ \bibnamefont
  {Legr\'e}}, \bibinfo {author} {\bibfnamefont {Nicolas}\ \bibnamefont
  {Gisin}}, \ and\ \bibinfo {author} {\bibfnamefont {Hugo}\ \bibnamefont
  {Zbinden}},\ }\bibfield  {title} {\enquote {\bibinfo {title} {Random
  variation of detector efficiency: a countermeasure against detector blinding
  attacks for quantum key distribution},}\ }\href {\doibase
  10.1109/JSTQE.2015.2389528} {\bibfield  {journal} {\bibinfo  {journal} {IEEE
  J. Sel. Top. Quantum Electron.}\ }\textbf {\bibinfo {volume} {21}},\ \bibinfo
  {pages} {6601305} (\bibinfo {year} {2015})}\BibitemShut {NoStop}%
\bibitem [{\citenamefont {da~Silva}\ \emph {et~al.}(2015)\citenamefont
  {da~Silva}, \citenamefont {do~Amaral}, \citenamefont {Xavier}, \citenamefont
  {Tempor\^{a}o},\ and\ \citenamefont {von~der Weid}}]{silva2015}%
  \BibitemOpen
  \bibfield  {author} {\bibinfo {author} {\bibfnamefont {Thiago~Ferreira}\
  \bibnamefont {da~Silva}}, \bibinfo {author} {\bibfnamefont {Gustavo~C.}\
  \bibnamefont {do~Amaral}}, \bibinfo {author} {\bibfnamefont {Guilherme~B.}\
  \bibnamefont {Xavier}}, \bibinfo {author} {\bibfnamefont {Guilherme~P.}\
  \bibnamefont {Tempor\^{a}o}}, \ and\ \bibinfo {author} {\bibfnamefont
  {Jean~Pierre}\ \bibnamefont {von~der Weid}},\ }\bibfield  {title} {\enquote
  {\bibinfo {title} {Safeguarding quantum key distribution through detection
  randomization},}\ }\href {\doibase 10.1109/JSTQE.2014.2361793} {\bibfield
  {journal} {\bibinfo  {journal} {IEEE J. Sel. Top. Quantum Electron.}\
  }\textbf {\bibinfo {volume} {21}},\ \bibinfo {pages} {6600309} (\bibinfo
  {year} {2015})}\BibitemShut {NoStop}%
\bibitem [{\citenamefont {Huang}\ \emph {et~al.}(2016)\citenamefont {Huang},
  \citenamefont {Sajeed}, \citenamefont {Chaiwongkhot}, \citenamefont
  {Soucarros}, \citenamefont {Legr{\' e}},\ and\ \citenamefont
  {Makarov}}]{huang2016}%
  \BibitemOpen
  \bibfield  {author} {\bibinfo {author} {\bibfnamefont {Anqi}\ \bibnamefont
  {Huang}}, \bibinfo {author} {\bibfnamefont {Shihan}\ \bibnamefont {Sajeed}},
  \bibinfo {author} {\bibfnamefont {Poompong}\ \bibnamefont {Chaiwongkhot}},
  \bibinfo {author} {\bibfnamefont {Mathilde}\ \bibnamefont {Soucarros}},
  \bibinfo {author} {\bibfnamefont {Matthieu}\ \bibnamefont {Legr{\' e}}}, \
  and\ \bibinfo {author} {\bibfnamefont {Vadim}\ \bibnamefont {Makarov}},\
  }\bibfield  {title} {\enquote {\bibinfo {title} {Testing
  random-detector-efficiency countermeasure in a commercial system reveals a
  breakable unrealistic assumption},}\ }\href {\doibase
  10.1109/JQE.2016.2611443} {\bibfield  {journal} {\bibinfo  {journal} {IEEE J.
  Quantum Electron.}\ }\textbf {\bibinfo {volume} {52}},\ \bibinfo {pages}
  {8000211} (\bibinfo {year} {2016})}\BibitemShut {NoStop}%
\bibitem [{\citenamefont {Tutt}\ and\ \citenamefont
  {Boggess}(1993)}]{tutt1993}%
  \BibitemOpen
  \bibfield  {author} {\bibinfo {author} {\bibfnamefont {Lee~W.}\ \bibnamefont
  {Tutt}}\ and\ \bibinfo {author} {\bibfnamefont {Thomas~F.}\ \bibnamefont
  {Boggess}},\ }\bibfield  {title} {\enquote {\bibinfo {title} {A review of
  optical limiting mechanisms and devices using organics, fullerenes,
  semiconductors and other materials},}\ }\href {\doibase
  10.1016/0079-6727(93)90004-S} {\bibfield  {journal} {\bibinfo  {journal}
  {Prog. Quantum Electron.}\ }\textbf {\bibinfo {volume} {17}},\ \bibinfo
  {pages} {299--338} (\bibinfo {year} {1993})}\BibitemShut {NoStop}%
\bibitem [{\citenamefont {DeRosa}\ and\ \citenamefont
  {Logunov}(2003)}]{derosa2003}%
  \BibitemOpen
  \bibfield  {author} {\bibinfo {author} {\bibfnamefont {Michael~E.}\
  \bibnamefont {DeRosa}}\ and\ \bibinfo {author} {\bibfnamefont {Stephan~L.}\
  \bibnamefont {Logunov}},\ }\bibfield  {title} {\enquote {\bibinfo {title}
  {Fiber-optic power limiter based on photothermal defocusing in an optical
  polymer},}\ }\href {\doibase 10.1364/AO.42.002683} {\bibfield  {journal}
  {\bibinfo  {journal} {Appl. Opt.}\ }\textbf {\bibinfo {volume} {42}},\
  \bibinfo {pages} {2683} (\bibinfo {year} {2003})}\BibitemShut {NoStop}%
\bibitem [{\citenamefont {Martincek}\ and\ \citenamefont
  {Pudis}(2012)}]{martincek2012}%
  \BibitemOpen
  \bibfield  {author} {\bibinfo {author} {\bibfnamefont {Ivan}\ \bibnamefont
  {Martincek}}\ and\ \bibinfo {author} {\bibfnamefont {Dusan}\ \bibnamefont
  {Pudis}},\ }\bibfield  {title} {\enquote {\bibinfo {title} {Fiber-optical
  power limiter and cut-off switch based on thermo-optical effect},}\ }\href
  {\doibase 10.1109/LPT.2011.2177655} {\bibfield  {journal} {\bibinfo
  {journal} {IEEE Photon. Technol. Lett.}\ }\textbf {\bibinfo {volume} {24}},\
  \bibinfo {pages} {297--299} (\bibinfo {year} {2012})}\BibitemShut {NoStop}%
\bibitem [{\citenamefont {Zhang}\ \emph {et~al.}(2021)\citenamefont {Zhang},
  \citenamefont {Primaatmaja}, \citenamefont {Haw}, \citenamefont {Gong},
  \citenamefont {Wang},\ and\ \citenamefont {Lim}}]{zhang2021}%
  \BibitemOpen
  \bibfield  {author} {\bibinfo {author} {\bibfnamefont {Gong}\ \bibnamefont
  {Zhang}}, \bibinfo {author} {\bibfnamefont {Ignatius~William}\ \bibnamefont
  {Primaatmaja}}, \bibinfo {author} {\bibfnamefont {Jing~Yan}\ \bibnamefont
  {Haw}}, \bibinfo {author} {\bibfnamefont {Xiao}\ \bibnamefont {Gong}},
  \bibinfo {author} {\bibfnamefont {Chao}\ \bibnamefont {Wang}}, \ and\
  \bibinfo {author} {\bibfnamefont {Charles Ci~Wen}\ \bibnamefont {Lim}},\
  }\bibfield  {title} {\enquote {\bibinfo {title} {Securing practical quantum
  communication systems with optical power limiters},}\ }\href {\doibase
  10.1103/PRXQuantum.2.030304} {\bibfield  {journal} {\bibinfo  {journal} {PRX
  Quantum}\ }\textbf {\bibinfo {volume} {2}},\ \bibinfo {pages} {030304}
  (\bibinfo {year} {2021})}\BibitemShut {NoStop}%
\bibitem [{\citenamefont {Jogenfors}\ \emph {et~al.}(2015)\citenamefont
  {Jogenfors}, \citenamefont {Elhassan}, \citenamefont {Ahrens}, \citenamefont
  {Bourennane},\ and\ \citenamefont {Larsson}}]{jogenfors2015}%
  \BibitemOpen
  \bibfield  {author} {\bibinfo {author} {\bibfnamefont {Jonathan}\
  \bibnamefont {Jogenfors}}, \bibinfo {author} {\bibfnamefont {Ashraf~Mohamed}\
  \bibnamefont {Elhassan}}, \bibinfo {author} {\bibfnamefont {Johan}\
  \bibnamefont {Ahrens}}, \bibinfo {author} {\bibfnamefont {Mohamed}\
  \bibnamefont {Bourennane}}, \ and\ \bibinfo {author} {\bibfnamefont
  {Jan-{\AA}ke}\ \bibnamefont {Larsson}},\ }\bibfield  {title} {\enquote
  {\bibinfo {title} {Hacking the {B}ell test using classical light in
  energy-time entanglement--based quantum key distribution},}\ }\href {\doibase
  10.1126/sciadv.1500793} {\bibfield  {journal} {\bibinfo  {journal} {Sci.
  Adv.}\ }\textbf {\bibinfo {volume} {1}},\ \bibinfo {pages} {e1500793}
  (\bibinfo {year} {2015})}\BibitemShut {NoStop}%
\bibitem [{Note1()}]{Note1}%
  \BibitemOpen
  \bibinfo {note} {The quartz fiber transparency range is estimated
  approximately, since the boundaries of this range significantly depend on the
  fiber type and the technology of its production. The long-wavelength cutoff
  is determined by the radius of curvature of installed fiber, which is
  device-specific. We are not aware of reliable data in the
  literature.}\BibitemShut {Stop}%
\bibitem [{Thorlabs, IO-H-1550 fiber-optic isolator isolation
  plot()}]{Thorlabs_IO-H-1550_spc_sheet}%
  \BibitemOpen
  Thorlabs, IO-H-1550 fiber-optic isolator isolation plot,\ \href@noop {}
  {}\bibinfo {note}
  {\url{https://www.thorlabs.com/images/tabImages/IO-H-1550_780.gif}, visited
  16 April 2022}\BibitemShut {NoStop}%
\bibitem [{Thorlabs, FA20T fiber-optic attenuator attenuation
  plot()}]{Thorlabs_FA20T_spc_sheet}%
  \BibitemOpen
  Thorlabs, FA20T fiber-optic attenuator attenuation plot,\ \href@noop {}
  {}\bibinfo {note}
  {\url{https://www.thorlabs.com/images/TabImages/FA20T_780.gif}, visited 16
  April 2022}\BibitemShut {NoStop}%
\bibitem [{Opneti, 100Ghz Dense Wavelength Division Multiplexer data
  sheet()}]{Opneti_fiber-optic_DWDM_specification_sheet_sheet}%
  \BibitemOpen
  Opneti, 100Ghz Dense Wavelength Division Multiplexer data sheet,\ \href@noop
  {} {}\bibinfo {note}
  {\url{http://www.opneti.com/uploadfile/20101208/20101208120646747.pdf},
  visited 16 April 2022}\BibitemShut {NoStop}%
\bibitem [{\citenamefont {Tan}\ \emph {et~al.}(2024)\citenamefont {Tan},
  \citenamefont {Petrov}, \citenamefont {Zhang}, \citenamefont {Han},
  \citenamefont {Liao}, \citenamefont {Makarov},\ and\ \citenamefont
  {Xu}}]{tan2024}%
  \BibitemOpen
  \bibfield  {author} {\bibinfo {author} {\bibfnamefont {Hao}\ \bibnamefont
  {Tan}}, \bibinfo {author} {\bibfnamefont {Mikhail}\ \bibnamefont {Petrov}},
  \bibinfo {author} {\bibfnamefont {Weiyang}\ \bibnamefont {Zhang}}, \bibinfo
  {author} {\bibfnamefont {Liying}\ \bibnamefont {Han}}, \bibinfo {author}
  {\bibfnamefont {Sheng-Kai}\ \bibnamefont {Liao}}, \bibinfo {author}
  {\bibfnamefont {Vadim}\ \bibnamefont {Makarov}}, \ and\ \bibinfo {author}
  {\bibfnamefont {Feihu}\ \bibnamefont {Xu}},\ }\href@noop {} {\enquote
  {\bibinfo {title} {Wide-spectrum security against attacks in quantum key
  distribution},}\ } (\bibinfo {year} {2024}),\ \bibinfo {note}
  {{u}npublished}\BibitemShut {NoStop}%
\end{thebibliography}
\end{document}